\documentclass[notitlepage,aps,prd,twocolumn,superscriptaddress,floatfix, nofootinbib]{revtex4-2}

\usepackage[utf8]{inputenc}
\pdfoutput=1

\usepackage{enumitem}
\usepackage{braket}
\usepackage{hyperref}
\usepackage{comment}
\usepackage{amsmath,amssymb,amsfonts,mathrsfs,amsthm}
\usepackage{booktabs}
\usepackage{mathtools}
\usepackage{bm}
\usepackage{dcolumn}
\usepackage{graphicx}   
\usepackage{latexsym}
\usepackage[dvipsnames]{xcolor}
\usepackage{colortbl}
\usepackage{soul}
\usepackage{manfnt,nicefrac}
\usepackage{scalerel, stackengine}
\usepackage{bbm}    
\usepackage{accents} 
\usepackage[makeroom]{cancel}
\usepackage[framemethod=tikz]{mdframed}
\usepackage[most]{tcolorbox}
\usepackage{cleveref}
\usepackage{color}
\usetikzlibrary{arrows}
\usepackage[acronym]{glossaries}

\usepackage[english]{babel}

\hypersetup{
  colorlinks=true
}

\definecolor{LBlue}{rgb}{0.5,0.9,0.5}
\definecolor{LCyan}{rgb}{0.6,1,1}
\definecolor{LRed}{rgb}{0.9,0.6,0.7}
\definecolor{LGreen}{rgb}{0.98, 0.93, 0.36}

\newcommand\be{\begin{equation}}
\newcommand\ba{\begin{eqnarray}}
\newcommand\ee{\end{equation}}
\newcommand\ea{\end{eqnarray}}
\newtheorem*{remark}{Conjecture}

\newcommand{\scri}{{\cal I}}
\newcommand{\RNum}[1]{\uppercase\expandafter{\romannumeral #1\relax}}

\newcommand{\dd}{\mathrm{d}}

\newcommand{\ud}[2]{^{#1}{}_{#2}}
\newcommand{\du}[2]{_{#1}{}^{#2}}

\newcommand{\lie}{\mathcal{L}}

\newcommand{\bt}{\boldsymbol{\theta}}
\newcommand{\bT}{\boldsymbol{\Theta}}
\newcommand{\bw}{\boldsymbol{\omega}}

\newcommand{\gd}{g_{\mu\nu}}
\newcommand{\hatgd}{\hat g_{\mu\nu}}

\newcommand{\orcid}[1]{\href{https://orcid.org/#1}{\includegraphics[width=8pt]{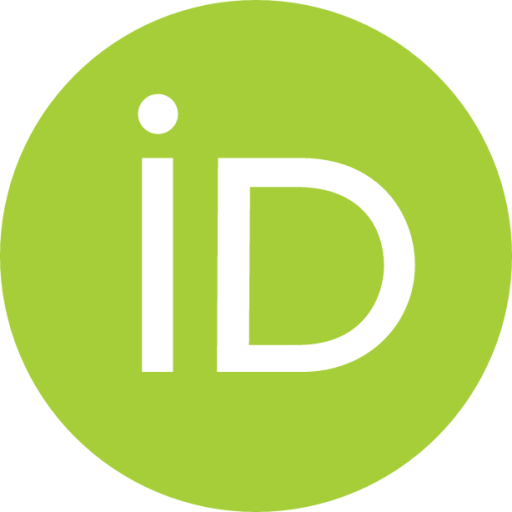}}}

\newcommand*{\rom}[1]{\expandafter\@slowromancap\romannumeral #1@}

\makeatletter
\newcommand{\dalembertian}{\mathop{\mathpalette\dalembertian@\relax}}
\newcommand{\dalembertian@}[2]{%
  \begingroup
  \sbox\z@{$\m@th#1\square$}%
  \dimen0=\fontdimen8
    \ifx#1\displaystyle\textfont\else
    \ifx#1\textstyle\textfont\else
    \ifx#1\scriptstyle\scriptfont\else
    \scriptscriptfont\fi\fi\fi3
  \makebox[\wd\z@]{%
    \hbox to \ht\z@{%
      \vrule width \dimen0
      \kern-\dimen0
      \vbox to \ht\z@{
        \hrule height \dimen0 width \ht\z@
        \vss
        \hrule height 2\dimen0
      }%
      \kern-2.5\dimen0
      \vrule width 2.5\dimen0
    }%
  }%
  \endgroup
}
\makeatother

\makeatletter
\newcommand*{\pgfunderleftarrow}{%
  \@ifstar
    {\let\ifpgf@depth\iftrue\mathpalette\@pgfunderleftarrow}
    {\let\ifpgf@depth\iffalse\mathpalette\@pgfunderleftarrow}%
}
\newcommand*{\@pgfunderleftarrow}[2]{%
  #2%
  \edef\pgf@math@fam{\the\fam}%
  \pgfpicture
    \pgfsetbaseline{0pt}
    \pgf@relevantforpicturesizefalse      
    \pgfsetroundcap                       
    \pgfsetarrowsend{left to}
    \pgfutil@tempdima=0.28pt%
    \advance\pgfutil@tempdima by.8\pgflinewidth%
    \pgfutil@tempdima-4\pgfutil@tempdima
    \sbox\pgfutil@tempboxa{$\m@th\fam\pgf@math@fam#1#2$}%
    \advance\pgfutil@tempdima-\dp\pgfutil@tempboxa
    \pgfutil@tempdimb\wd\pgfutil@tempboxa
    \pgfpathmoveto{\pgfqpoint{0pt}{\pgfutil@tempdima}}%
    \pgfpathlineto{\pgfqpoint{-\pgfutil@tempdimb}{\pgfutil@tempdima}}%
    \pgfusepath{stroke}
    \ifpgf@depth
      \pgf@relevantforpicturesizetrue
      \pgfpathmoveto{\pgfqpoint{0pt}{-\pgfutil@tempdimb}}%
      \pgfusepath{use as bounding box}%
    \fi
  \endpgfpicture
}
\makeatother

\makeatletter
\newcommand*{\pgfunderrightarrow}{%
  \@ifstar
    {\let\ifpgf@depth\iftrue\mathpalette\@pgfunderrightarrow}
    {\let\ifpgf@depth\iffalse\mathpalette\@pgfunderrightarrow}%
}
\newcommand*{\@pgfunderrightarrow}[2]{%
  #2%
  \edef\pgf@math@fam{\the\fam}%
  \pgfpicture
    \pgfsetbaseline{0pt}
    \pgf@relevantforpicturesizefalse      
    \pgfsetroundcap                       
    \pgfsetarrowsend{right to}
    \pgfutil@tempdima=0.28pt%
    \advance\pgfutil@tempdima by.8\pgflinewidth%
    \pgfutil@tempdima-4\pgfutil@tempdima
    \sbox\pgfutil@tempboxa{$\m@th\fam\pgf@math@fam#1#2$}%
    \advance\pgfutil@tempdima-\dp\pgfutil@tempboxa
    \pgfutil@tempdimb\wd\pgfutil@tempboxa
    \pgfpathmoveto{\pgfqpoint{-\pgfutil@tempdimb}{\pgfutil@tempdima}}%
    \pgfpathlineto{\pgfqpoint{0pt}{\pgfutil@tempdima}}%
    \pgfusepath{stroke}
    \ifpgf@depth
      \pgf@relevantforpicturesizetrue
      \pgfpathmoveto{\pgfqpoint{0pt}{-\pgfutil@tempdimb}}%
      \pgfusepath{use as bounding box}%
    \fi
  \endpgfpicture
}
\makeatother

\newcommand{\newpb}[1]{\pgfunderleftarrow{#1}}

\makeglossaries
\newacronym{gw}{GW}{Gravitational Wave}

\def\scri{\mathscr{I}}
\def\scrip{\scri^{+}}

\newcommand{\equaldot}{\,\dot{=}\,}

\newcommand{\jz}[1]{\textcolor{orange}{[jz: #1]}}

\newacronym{gr}{GR}{general relativity}

\newif\ifcollab

\collabtrue

\ifcollab
\newcommand{\fadeout}[1]{{\small\color{black!40}#1}}

\else

\newcommand{\fadeout}[1]{} 
\newcommand{\st}[1]{} 

\fi

\begin{document}

\newpage


\title{Balance flux laws beyond general relativity}

\author{David Maibach \orcid{0000-0002-5294-464X}}
\email{d.maibach@thphys.uni-heidelberg.de}
\affiliation{Institute for Theoretical Physics, University of Heidelberg, Philosophenweg 16,
D-69120	Heidelberg,
Germany} 

\author{Jann Zosso \orcid{0000-0002-2671-7531}}
\email{jann.zosso@nbi.ku.dk}
\affiliation{Center of Gravity, Niels Bohr Institute, Blegdamsvej 17, 2100 Copenhagen, Denmark}

\date{\today}


\begin{abstract}
Balance flux laws of asymptotic symmetries in general relativity provide fully non-perturbative constraint equations on gravitational strain. They have proven useful for constructing numerical gravitational waveforms and for characterizing gravitational memory. As the precision of current and future detectors continues to improve, such constraints become increasingly important for high-precision tests of gravity, including searches for deviations from general relativity. This motivates a systematic understanding of analogous balance laws in theories beyond general relativity.
In this work, we investigate the existence and structure of flux laws at null infinity in diffeomorphism-invariant extensions of general relativity. Our analysis is based on the covariant phase space formalism and the definition of conserved quantities, as presented by Wald and Zoupas. For a particularly relevant class of Horndeski theories, we derive a general expression for the flux and formulate the corresponding balance equation via the associated non-conserved charges. We cross-check our general results by comparing with previous studies of Brans–Dicke gravity. Furthermore, we demonstrate that the employed methods extend straightforwardly to scalar-Gauss-Bonnet gravity and provide a conjecture of the flux balance laws for full massless Horndeski theory.
The null part of the resulting flux laws associated with null memory is compared with and validated against the alternative derivation based on the Isaacson approach to gravitational radiation. Beyond the specific results obtained, this work is intended to serve as a practical guide for computing balance laws in generic diffeomorphism-invariant theories of gravity and paves the way for an in-depth comparison between the Isaacson approach and the covariant phase space formalism.

\end{abstract}



\maketitle


\newcommand{\myhyperref}[1]{\hyperref[#1]{\ref{#1}}}




\section{Introduction} 
\label{sec:intro}
The first direct detection of gravitational waves (GWs) from the merger of two black holes in 2015~\cite{PhysRevLett.116.061102} marked a key breakthrough for gravitational physics. It confirmed a central prediction of Einstein's general relativity (GR) and not only ushered in a new era of GW astronomy, but also provides a powerful observational probe of the strong-field, highly dynamical regime of gravity. GW observations encode detailed information about the source, including the masses and spins of compact objects, their orbital dynamics, and the distance and orientation of the system. Extracting this information with high fidelity provides both new information on the astrophysical environments of the cosmos and enables precision tests of GR that opens a new window onto fundamental physics.

Parameter estimation and theory testing in GW physics rely on the comparison of detector data with accurate theoretical waveform templates. These templates must span a high-dimensional parameter space and faithfully represent the dynamics predicted by the underlying theory of gravity. The accuracy of inferred source parameters depends sensitively on the precision of the waveform models, the signal-to-noise ratio (SNR) of the detected signal, and the complexity of the parameter space explored in Bayesian inference. This demand for accuracy will become increasingly stringent with next-generation detectors such as LISA~\cite{colpi_lisa_2024}, the Einstein Telescope~\cite{maggiore_science_2020}, and the Cosmic Explorer~\cite{evans_cosmic_2023}, whose enhanced sensitivity will allow for the observation of new fundamental physics signals, including gravitational memory~\cite{zeldovichRadiationGravitationalWaves1974, christodoulouNonlinearNatureGravitation1991a, thorneGravitationalwaveBurstsMemory1992, PhysRevD.44.R2945, PhysRevD.80.024002, favataNONLINEARGRAVITATIONALWAVEMEMORY2009, Inchauspe:2024ibs,Cogez:2026frh,Zosso:2026czc}, and potentially signatures of physics beyond GR~\cite{Heisenberg:2018vsk}.

Within GR, balance flux laws have proven to be particularly valuable tools for both analytical understanding and numerical waveform modeling. These laws relate changes in conserved quantities, such as energy, momentum, and angular momentum, to fluxes through null infinity, thereby providing exact constraint equations for the emitted radiation. A major breakthrough in their formulation was achieved by Ashtekar and Streubel~\cite{Ashtekar:1981}, who identified well-defined radiative degrees of freedom at null infinity. This work was later placed on a rigorous and general footing through the covariant phase space formalism~\cite{Lee:1990nz, Wald:1993nt, Iyer_1994, Iyer_1995} and the construction of conserved quantities with fluxes by Wald and Zoupas~\cite{waldGeneralDefinitionConserved2000}. When applied to null infinity, their framework reproduces the Ashtekar–Streubel results and provides a systematic method for deriving balance laws in diffeomorphism-invariant theories.

In practical applications, balance flux laws play an important role in GW data analysis and numerical relativity. They enable the post-processing of numerical waveforms, allowing for the consistent inclusion of gravitational memory effects~\cite{Mitman:2020bjf, Khera:2020mcz} and for detailed, mode-by-mode consistency checks of simulated strain data against the full nonlinear theory~\cite{Borchers:2021vyw, Borchers_2023, DAmbrosio:2024zok}. Such techniques enhance the physical fidelity of waveform templates and help control systematic errors in parameter estimation~\cite{Ashtekar:2019viz, Heisenberg:2023mdz, Deppe:2025pvd}. To date, however, the overwhelming majority of waveform modeling and balance-law applications have been developed within GR.

As observational precision improves, this GR-centric focus becomes a potential source of fundamental bias. If the true theory of gravity deviates even slightly from GR, parameter estimation based exclusively on GR templates may systematically misinterpret the data. Robust tests of gravity therefore require waveform templates of comparable accuracy in both GR and well-motivated beyond-GR theories. Yet, constructing such templates is challenging: modified gravity theories often introduce additional degrees of freedom and a large number of free coupling parameters, making systematic numerical simulations across the full theory space impractical. Recent progress in numerical relativity beyond GR~\cite{PhysRevD.107.044003, PhysRevD.108.024006, Luna_2024, AresteSalo:2025sxc, Brady:2023dgu, Fell:2024wdd} represents an important step forward, but scalable strategies for exploring extended parameter spaces remain urgently needed.

In this context, balance flux laws beyond GR offer a promising avenue. For a broad class of dynamical metric theories, conserved quantities can again be defined in the presence of radiation, leading to generalized balance laws that typically decompose into a GR-like metric contribution and additional terms associated with new fields~\cite{Hou:2021bxz, hou_gravitational_2021, hou_gravitational_2021_2}. These additional contributions imprint characteristic corrections on both the waveform and the associated gravitational memory~\cite{Heisenberg:2023prj,heisenberg2025unifyingordinarynullmemory}. Such relations can serve as powerful constraint equations, enabling the post-processing of waveform data and potentially reducing the need for exhaustive numerical simulations across large theory spaces. In this way, balance flux laws may enhance the discriminatory power of GW observations and help identify signatures of new gravitational physics.

Beyond their practical utility, balance flux laws are also of conceptual interest. In GR, they provide direct access to non-conserved charges and gravitational memory at null infinity and play a central role in the infrared structure of gravity, flat-space holography, and the so-called infrared triangle~\cite{Strominger:2014pwa, Pasterski:2015tva, strominger2018lecturesinfraredstructuregravity}. For beyond-GR theories, analogous structures are largely unexplored and may reveal qualitatively new features that lie outside the traditional Einsteinian framework.

In this work, we investigate how balance flux laws arise in beyond-GR theories in asymptotically flat spacetimes. We do so by explicitly performing computations within a broad subclass of Horndeski gravity \cite{Horndeski:1974wa} and discuss an extension of our results to the full Horndeski theory. We thereby generalize earlier studies that focused on restricted classes of theories or isolated aspects of gravitational memory~\cite{hou_conserved_2021, Tahura_2025, hou_gravitational_2021_2, Hou:2021bxz, Heisenberg:2023prj, heisenberg2025unifyingordinarynullmemory,Tahura:2025ebb}. Our analysis is based on the covariant phase space formalism~\cite{Lee:1990nz, Wald:1993nt, Iyer_1994, Iyer_1995} and the definition of conserved Hamiltonians with fluxes developed by Wald and Zoupas~\cite{waldGeneralDefinitionConserved2000}.

The paper is organized as follows. In Sec.~\ref{sec:intuition}, we review the covariant phase space formalism and its relation to conserved quantities in diffeomorphism-invariant theories, with particular emphasis on spacetimes with boundaries. The Wald–Zoupas construction is summarized in Sec.~\ref{subsec:WZ_I} and \ref{subsec:WZ_II}. In Sec.~\ref{sec:BL_all}, we introduce the Horndeski theory under consideration and compute the relevant symplectic structures. The asymptotic framework at null infinity, based on conformal compactification and an appropriate foliation of $\scrip$, is developed in Sec.~\ref{subsec:null_infty_shit}, with further details provided in Appendix~\ref{app:asymptotic_Horn}. We then derive the flux formula for luminal Horndeski theory and formulate the corresponding balance flux law in Sec.~\ref{subsec:balance_law_equation}. Finally, Sec.~\ref{sec:BDT} presents a concrete example within the class of Horndeski-type theories and we discuss the physical implications of our results in Sec.~\ref{sec:discuss}.

\section{Intuition and Notation}
\label{sec:intuition}

Since the subject of flux balance laws in asymptotically flat spacetimes is traditionally
heavy in mathematical definitions, we begin by developing an intuitive understanding of
the underlying structures. This section also serves to introduce the notation and clarify
the geometric meaning of conserved quantities in both classical mechanics and
diffeomorphism-invariant field theories.

In theories defined on a fixed background spacetime, such as Minkowski or Euclidean space,
continuous spacetime symmetries give rise to conservation laws through Noether’s theorem.
Each global symmetry corresponds to a conserved current, whose integral over a spatial
slice yields an associated conserved charge. In this setting, the symmetries are
properties of the background geometry itself and exist independently of the dynamical
state of the fields.
These statements change fundamentally in the context of gravity, where the spacetime
metric is itself a dynamical field. To motivate the covariant phase space formalism used
throughout this work, it is therefore useful to recall how symplectic geometry and
conserved quantities already arise in classical mechanics, where the relevant structures
appear in a finite-dimensional and conceptually transparent setting.

\subsection{Symplectic structure in classical mechanics}\label{sSec: CM intuition}

In classical mechanics, a system with $N$ degrees of freedom is described by a
$2N$-dimensional phase space spanned by position and momentum coordinates $(q_i,p_i)$. At
a given time $t_0$, specifying all $q_i(t_0)$ and $p_i(t_0)$ uniquely determines the state
of the system. Yet, the variables $q_i$ and $p_i \sim \dot q_i$ are not independent and their
relation is encoded in the Lagrangian and its variation, leading to the Euler-Lagrange
equations and the associated Noether currents.

This interdependence admits a natural geometric reformulation. The evolution of the system
is described by a phase space trajectory
\[
  \gamma(t) = \big(q_i(t),p_i(t)\big),
\]
which is an integral curve of a vector field $X_H$ on phase space. If the Hamiltonian
$H(q,p)$ is conserved, then $X_H$ is tangent to the level surfaces of constant $H$, and the
trajectory $\gamma(t)$ evolves on a surface of fixed energy. The Hamiltonian thus
generates the flow, while $X_H$ determines the velocity of the trajectory in phase space.

To make this relation precise, phase space must be equipped with a geometric structure
that determines how a function $H$ gives rise to a vector field $X_H$. This structure is
the \emph{symplectic form},
\begin{equation}\label{eq:SymplecticForm1}
  \Omega = \delta p_i \wedge \delta q^i ,
\end{equation}
a closed, non-degenerate two-form encoding the canonical Poisson brackets
$\{q_i,p_j\}=\delta_{ij}$. Here, $\delta$ denotes the exterior derivative on phase space,
acting on functions of the coordinates $(q_i,p_i)$.

The Hamiltonian vector field $X_H$ is defined implicitly by
\begin{equation}\label{eq:HamiltonianVectorField}
  \iota_{X_H}\Omega = -\,\delta H ,
\end{equation}
where $\iota_{X_H}$ denotes the \emph{interior contraction} of the vector field $X_H$ with
the differential form $\Omega$. Explicitly, $\iota_{X_H}\Omega$ is the one-form obtained by
inserting $X_H$ into the first argument of $\Omega$, thereby mapping the two-form
$\Omega$ to a one-form.

The symplectic form also encodes the fundamental conservation property of Hamiltonian
dynamics. Using Cartan’s identity,
\[
  \mathcal L_{X_H} = \iota_{X_H}\mathrm d + \mathrm d \iota_{X_H},
\]
where $\mathcal L_{X_H}$ denotes the Lie derivative along $X_H$ and $\mathrm d$ the exterior
derivative on phase space, one finds
\begin{equation}
  \mathcal L_{X_H}\Omega
  = \mathrm d(\iota_{X_H}\Omega)
  = -\,\mathrm d(\delta H)
  = 0 .
\end{equation}
This expresses Liouville’s theorem: Hamiltonian flows preserve the symplectic form and,
consequently, the phase space volume element.

More generally, a vector field $X$ on phase space is said to be \emph{Hamiltonian} if
$\iota_X\Omega$ is an exact one-form, $\iota_X\Omega = -\delta H$ for some globally defined
function $H$. If $\iota_X\Omega$ is closed but not exact, $X$ is called \emph{symplectic}
but does not admit a globally defined Hamiltonian. This distinction between Hamiltonian
and merely symplectic flows will play a central role in the field-theoretic context.

\subsection{Covariant phase space formalism}\label{subsec:SymplecticForms}
\label{subsec:WZ_I}

\subsubsection{From mechanical phase space to covariant phase space}

The covariant phase space formalism generalizes this geometric picture to classical field
theories, in which the spacetime metric is a dynamical
field. In this setting, the analogue of phase space is the infinite-dimensional space of
solutions to the field equations, and the symplectic form is replaced by a closed two-form
constructed covariantly from the Lagrangian. Note that while any field theory can be written in a generally covariant form, the terminology of covariance more specifically refers to a theory in which the spacetime metric $\hat g_{\mu\nu}$ itself is dynamical rather than
a fixed auxiliary structure. In this case the dynamics are formulated on a differentiable spacetime
manifold $\hat{\mathcal M}$ without any fixed background metric, with diffeomorphisms
acting as gauge redundancies of the description.

A key consequence of treating spacetime geometry as dynamical is that conserved
quantities associated with spacetime transformations cannot be defined locally in the
bulk. In particular, GR does not admit a covariant, local
energy--momentum tensor for the gravitational field. As a result, conserved quantities in
diffeomorphism-invariant theories arise as surface integrals rather than volume integrals.
Local conservation laws of the form $\nabla_\mu T^{\mu\nu}=0$ apply only to matter
energy--momentum tensors and do not provide a notion of gravitational energy density.

Nevertheless, vector fields $\xi^\mu$ on spacetime act on the dynamical fields via Lie
derivatives,
\[
  \delta_\xi \hat\psi = \mathcal L_\xi \hat\psi ,
\]
where $\hat\psi$ collectively denotes all dynamical fields. This induces vector fields on
covariant phase space, playing a role directly analogous to the Hamiltonian vector fields
$X_H$ of classical mechanics.

In analogy to Eq.~\eqref{eq:HamiltonianVectorField}, it is then possible to define a field-theoretic symplectic form $\Omega_\Sigma$ on covariant phase space on each Cauchy surface $\Sigma$ related to a Hamiltonian generator $H_\xi$, if it
exists, through
\begin{equation}\label{eq:field_hamiltonian_def}
  \delta H_\xi
  = \iota_{\delta_\xi}\Omega_\Sigma
  = \Omega_\Sigma(\delta\hat\psi,\mathcal L_\xi\hat\psi) .
\end{equation}
Here, $\delta$ denotes the exterior derivative on covariant phase space, and
$\iota_{\delta_\xi}$ is the interior contraction with the vector field on phase space
generated by $\delta_\xi\hat\psi=\mathcal L_\xi\hat\psi$. Hence, if $\xi^\mu$ generates an isometry of the (asymptotic) solution, the one-form
$\delta H_\xi$ on covariant phase space is generally integrable, and its integral defines
a conserved quantity $H_\xi$ associated with $\xi$ on the Cauchy surface $\Sigma$. 
Equivalently, $H_\xi$ generates the flow on covariant phase space corresponding to the
symmetry generated by $\xi^\mu$, in direct analogy with Hamiltonian evolution in
classical mechanics. The conservation of $H_\xi$ then follows from the fact that the
associated Hamiltonian flow preserves the symplectic structure. In the following we will offer a general understanding of the details of this relation.

\subsubsection{Symplectic form}

The symplectic structure of a covariant theory on a spacetime manifold
$\hat{\mathcal M}$ is constructed from the Lagrangian that encodes the dynamics. Consider thus a generic Lagrangian $4$-form
\begin{align}
  \mathbf L
  = \mathbf L(\hat g_{\mu\nu}, \hat R_{\mu\nu\rho\sigma},
  \hat\nabla_\alpha \hat R_{\mu\nu\rho\sigma}, \ldots,
  \hat\Phi, \hat\nabla_\mu \hat\Phi, \ldots) ,
\end{align}
where $\hat\nabla_\mu$ is the torsion- and non-metricity-free covariant derivative
associated with $\hat g_{\mu\nu}$. We collectively denote all dynamical fields by
$\hat\psi\in\{\hat g_{\mu\nu},\hat\Phi\}$.
Variation yields
\begin{align}\label{eq:var_L}
  \delta \mathbf L
  = \mathbf E(\hat\psi)\,\delta\hat\psi
  + \mathrm d \bar{\boldsymbol{\theta}}(\hat\psi,\delta\hat\psi) ,
\end{align}
where $\mathbf E(\hat\psi)=0$ are the equations of motion and
$\bar{\boldsymbol{\theta}}$ is the presymplectic potential. 

The associated presymplectic
current is given by
\begin{align}
  \bar{\boldsymbol{\omega}}(\hat\psi,\delta\hat\psi,\delta'\hat\psi)
  = \delta\bar{\boldsymbol{\theta}}(\hat\psi,\delta'\hat\psi)
  - \delta'\bar{\boldsymbol{\theta}}(\hat\psi,\delta\hat\psi) .
\end{align}
Integrating over a Cauchy surface $\Sigma$ then defines the presymplectic form
\begin{align}\label{eq:presymp_form}
  \bar\Omega_\Sigma = \int_\Sigma \bar{\boldsymbol{\omega}} .
\end{align}
The covariance of the underlying theory here guarantees that the latter expression is valid for any slice, as long as it defines a Cauchy surface.

However, because gauge-redundancies introduced by diffeomorphism invariance, $\bar\Omega_\Sigma$ is degenerate and possesses zero
modes corresponding to gauge directions in field space, hence the ``pre'' terminology. These zero modes can be easily identified at the level of the presymplectic potential by acknowledging that it is defined only up to the addition of an exact form.
Factoring out these zero modes \cite{waldGeneralDefinitionConserved2000} finally
yields the covariant phase space $\Gamma$ equipped with a non-degenerate symplectic form
$\Omega_\Sigma$ used in Eq.~\eqref{eq:field_hamiltonian_def}.

\subsubsection{Relation to Noether current}

It is instructive to rewrite the relation in Eq.~\eqref{eq:field_hamiltonian_def} by using the $3$-form Noether current associated with $\xi^\mu$
\begin{align}
  \mathbf j
  = \bar{\boldsymbol{\theta}}(\hat\psi,\mathcal L_\xi\hat\psi)
  - \xi\cdot\mathbf L ,
\end{align}
such that
\begin{align}
  \delta\mathbf j
  = \boldsymbol{\omega}(\hat\psi,\delta\hat\psi,\mathcal L_\xi\hat\psi)
  + \mathrm d(\xi\cdot\bar{\boldsymbol{\theta}}) .
\end{align}
For diffeomorphism-covariant theories one has \cite{Iyer_1994}
\begin{align}\label{equ:Noether_current}
  \mathbf j = \mathrm d\mathbf Q + \xi^\mu C_\mu ,
\end{align}
where $C_\mu$ are the constraints and $\mathbf Q$ is the Noether charge two-form. On-shell one arrives at
\begin{align}\label{equ:hamiltonian_cps}
  \delta H_\xi
  = \int_{\partial\Sigma}
  \left( \delta\mathbf Q - \xi\cdot\bar{\boldsymbol{\theta}} \right) .
\end{align}
Thus, in diffeomorphism-invariant theories, Hamiltonian generators, and hence conserved
quantities, are determined entirely by boundary data.

\subsection{Radiative boundaries and Wald--Zoupas formalism}
\label{subsec:WZ_II}

Generally, it is also possible to define $\delta H_\xi$ for non-integrable charges. This non-integrability is the field-theoretic analogue of a symplectic but non-Hamiltonian flow in classical mechanics and precisely signals the presence of radiation. The systematic treatment of the associated boundary charges is provided by the Wald--Zoupas (WZ) formalism, which we now briefly sketch.

The WZ formalism is particularly relevant for physically realistic
spacetimes describing gravitational radiation emitted by localized sources, where
non-trivial fluxes arise at the asymptotic boundary of null infinity $\scri$. In this
setting, the variation of the Hamiltonian fails to be integrable due to the presence of
radiative fluxes
\begin{align}
\label{eq:flux_obstruction}
\delta H_\xi\big|_{\partial_2\scri}
-
\delta H_\xi\big|_{\partial_1\scri}
=
-
\int_{\Delta\scri}
\boldsymbol{\omega}
\neq 0 \, .
\end{align}
Here, the $\Delta\scri$ denotes the portion of $\scri$
enclosed by the cross sections $\partial_1\scri$ and $\partial_2\scri$.
The WZ prescription resolves this obstruction by introducing an additional
boundary term,
\begin{align}
\label{equ:why_not}
\int_{\partial\scri} \xi \cdot \bT \, ,
\end{align}
constructed from a symplectic potential $\bT$ intrinsic to $\scri$ and chosen such that
\begin{align}\label{equ:current_PB}
\newpb{\boldsymbol{\omega}}
=
\delta \bT(\hat\psi,\delta'\hat\psi)
-
\delta' \bT(\hat\psi,\delta\hat\psi) \, .
\end{align}
Here, $\newpb{\,\cdot\,}$ denotes the pullback to the asymptotic boundary $\scri$ and
represents one of the crucial ingredients in the practical implementation of the WZ
formalism.

As shown by Wald and Zoupas, the modified Hamiltonian variation
\begin{align}
\label{eq:WZ_Hamiltonian}
\delta \mathcal H_\xi
=
\int_{\partial\scri}
\left(
\delta \mathbf Q_\xi
-
\xi \cdot \bar{\boldsymbol{\theta}}
\right)
-
\int_{\partial\scri}
\xi \cdot \bT
\end{align}
is integrable on covariant phase space and defines a conserved charge, up to an additive
constant.\footnote{This additive constant is fixed upon a suitable choice of a reference solution $\psi_0$ as shown in Ref.~\cite{waldGeneralDefinitionConserved2000}.} The associated flux is given by
\begin{align}\label{equ:Jann_Lfux}
\boldsymbol{F}_\xi
=
\bT(\hat\psi,\mathcal L_\xi \hat\psi) \, ,
\qquad
\mathcal F_\xi[\Delta\scri]
=
\int_{\Delta\scri}
\boldsymbol{F}_\xi \, .
\end{align}
Thus, when an asymptotic symmetry generated by $\xi$ is symplectic but not Hamiltonian,
that is, when $\delta H_\xi$ cannot be integrated to a single-valued function on phase
space, the WZ formalism provides a systematic prescription for constructing an extended,
integrable Hamiltonian that consistently accounts for the presence of non-trivial flux
through the boundary.

While the flux in Eq.~\eqref{equ:Jann_Lfux} is interesting in its own right, in this form it solely describes a property of the (gravitational) radiation. To constrain or test GR or any beyond-GR theory in a meaningful way, the flux has to be converted into a constraint equation, i.e., a flux–balance relation. Furthermore, acknowledging that the flux is associated to a asymptotic symmetry $\xi$ this flux-balance law should describe the non-conservation of a certain charge $Q_\xi$, generated by $\xi$. Given Eqs.~\eqref{eq:flux_obstruction}, formulating such a constraint equation that balances the flux leakage across $\scrip$ with the change in $ \delta \mathcal H_\xi$ is straightforward, provided that $F_{\xi,\Delta\scrip}$ can be written as the integral of an exact $3$--form on $\scrip$. In this case, it follows from Eqs.~\eqref{eq:WZ_Hamiltonian} and \eqref{equ:Jann_Lfux} that 
\begin{align}
   \mathcal F_{\alpha(\theta,\phi)n}[\Delta\scrip]=\int_{\Delta\scrip}\delta \boldsymbol{F}_\xi=-\big[ \delta \mathcal H_\xi |_{\partial_2\scrip}-\delta \mathcal H|_{\partial_1\scrip} \big]\,.
\end{align}
In turn, Eq.~\eqref{equ:Jann_Lfux} relates the physical flux with the symplectic current at null infinity $\scrip$.\footnote{Integrating over a section of $\scrip$ with topology $\mathbb R \times \mathcal S^2$ amounts to $\int_{\Delta\scrip} \equiv \int du \int_{\mathcal{S}^2}$, while a cross section corresponds to a $2$--sphere, i.e., $\int_{\partial\scrip}\equiv \oint_{\mathcal{S}^2}$.} 

Thus, the asymptotic charges $Q_\xi$ that we seek are indeed closely connected to $\delta \mathcal H_\xi$. Indeed, as anticipated in Eq.~\eqref{equ:hamiltonian_cps}, for $\xi$ in which the flux can be written as an exact 3--form, and in particular in cases where $\xi$ is everywhere tangent to $\partial \scrip$, one has \cite{waldGeneralDefinitionConserved2000}
\begin{align}
\delta \mathcal H_\xi= \int_{\partial\scrip} \boldsymbol Q_\xi\,.
\end{align}
The balance equation foliated by retarded time $u$, which relates the flux during a given $u$--interval to the difference of the asymptotic charges defied as integrals over constant-$u$ cross
sections (i.e., over two $2$-spheres) then takes the form 
\begin{align}\label{equ:flalal}
    \mathcal F_{\alpha(\theta,\phi)n}[\Delta\scrip]
    = -\Big[
    Q_{\xi}[\partial_2\scrip]\big|_{u=u_2}
    - Q_{\xi}[\partial_1\scrip]\big|_{u=u_1}
    \Big]\,.
\end{align}
The overall sign reflects the interpretation of the flux as outgoing leakage across null infinity. 
In the following sections, we apply this framework to diffeomorphism-invariant extensions
of GR, focusing on subclasses of Horndeski theories, with the goal of
deriving balance laws relevant for gravitational radiation beyond GR.

\section{Asymptotic flux of luminal Horndeski gravity}\label{sec:BL_all}
\label{sec:Horndeski}
In the following, by hindsight, we compute the balance flux laws using the WZ formalism for a subset of the Horndeski family \cite{Deffayet:2010qz, Deffayet_2011, Kobayashi:2010cm}. Horndeski gravity defines the most general scalar-tensor theory with second-order equations of motion, thus allowing for higher-order derivative operators while avoiding Ostrogradski instabilities \cite{Ostro_Theo}. The selection of the subset is motivated by computational simplicity and relevance in recent literature.

\subsection{Luminal Horndeski gravity}

Consider the particularly well-motivated subset of luminal Horndeski gravity \cite{Kase:2018aps,Zosso:2024xgy}
\begin{equation}\label{equ:luminal_horn}
    S^\ell=\frac{1}{2\kappa_0}\int_{\mathcal M} d^4x\sqrt{-\hat g} L^\ell+S_\text{m}[\hatgd,\hat \Psi_\text{m}]\,,
\end{equation}
where $\kappa_0\equiv 8\pi G$ defines the bare Newton constant, $\hat \Psi_\text{m}$ collectively denotes matter fields universally coupled to the physical metric, and
\begin{equation}\label{equ:general_form}
   L^\ell=G_2( \hat\Phi,\hat X)-G_3( \hat\Phi,\hat X)\hat \dalembertian\hat \Phi+G_4(\hat \Phi)\hat R\,,
\end{equation}
with $\hat X\equiv -\frac{1}{2}\hat g^{\mu\nu}\hat \nabla_\mu \hat\Phi\hat \nabla_\nu \hat\Phi$. All quantities marked by a hat are defined on the physical manifold $\hat{\mathcal M}$. Moreover, we assume that the general functionals $G_i$ are such that the theory remains regular, in particular within the asymptotic limit considered in Sec~\ref{ssSec:AsymptoticFlatness}. A precise definition of this regularity if provided in Appendix~\ref{app:assumptionsonG}. Finally, to also ensure a luminal propagation of the scalar field, we assume its massless-ness through $G_{2,\hat\Phi\hat \Phi}=0$, where the notation defines a second derivative of $G_2$ with respect to the argument $\hat \Phi$. 

The action \eqref{equ:luminal_horn} is obtained from the full Horndeski action,
\begin{align}\label{eq:fullHorndeski}
    S^\text{SVT} = \frac{1}{2\kappa_0}\int \dd^4  x \sqrt{-\hat g} \left(\sum_{i=2}^5L_i\right)\,,
\end{align}
with
\begin{align}
    L_2=& G_2(\hat{\Phi},\hat{X})\,,\\
    L_3=& -G_3(\hat{\Phi},\hat{X})\hat \dalembertian \hat{\Phi} \,,\\
    L_4=&  G_4(\hat{\Phi},\hat{X}) R + G_{4\hat{X}}\left[(\hat{\dalembertian} \hat{\Phi})^2 - \hat{\Phi}^{\mu\nu}\hat{\Phi}_{\mu\nu}\right]\,,\\
    L_5=& G_5(\hat{\Phi},\hat{X})\hat G^{\mu\nu} \hat{\Phi}_{\mu\nu}-\frac{G_{5\hat{X}}}{6}\left[(\hat{\dalembertian} \hat{\Phi})^2\right. \notag\\
    &\left.-3\hat{\dalembertian} \hat{\Phi} \hat{\Phi}^{\mu\nu} \hat{\Phi}_{\mu\nu} + 2 \hat{\Phi}_{\mu\nu}\hat{\Phi}^{\nu\lambda}\hat{\Phi}\du{\lambda}{\mu}\right]\,,    
\end{align}
where $\hat{\Phi}_{\mu\nu}\equiv\hat{\nabla}_\mu\hat{\nabla}_\nu\hat{\Phi}$, $G_{iZ}\equiv \partial G_i/\partial Z$,
by imposing the constraints $ G_{4,X}=0$ and $G_5=\text{const.}$. These constraints preserve a luminal speed of cosmological propagation of tensor modes.

The Lagrangian in Eq.~\eqref{equ:general_form} contains a term that resembles the Einstein--Hilbert term, up to a non-minimal coupling to the scalar field. For the subsequent computations, it is therefore desirable to apply a transformation that isolates a pure Ricci scalar term, allowing one to recycle well-established results from GR.  
Indeed, this is achieved through a Weyl rescaling of the form
\begin{equation}\label{equ:rescaling_horndeski}
    \hat g_{\mu\nu}(x)=\frac{1}{G_4(\hat\Phi)}\tilde g_{\mu\nu}(x)\,,
\end{equation}
such that one arrives at the so-called Einstein frame action,
\begin{equation}\label{equ:einstein_frame_lagrangian}
    S^\ell_E=\frac{1}{2\kappa_0}\int_{\mathcal M} d^4x\sqrt{-\tilde g}\,\tilde L^\ell
    +S_\text{m}\!\left[\frac{\tilde g_{\mu\nu}}{G_4(\hat \Phi)},\hat \Psi_\text{m}\right]\,,
\end{equation}
with
\begin{align}\label{equ:Horndeski_conformal}
   \tilde L^\ell
   =&\,\frac{1}{G_4(\hat\Phi)^2}\Big[G_2(\hat\Phi, G_4\cdot \tilde X)
   -\Big\{2G_3(\hat\Phi, G_4\cdot \tilde X)G'_{4}(\hat \Phi)\nonumber\\
   &\hspace{2.3cm}-9 G'^2_{4}(\hat\Phi)+6G_{4}(\hat\Phi)G''_{4}(\hat\Phi)\Big\}\tilde X\Big]\nonumber\\
   &+\frac{1}{G_4(\hat\Phi)}\Big[-G_3(\hat\Phi, G_4\cdot \tilde X)+3G'_{4}(\hat\Phi)\Big]
   \tilde\dalembertian\hat\Phi+\tilde R\,,\nonumber\\
   =&\,\tilde R + \tilde{\mathfrak{L}}^\ell[\hat\Phi,\tilde\nabla \hat\Phi]\,.
\end{align}
Here,
\begin{equation}
    \tilde X=-\frac{1}{2}\tilde g^{\mu\nu}\tilde\nabla_\mu\hat\Phi \tilde\nabla_\nu\hat\Phi\,.
\end{equation}
As the scalar does not transform under \eqref{equ:rescaling_horndeski}, it retains its notation. 

Note that, in terms of the unphysical Einstein frame metric, universal coupling is lost. Indeed, we emphasize that the transformation leading to the Einstein frame, despite the name suggesting otherwise, does not correspond to a coordinate transformation but instead results in a change of the physical theory. Therefore, it is crucial to reverse this transformation prior to making physical predictions, in particular when discussing physical effects of gravity on matter, which should be described by the physical spacetime metric.\footnote{Intuition for the Einstein frame transformation is nicely provided using the example of Brans--Dicke theory in \cite{poisson2014gravity}. For a concrete example of how the Einstein frame theory is transformed back into the Jordan frame, we refer to \cite{tahura_brans-dicke_2021}. Generally, solutions in the Jordan frame can be obtained via a conformal transformation of Einstein frame solutions, reversing the Weyl rescaling---in our case Eq.~\eqref{equ:rescaling_horndeski}---i.e., by replacing $\tilde g_{\mu\nu}\rightarrow G_4(\hat\Phi)\hat g_{\mu\nu}$.}
In principle, for the analysis below, switching to the Einstein frame is not strictly necessary, yet it is desirable due to computational ease and the availability of GR results.  
However, note that for a broader class of Horndeski theories, a transformation to the Einstein frame is only available perturbatively, and more care is required.

\subsection{Symplectic potential and current}

Following the prescription by Wald and Zoupas \cite{waldGeneralDefinitionConserved2000} described in Sec.~\ref{sec:intuition} (see also Appendix~\ref{app:GR_mod}), we compute the symplectic potential of the Horndeski Lagrangian in the Einstein frame [Eq.~\eqref{equ:Horndeski_conformal}]. For simplicity, in subsequent computations, we omit the common prefactor of $(2\kappa_0)^{-1}$ in Eq.~\eqref{equ:einstein_frame_lagrangian} by choosing suitable geometric units. 

Due to the natural split between a pure Ricci scalar $\tilde R$ and a beyond GR Lagrangian $\tilde {\mathfrak{L}}^\ell[\hat\Phi,\tilde \nabla \hat\Phi]$ in terms of the unphysical metric $\tilde g_{\mu\nu}$ the total potential has the simple linear decomposition $\tilde\bt^\ell =\tilde \bt_\text{Horndeski}+\tilde\bt_\text{GR}$, where the latter is equivalent to the standard GR symplectic potential, Eq.~\eqref{equ:idk_KKKK} in Appendix \ref{app:GR_mod}. On the other hand, the beyond GR contribution to luminal Horndeski theory, i.e., $\tilde {\mathfrak{L}}^\ell[\hat\Phi,\tilde \nabla \hat\Phi]$ reads 
\begin{align}\label{equ:symplectic_potential_horndeski}
    &\tilde\bt_\text{Horndeski} =\tilde \epsilon_{\mu\alpha\beta\gamma}\delta \hat\Phi \tilde \nabla^\mu \hat \Phi \Bigg[ \Big(-\frac{G_{2\tilde{X}}(\hat \Phi, G_4\cdot \tilde{X})}{G_4(\hat \Phi)^2}\notag \\&+\frac{1}{G_4(\hat \Phi)^2}\Big\{2G_3(\hat \Phi, G_4\cdot \tilde{X})G'_{4}(\hat \Phi)-9 G'_{4}(\hat \Phi)^2\notag \\&+6G_{4}(\hat \Phi)G''_{4}(\hat \Phi)\Big\}+2G_{3\tilde{X}}(\hat \Phi,  G_4\cdot \tilde{X})G_{4\hat \Phi}(\hat \Phi)\tilde{X}\Big) \notag\\&
-2\bigg[-\frac{G_{3\hat \Phi}(\hat \Phi, G_4\cdot \tilde{X}) - \frac{3}{2}G_{3\tilde{X}}(\hat \Phi, G_4 \cdot \tilde{X})\tilde\dalembertian \hat \Phi}{G_4(\hat \Phi)}\nonumber \\&+\frac{G_3(\hat \Phi, G_4\cdot \tilde{X})}{G_4(\hat \Phi)^2}G_{4\hat \Phi}(\hat \Phi)
    +3\frac{G_{4\hat \Phi}'(\hat \Phi)}{G_4(\hat \Phi)}-3\frac{G_{4\hat \Phi}(\hat \Phi)^2}{G_4(\hat \Phi)^2}\bigg]
    \Bigg]\nonumber\\
    &=: \tilde \epsilon_{\mu\alpha\beta\gamma}\delta \hat \Phi \tilde \nabla^\mu \hat \Phi \cdot \Xi[\hat \Phi, \tilde{X}]
\end{align}
Here, derivatives of the functionals $G_i[Z]$, with $i\in\{2,3,4\}$ and $Z\in \{\hat \Phi, \tilde X\}$, are denoted as $G_{iZ}:=\partial G_i/\partial Z$. Further, in Eq.~\eqref{equ:symplectic_potential_horndeski}, the expression in the large square brackets has been abbreviated by $\Xi[\hat \Phi, \tilde{X}]$ for better readability and due to frequent appearance in the subsequent analysis.

To arrive at Eq.~\eqref{equ:symplectic_potential_horndeski}, we made use of the fact that the presymplectic potential is defined only up to boundary terms. Concretely, one has the freedom
\[
\bt \;\rightarrow\; \bt + \dd Y + \delta K \,, \qquad 
\boldsymbol{L}\;\rightarrow\;\boldsymbol{L}+\dd K\,,
\]
as shown in \cite{waldGeneralDefinitionConserved2000}, where $Y$ is an arbitrary $(d-2)$-form and $\dd K$ denotes a total derivative at the level of the Lagrangian. This ambiguity reflects the fact that the Lagrangian itself is only defined up to exact forms, which do not affect the equations of motion.

This freedom has a direct geometric interpretation. The presymplectic form constructed from $\bt$ on the space of field configurations is, in general, degenerate: it possesses null directions corresponding to variations that do not change the physical state. These null directions include gauge redundancies as well as variations induced by boundary terms in the action. Consequently, the presymplectic structure defines a presymplectic, rather than a true symplectic, manifold. Shifts of $\bt$ by $\delta K$ or $\dd Y$ correspond precisely to moving along these null directions and therefore do not alter the physical symplectic structure once the degeneracies are removed.

In the present case, this ambiguity is particularly useful for handling higher-derivative terms. The Einstein-frame Horndeski Lagrangian \eqref{equ:Horndeski_conformal} contains contributions proportional to $\tilde\dalembertian\hat \Phi$, namely
\begin{align}\label{equ:trick}
    &\frac{1}{G_4(\hat \Phi)}\Big[-G_3(\hat \Phi, G_4\cdot \tilde{X})+3G'_{4}(\hat \Phi)\Big]\tilde\dalembertian\hat \Phi\,.
\end{align}
Upon partial integration at the level of the Lagrangian, such terms generate boundary contributions of the form $\boldsymbol{L}\rightarrow\boldsymbol{L}+\dd K$, where
\begin{align}\label{equ:trick_II}
     K_\mu \sim \frac{1}{G_4(\hat \Phi)}\Big[-G_3(\hat \Phi, G_4\cdot \tilde{X})+3G'_{4}(\hat \Phi)\Big]\tilde\nabla_\mu\hat \Phi\,.
\end{align}
These boundary terms appear in the presymplectic potential as total variations $\delta K$. Exploiting the intrinsic ambiguity of $\bt$, such contributions can be discarded without loss of physical information. In fact, fixing this freedom precisely amounts to selecting a representative of the equivalence class of presymplectic potentials that is adapted to the reduced phase space $\Gamma$. In other words, we promote the presymplectic potential to a symplectic potential and the associated symplectic current yields a non-degenerate symplectic form when evaluated on tangent vectors to phase space.\footnote{See \cite{waldGeneralDefinitionConserved2000} for a detailed discussion of this reduction.} These tangent vectors, in turn, are nothing else than variations, $\delta\hat\psi$, of the dynamical fields,  $\hat\psi\in\{\hat g_{\mu\nu},\hat\Phi\}$, on phase space $\Gamma$. This picture is particularly helpful when dealing with two distinct variations, $\delta,\delta'$, as in the definition of the symplectic current: Let $\hat\psi=\hat\psi(\lambda,\lambda')$ be a two-parameter family of fields with $\hat\psi(0,0)=\hat\psi$ on $\Gamma$. Then, $\delta\hat\psi:=\partial\hat\psi/\partial\lambda$ and $\delta'\hat\psi:=\partial\hat\psi/\partial\lambda'$, i.e., two variations correspond to two distinct tangent vectors on phase space $\Gamma$.

With the ``selected'' symplectic potential and the corresponding phase space (and associated tangent vectors) at hand, the next step in the WZ formalism is the computation of the symplectic current from the symplectic potential,
\begin{align}\label{equ:def_current}
    \tilde\bw^\ell = \delta'\tilde\bt^\ell(\delta g,\delta\hat \Phi) - \braket{\delta'\leftrightarrow\delta}\,.
\end{align}
Owing to the convenient form of the Lagrangian in the Einstein frame in Eq.~\eqref{equ:Horndeski_conformal}, the symplectic current can be decomposed as
\begin{align}\label{equ:full_sympl_current}
    \tilde\bw^\ell = \tilde\bw_\text{GR} + \tilde\bw_\text{Horndeski} + \tilde\bw_\times\,,
\end{align}
where the tilde indicates quantities evaluated in the Einstein frame.

The term $\tilde\bw_\text{GR}$ is structurally identical to the familiar GR symplectic current, Eq.~\eqref{equ:symp_current_GR}, but expressed in terms of the Einstein frame metric $\tilde g_{\mu\nu}$. The contribution $\tilde\bw_\text{Horndeski}$ corresponds to the symplectic current of the scalar sector on a ``flat'' background, in the sense that it involves only variations with respect to $\hat \Phi$. It is given by
\begin{align}\label{equ:symplectic_current_Horndeski}
    \tilde\bw_\text{Horndeski}
    =&\, \tilde \epsilon_{\mu\alpha\beta\gamma}\,\delta\hat \Phi\, \tilde \nabla^\mu \delta' \hat \Phi\cdot\Xi[\hat \Phi, \tilde{X}]\nonumber\\
    &+ \tilde \epsilon_{\mu\alpha\beta\gamma}\,\delta \hat \Phi\, \tilde \nabla^\mu \hat \Phi \,\delta'\Xi[\hat \Phi, \tilde{X}]
    - \braket{\delta'\leftrightarrow\delta}\,,
\end{align}
with
\begin{align}
    \tilde \epsilon_{\mu\alpha\beta\gamma}\,\delta \hat \Phi \tilde \nabla^\mu \hat \Phi \,\delta'\Xi[\hat \Phi, \tilde{X}]
    := & \tilde \epsilon_{\mu\alpha\beta\gamma}\,\delta \hat \Phi \tilde \nabla^\mu \hat \Phi\Bigg[
    \,\frac{\delta}{\delta\hat \Phi}\Xi[\hat \Phi, \tilde{X}]\,\delta'\hat \Phi \nonumber\\
    &+ \frac{\delta}{\delta\tilde\nabla_\delta\hat \Phi}\Xi[\hat \Phi, \tilde{X}]\,\delta'\tilde\nabla_\delta\hat \Phi \nonumber\\
    &+ \frac{\delta}{\delta\tilde\nabla_\delta\tilde\nabla_\sigma\hat \Phi}\Xi[\hat \Phi, \tilde{X}]\,\delta'\tilde\nabla_\delta\tilde\nabla_\sigma\hat \Phi
    \Bigg]\,,\label{equ:variation}
\end{align}
where variations appear only up to second order, as the Lagrangian contains at most second derivatives of the scalar field. Consequently, the number of terms is finite and no truncation is required.

The remaining contribution $\tilde\bw_\times$ contains mixed variations with respect to $\tilde g_{\mu\nu}$ and $\hat \Phi$ and arises from applying the variation $\delta'$ in Eq.~\eqref{equ:def_current} to the metric dependence in Eq.~\eqref{equ:symplectic_potential_horndeski}. This includes the metric determinant $\sqrt{-\tilde g}$, absorbed into $\tilde \epsilon_{\mu\alpha\beta\gamma}$, as well as the metric entering implicitly through contractions of scalar derivatives, such as $\tilde X = -\frac12 \tilde g^{\alpha\delta}\tilde \nabla_\alpha\hat \Phi\tilde \nabla_\delta\hat \Phi$. One finds
\begin{align}\label{equ:symplectic_current_cross}
&    \tilde\bw_\times = \,\,\underbrace{\tilde \epsilon_{\mu\alpha\beta\gamma}\delta\hat \Phi \tilde\nabla_\delta\hat \Phi\Big(\frac{1}{2}\tilde g^{\mu\delta}\tilde g^{\rho\sigma}\delta' \tilde g_{\rho\sigma} + \delta'\tilde g^{\mu\delta}\Big)\Xi[\hat \Phi, \tilde{X}]}_{=:\bw_\times^{I}}\nonumber\\
    &+\underbrace{\tilde \epsilon_{\mu\alpha\beta\gamma}\tilde g^{\mu\delta}\delta\hat \Phi \tilde\nabla_\delta\hat \Phi\, \frac{\delta}{\delta \tilde{X}}\Xi[\hat \Phi, \tilde{X}]\left(-\frac{1}{2}\delta' \tilde g^{\rho\sigma}\tilde \nabla_\rho \hat \Phi \tilde \nabla_\sigma \hat \Phi\right)}_{=:\bw_\times^{II}}\nonumber \\
    &+\underbrace{\tilde \epsilon_{\mu\alpha\beta\gamma}\tilde g^{\mu\delta}\delta\hat \Phi \tilde\nabla_\delta\hat \Phi\, \frac{\delta}{\delta \tilde \dalembertian \hat \Phi}\Xi[\hat \Phi, \tilde{X}]\left(\delta' \tilde g^{\rho\sigma}\tilde \nabla_\rho \tilde \nabla_\sigma \hat \Phi\right)}_{=:\bw_\times^{III}}\notag\\
    &- \braket{\delta'\leftrightarrow\delta}\,.
\end{align}
For clarity, we decompose $\tilde\bw_\times$ into three contributions, $\tilde\bw_\times^{I,II,III}$, each understood to include the antisymmetrization in the variations $\braket{\delta'\leftrightarrow\delta}$. For example,
\begin{align}
\tilde\bw_\times^{I}
:= &\tilde \epsilon_{\mu\alpha\beta\gamma}\,\delta\hat \Phi \tilde\nabla_\delta\hat \Phi
\Big(\frac{1}{2}\tilde g^{\mu\delta}\tilde g^{\rho\sigma}\delta' \tilde g_{\rho\sigma}
+ \delta'\tilde g^{\mu\delta}\Big)\Xi[\hat \Phi, \tilde{X}]\nonumber\\
&- \braket{\delta'\leftrightarrow\delta}\,.
\end{align}
In the following, we analyze the behavior of each contribution in the limit toward future null infinity, $\scrip$.

\subsection{The Wald--Zoupas flux in the limit to $\scrip$}
\label{subsec:null_infty_shit}

Our analysis focuses on spacetimes that are asymptotically flat at future null infinity, $\scrip$, owing both to their direct observational relevance and to their natural suitability for describing radiative phenomena. Such spacetimes are ubiquitous in the literature, particularly in studies of compact binary coalescences and gravitational radiation (see, e.g., \cite{Ashtekar:2019viz,Mitman:2020bjf,DAmbrosio:2024zok,Borchers:2021vyw}). Although future null infinity is, strictly speaking, not part of the four-dimensional physical spacetime manifold $(\mathcal M,\hat g_{\mu\nu})$ (cf.~Fig.~\ref{fig:scrip}), it nevertheless provides a rigorous mathematical setting in which fluxes and conserved quantities associated with radiation can be defined and computed.

\subsubsection{Asymptotics in luminal Horndeski theory}\label{ssSec:AsymptoticFlatness}

In GR, the framework describing asymptotically flat spacetimes at $\scrip$ is well established. For theories beyond GR, however, the corresponding structures, in particular the identities governing the asymptotic behavior of the metric and its perturbations, need not carry over unchanged. For the subclass of luminal Horndeski theories considered here, we argue in Appendix~\ref{app:asymptotic_Horn} that the asymptotic results familiar from GR remain directly applicable. Before motivating this claim, we briefly summarize the standard GR construction and introduce the geometric tools required to compute the pullback of the symplectic current to the boundary.

\begin{figure}[t!]
    \centering
    \includegraphics[width=0.45\textwidth]{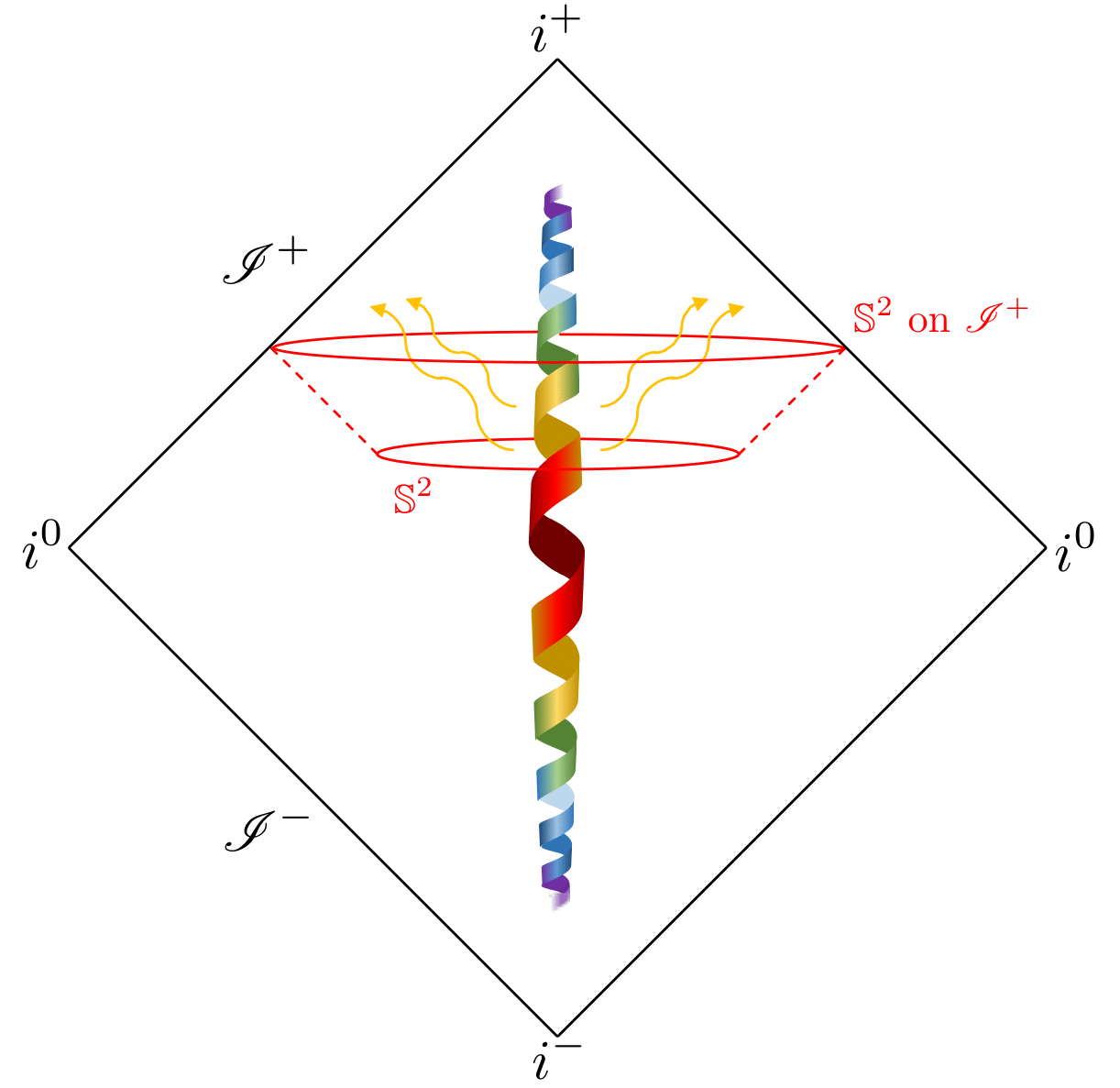}
    \caption{Sketch of $\scrip$ as the future boundary of the Penrose diagram of Minkowski spacetimes \cite{Maibach:2025iku}. In red, we denote cross sections of constant $u$. The spiral in the center of the diagram represents a event emitting GWs in null direction toward $\scrip$.}
    \label{fig:scrip}
\end{figure}

The standard definition of asymptotic flatness in GR (see, for example, Sec.~2.1.3 of \cite{Maibach:2025iku}) involves an extension of spacetime all the way to $\scrip$ by means of a conformal compactification. This provides a smooth, unphysical metric $g_{\mu\nu}$ defined on $\mathcal M\cup\scrip$, related to the physical metric by
\begin{align}\label{equ:conformal_compact}
    g_{\mu\nu} = \Omega^2 \hat g_{\mu\nu}\,.
\end{align}
Here, $\Omega$ is a smooth function on $\mathcal M\cup\scrip$ such that $\Omega>0$ on $\mathcal M$ and $\Omega=0$ on $\scrip$.\footnote{The conformal factor $\Omega$ is not to be confused with the symplectic form $\boldsymbol\Omega$ defined earlier.}

Defining
\[
n_\mu := \nabla_\mu \Omega\,,
\]
one finds that $n_\mu$ is null at $\scrip$. Without loss of generality, the conformal factor may be chosen such that the Bondi condition $\nabla_\mu n_\nu \equaldot 0$ holds, where ``$\equaldot$'' denotes equality at $\scrip$. An immediate consequence is that $n^\mu n_\mu = \mathcal O(\Omega^2)$.
It is furthermore convenient to foliate $\scrip$ by $u=\text{constant}$ cross sections, where $u$ is a coordinate along the null generators of $\scrip$ (see Fig.~\ref{fig:scrip}). With this choice, one has $n^\mu\partial_\mu \equaldot \partial_u$. Associated with this foliation, there exists a unique auxiliary null vector field $\ell_\mu$ normal to $\scrip$ such that $\ell_\mu n^\mu \equaldot -1$. The pullback of the unphysical metric to $\scrip$,
\[
q_{\mu\nu} := \newpb{g_{\mu\nu}}\,,
\]
defines an induced metric whose spatial part corresponds to the standard metric on the unit two-sphere.\footnote{Throughout this work, we assume that the pair $(g_{\mu\nu},\Omega)$ belongs to the universal asymptotic structure described, for instance, in \cite{Maibach:2025iku}.}

We further require that both the unphysical metric and its perturbations extend smoothly to $\scrip$. By definition of asymptotic flatness, the perturbation vanishes at $\scrip$, i.e.,
\[
\delta g_{\mu\nu} \equaldot \Omega^2 \delta \hat g_{\mu\nu} \equaldot 0\,,
\]
which allows one to introduce an auxiliary tensor $\tau_{\mu\nu}$ such that $\tau_{\mu\nu}:=\Omega^{-1}\delta g_{\mu\nu}$ extends smoothly to $\scrip$ and is non-vanishing there. Note further that $\tau_{\mu\nu}n^\nu = \Omega \tau_\mu$ where $\tau_\mu$ is smooth at $\scrip$ \cite{waldGeneralDefinitionConserved2000}. Finally, for a physical matter stress-energy tensor $\hat T_{\mu\nu}$ appearing in the standard Einstein equations with respect to the physical metric $\hat g_{\mu\nu}$, asymptotic flatness requires $\hat T_{\mu\nu}=\mathcal O(\Omega^2)$. This condition ensures that the metric components determined by Einstein’s equations exhibit the correct $1/r$ fall-off and asymptotically approach Minkowski spacetime.\footnote{See also Appendix~B of \cite{Bieri_2014} for further discussion.}

We now turn to luminal Horndeski theory and ask what changes relative to GR. The key difference lies in the equations of motion determining the metric, which now include an effective stress-energy tensor sourced by the scalar field. While the field equations of the luminal Horndeski theories considered here retain a GR-like structure, this is not true for generic Horndeski models, and the definition of asymptotic flatness may then require additional care. In our case, however, the equations of motion take the form
\begin{align}\label{equ:New_Einstein_II}
    \hat R_{\mu\nu}-\frac{1}{2}\hat g_{\mu\nu}\hat R
    =  \underbrace{\frac{\kappa_0}{G_4(\hat \Phi)}\hat{\mathcal{T}}_{\mu\nu}}_{:= \kappa_0 \hat \Sigma_{\mu\nu}}\,,
\end{align}
where $\hat{\Sigma}_{\mu\nu}$ denotes the effective scalar-field stress-energy tensor.\footnote{In the presence of matter, the right-hand side of Eq.~\eqref{equ:New_Einstein_II} acquires an additional contribution from the matter stress-energy tensor $\hat T_{\mu\nu}=-\frac{2}{\sqrt{-\hat g}}\frac{\delta S_\text{m}}{\delta \hat g^{\mu\nu}}$, such that $\hat{\mathcal T}_{\mu\nu}$ is replaced by $\hat{\mathcal T}_{\mu\nu}+\hat T_{\mu\nu}$. For asymptotically flat spacetimes, one typically assumes $\hat T_{\mu\nu}=\mathcal O(1/r^2)$; see, e.g., \cite{Flanagan_2017}.} 

Since $\hat{\Sigma}_{\mu\nu}$ constitutes a new contribution compared to vacuum GR, the asymptotic fall-off of the metric must be verified explicitly. This requires solving Eq.~\eqref{equ:New_Einstein_II} using an appropriate parametrization of the metric together with suitable gauge conditions (see, for example, \cite{tahura_brans-dicke_2021,hou_conserved_2021}). The corresponding analysis for luminal Horndeski theory is outlined in Appendix~\ref{app:asymptotic_Horn}.

Based on this analysis, we find that the functions $G_i$ defining the theory, subject to the assumptions in Eq.~\eqref{equ:assumption}, ensure that the effective scalar-field stress-energy tensor $\hat{\Sigma}_{\mu\nu}$ does not spoil the required metric fall-off conditions.\footnote{Note however that the Peeling theorem may be violated \cite{Zosso:2024xgy}.} Consequently, the metric remains asymptotically flat in the same sense as in GR. This validates the use of conformal compactification and the associated unphysical metric $g_{\mu\nu}$ with $\delta g_{\mu\nu}\equaldot 0$, the induced metric $q_{\mu\nu}\equaldot \newpb{g_{\mu\nu}}$, the conformal factor $\Omega$, and the chosen foliation of $\scrip$. As a result, all geometric properties of $\scrip$, including those of the null normals $n^\mu$ and $\ell^\mu$, familiar from GR can be directly applied to luminal Horndeski theory.

We note at this point that while $\hat \Sigma_{\mu\nu}$ does not affect the lowest-order expansion in $1/r$ of the metric components, it does change the selected expansion coefficien‚ts. For the analysis presented in this work, this alteration is irrelevant with exception of the metric component corresponding to the shear which carries the metric's radiative degrees of freedom. Below in Sec.~\ref{subsec:balance_law_equation} as well as Appendix~\ref{app:asymptotic_Horn}, we delineate the consequences in more detail.

Given the introduction of several distinct metrics, arising from the conformal completion at $\scrip$ and from the transition to the Einstein frame, it is useful to summarize the geometric structures employed throughout the subsequent analysis. We therefore list the relevant metrics as follows:
\begin{itemize}
    \item $\hat g_{\mu\nu}$ denotes the physical spacetime metric defined on $\mathcal M$.
    \item $\tilde g_{\mu\nu}$ denotes the Einstein-frame metric on $\mathcal M$, obtained from $\hat g_{\mu\nu}$ by a Weyl (conformal) rescaling.
    \item $g_{\mu\nu}$ denotes the unphysical metric defined on the conformally completed spacetime $\mathcal M \cup \scrip$, which extends smoothly to $\scrip$.
    \item $q_{\mu\nu}$ denotes the (degenerate) metric induced on $\scrip$, defined as the pullback of the unphysical metric, $q_{\mu\nu}:=\newpb{g_{\mu\nu}}$.
\end{itemize}
The corresponding derivative operators associated with these metrics are labeled, respectively, as $\hat\nabla$, $\widetilde\nabla$, $\nabla$, and $D$. The same notational convention applies to all tensorial quantities and fields that transform non-trivially under the conformal completion or under the passage to the Einstein frame.
\subsubsection{Pulling back to $\scrip$}

Starting from the symplectic current constructed in the previous subsection, Eq.~\eqref{equ:full_sympl_current}, the asymptotic symplectic potential, i.e., the asymptotic flux, Eq.~\eqref{equ:Jann_Lfux}, can be computed by pulling back the full symplectic current to future null infinity $\scrip$. 

As we now show, this pullback simplifies the form of the symplectic current substantially. For Horndeski theories satisfying the assumptions stated in Eq.~\eqref{equ:assumption}, all cross terms $\tilde{\boldsymbol{\omega}}_{\times}$ vanish in the limit to $\scrip$. Consequently, only the pure gravitational contribution $\tilde{\boldsymbol{\omega}}_{\text{GR}}$ and the pure Horndeski contribution $\tilde{\boldsymbol{\omega}}_{\text{Horndeski}}$ enter the computation of $\newpb{\tilde{\boldsymbol{\omega}}}^{\ell}$.

\paragraph{Conformal completion and field expansions.}
Taking the limit to $\scrip$ requires passing from the physical Einstein-frame metric $\tilde g_{\mu\nu}$ to a conformally compactified metric $g_{\mu\nu}$ defined smoothly on $\mathcal M\cup\scrip$ via Eq.~\eqref{equ:conformal_compact}. Since the transformation to the Einstein frame does not affect the asymptotic decay of the metric in the class of theories considered here, the relation remains
\begin{equation}
    \tilde g_{\mu\nu} = \Omega^{-2} g_{\mu\nu}\,.
\end{equation}
We therefore use $g_{\mu\nu}$ to denote the unphysical metric on $\mathcal M\cup\scrip$ throughout. The associated Levi--Civita tensor and metric perturbations transform as
\begin{equation}
    \tilde\epsilon_{\mu\alpha\beta\gamma} = \Omega^{-4}\epsilon_{\mu\alpha\beta\gamma}\,, 
    \qquad 
    \delta \tilde g_{\mu\nu} = \Omega^{-2}\delta g_{\mu\nu}\,.
\end{equation}

Near $\scrip$, the massless scalar field admits an expansion in powers of $\Omega$\footnote{Note that this expansion requires the field to be regular at $\scrip$, i.e., finite and smooth. This constitutes an additional assumption on the field.},
\begin{equation}\label{equ:expansssion}
    \hat\Phi = \varphi_0 + \Omega \varphi_1 + \mathcal O(\Omega^2)\,,
\end{equation}
where $\varphi_0\neq 0$ is a model-dependent constant.%
\footnote{For massive scalars, the asymptotic behavior is Yukawa-suppressed, $\hat\Phi\sim e^{-mr}/r$. The arguments below rely only on the fact that the leading falloff is at least $1/r$, together with a non-vanishing constant background value, and therefore extend straightforwardly to that case.}

\paragraph{Vanishing of the cross terms.}
We begin with the common prefactor appearing in the cross terms $\tilde{\boldsymbol{\omega}}_{\times}$, cf. Eq.~\eqref{equ:symplectic_current_cross}. Using the tetrad decomposition
\begin{equation}
    \epsilon_{\mu\alpha\beta\gamma}=4\,\epsilon_{[\alpha\beta\gamma}n_{\mu]}\,,
\end{equation}
one finds
\begin{align}\label{equ:prefactor}
    \epsilon_{\mu\alpha\beta\gamma}\,\delta\hat\Phi\,\tilde\nabla_\delta\hat\Phi
    = -\Omega^{-4}\epsilon_{\alpha\beta\gamma}n_\mu
    \,\delta(\Omega\varphi_1)\,\nabla_\delta(\Omega\varphi_1)\,.
\end{align}
Consequently,
\begin{equation}\label{equ:nkjnsadf}
    \tilde{\boldsymbol{\omega}}_{\times}^{I}
    \sim \Omega^{-3} n_\mu \delta\varphi_1 \nabla_\delta(\Omega\varphi_1)
    \left(\tfrac{1}{2}\Omega^3 g^{\mu\delta}g^{\rho\sigma}\tau'_{\rho\sigma}
    +\Omega^3\tau'^{\mu\delta}\right)\,.
\end{equation}
Since
\begin{equation}
    n_\mu \delta\varphi_1\nabla_\delta(\Omega\varphi_1)
    = n_\mu n_\delta \delta\varphi_1\varphi_1
    +\Omega n_\mu \delta\varphi_1\nabla_\delta\varphi_1
    \,,
\end{equation}
the contraction with the bracket in Eq.~\eqref{equ:nkjnsadf} then yields that the contribution $\tilde{\boldsymbol{\omega}}_{\times}^{I}$ vanishes at $\scrip$.

The same conclusion holds for $\tilde{\boldsymbol{\omega}}_{\times}^{III}$. Indeed,
\begin{equation}
    \frac{\delta}{\delta(\tilde\Box\hat\Phi)}\Xi[\hat\Phi,\tilde X]
    \sim G_{3\tilde X}(\hat\Phi,G_4\tilde X)
    = \mathcal O(\Omega^{\varepsilon})\,,
\end{equation}
while
\begin{equation}
    \delta'\tilde g^{\rho\sigma}\tilde\nabla_\rho\tilde\nabla_\sigma\hat\Phi
    = \mathcal O(\Omega^2)\,.
\end{equation}
Using
\begin{equation}\label{equ:prefactorr}
    \tilde\epsilon_{\mu\alpha\beta\gamma}\tilde g^{\mu\delta}
    \delta\hat\Phi\,\tilde\nabla_\delta\hat\Phi
    \sim \Omega^{-1} n_\mu g^{\mu\delta}
    \delta\varphi_1\nabla_\delta(\Omega\varphi_1)\,,
\end{equation}
one sees that $\tilde{\boldsymbol{\omega}}_{\times}^{III}$ scales as $\mathcal O(\Omega^2)$ and therefore vanishes at $\scrip$.

Finally, consider $\tilde{\boldsymbol{\omega}}_{\times}^{II}$. The prefactor is again given by Eq.~\eqref{equ:prefactor}. The remaining factor
\begin{equation}
    -\tfrac{1}{2}\delta'\tilde g^{\rho\sigma}
    \tilde\nabla_\rho\hat\Phi\,\tilde\nabla_\sigma\hat\Phi
\end{equation}
scales as $\Omega\,\tilde X$. Hence, for any $G_i(\tilde X)$,
\begin{equation}
    \frac{\delta G_i}{\delta\tilde X}
    \left(-\tfrac{1}{2}\delta'\tilde g^{\rho\sigma}
    \tilde\nabla_\rho\hat\Phi\,\tilde\nabla_\sigma\hat\Phi\right)
    \sim \Omega\,\mathcal O(\tilde X^{\varepsilon})
\end{equation}
by Eq.~\eqref{equ:assumption}. A term-by-term scaling analysis%
\footnote{Several contributions in $\Xi[\hat\Phi,\tilde X]$ contain additional factors of $\Omega$, e.g.\
$\tilde\Box\hat\Phi=\mathcal O(\Omega^2)$ and $\tilde X=\mathcal O(\Omega^3)$, further accelerating the decay.}
shows that $\tilde{\boldsymbol{\omega}}_{\times}^{II}=\mathcal O(\Omega)$ and thus also vanishes at $\scrip$.
Collecting these results, the entire cross contribution $\tilde{\boldsymbol{\omega}}_{\times}$, Eq.~\eqref{equ:symplectic_current_cross}, drops out upon pullback to $\scrip$.

\paragraph{Horndeski contribution to the flux.}
We therefore turn to the remaining symplectic current $\tilde{\boldsymbol{\omega}}_{\text{Horndeski}}$, Eq.~\eqref{equ:symplectic_current_Horndeski}. As follows from Eqs.~\eqref{equ:prefactor} and~\eqref{equ:prefactorr}, the prefactors of both terms scale as $\mathcal O(\Omega^0)$ at $\scrip$. Hence, only those parts of $\Xi[\hat\Phi,\tilde X]$ and $\delta\Xi[\hat\Phi,\tilde X]$ that remain finite and non-trivial as $\Omega\to0$ contribute.

For the first term in Eq.~\eqref{equ:symplectic_current_Horndeski}, one finds
\begin{equation}
    \epsilon_{\alpha\beta\gamma}n_\mu
    \delta'\varphi_1\nabla^\mu\delta\varphi_1\,
    \Xi[\hat\Phi,\tilde X]
    -\braket{\delta'\leftrightarrow\delta}\,.
\end{equation}
Taking the limit to $\scrip$ yields
\begin{align}\label{equ:limit_horndeski}
    \Xi[\hat\Phi,\tilde X]\equaldot
    \Bigg[
    \frac{2\bar G_{3\hat\Phi}}{\bar G_4}
    -\frac{\bar G_{2\tilde X}}{\bar G_4^2}
    -3\frac{\bar G_{4\hat \Phi}^2}{\bar G_4^2}
    \Bigg]
    =:\overline{\Xi}[\varphi_0]\,,
\end{align}
where barred quantities denote evaluation at the background
$\hat\Phi\to\varphi_0$, $\tilde X\to0$.

The second term in $\tilde{\boldsymbol{\omega}}_{\text{Horndeski}}$ involves $\delta\Xi[\hat\Phi,\tilde X]$. Variations with respect to $\hat\Phi$ and $\tilde\nabla_\mu\tilde\nabla_\nu\hat\Phi$ are suppressed by at least one power of $\Omega$ and therefore vanish at $\scrip$. The remaining variation,
$\delta/\delta(\tilde\nabla_\mu\hat\Phi)$, acts only on functions of $\tilde X$ and satisfies
\begin{equation}
    \frac{\delta}{\delta\tilde\nabla_\mu\hat\Phi}f(\tilde X)\,
    \delta\tilde\nabla_\mu\hat\Phi
    = f'(\tilde X)\,
    \big(-\tilde g^{\mu\nu}\tilde\nabla_\mu\delta\hat\Phi
    \tilde\nabla_\nu\hat\Phi\big)
    = \mathcal O(\Omega^3)\,.
\end{equation}
Using Eq.~\eqref{equ:assumption}, this implies
\begin{equation}
    \frac{\delta}{\delta\tilde\nabla_\mu\hat\Phi}
    \Xi[\hat\Phi,\tilde X]
    = \mathcal O(\Omega^{3+\varepsilon})\,,
\end{equation}
so that the entire contribution proportional to $\delta\Xi$ vanishes at $\scrip$.

We therefore obtain
\begin{align}
    \newpb{\tilde{\boldsymbol{\omega}}}_{\text{Horndeski}}
    &=
    \epsilon_{\alpha\beta\gamma}
    \delta\varphi_1\,n_\mu\nabla^\mu\delta'\varphi_1\,
    \overline{\Xi}[\varphi_0]
    -\braket{\delta'\leftrightarrow\delta}\,.
\end{align}
Since $\overline{\Xi}[\varphi_0]$ is constant on $\scrip$, the associated flux density defined by Eq.~\eqref{equ:current_PB} reads
\begin{align}\label{equ:flux_Horndeski}
    \tilde{\bT}_{\text{Horndeski}}(\delta\varphi_1)
    &=
    \epsilon_{\alpha\beta\gamma}\,
    \delta\varphi_1\,n_\mu\nabla^\mu\varphi_1\,
    \overline{\Xi}[\varphi_0]\,.
\end{align}

\paragraph{Transformation back to the Jordan frame.}
What is left to do is to transform the results within the Einstein frame back to the Jordan frame, both for the pure scalar flux density $\tilde{\bT}_{\text{Horndeski}}$ in Eq.~\eqref{equ:flux_Horndeski}, and for the GR-like contribution $\tilde\bT_\text{GR}$ given by Eq.~\eqref{equ:last_step} in Appendix~\ref{app:GR_mod}. Let us start with  $\tilde{\bT}_{\text{Horndeski}}$. To that end, we explicitly write out the volume element such that 
\begin{equation}
    \tilde{\bT}_{\text{Horndeski}}
    = \sqrt{-q}\,q^{\mu\nu}\epsilon_{\alpha\beta\gamma}
    \delta\varphi_1 n_\mu\nabla_\nu\varphi_1\,
    \overline{\Xi}[\varphi_0]\,.
\end{equation}
Reversing the conformal transformation, one finds
$\sqrt{-q}\,q^{\mu\nu}\to \bar G_4^{1/2}\sqrt{-q}\,q^{\mu\nu}$ and
$n_\mu\to\bar G_4^{1/2}n_\mu$. Accounting also for the induced change in
$\bar G_{2\tilde X}$, the Jordan-frame result becomes
\begin{align}\label{equ:flux_Horndeski_final}
    \bT_{\text{Horndeski}}(\delta\varphi_1)
    =&-
    \frac{\bar G_4}{2\kappa_0}\,\delta\varphi_1\,\dot\varphi_1
    \Bigg[
\frac{\bar G_{2\tilde X}-2\bar G_{3\hat\Phi}}{\bar G_4}\nonumber \\ &+
    3\frac{\bar G_{4\hat \Phi}^2}{\bar G_4^2}
    \Bigg]
    {}^{(3)}\boldsymbol{\epsilon}\,,
\end{align}
where $n_\mu\equaldot\partial_u$, ${}^{(3)}\boldsymbol{\epsilon}_{\alpha\beta\gamma}= \sqrt{-q}\, \epsilon_{\alpha\beta\gamma}$ is the positively orientated volume three-form at $\scrip$, and the prefactor $(2\kappa_0)^{-1}$ in natural units was reintroduced.

An analogous analysis applies to the $\tilde R$-term in the Einstein-frame Lagrangian, Eq.~\eqref{equ:einstein_frame_lagrangian}. The details are sketched in Appendix~\ref{app:GR_mod}. Transforming back to the Jordan frame yields
\begin{align}\label{eq:FluxDensityGR}
    \bT_{\text{GR}}(\delta g_{\mu\nu})
    =
    -\frac{\bar G_4}{4\kappa_0}\,
    \Omega^{-1}\delta g_{\mu\nu}N^{\mu\nu}
    {}^{(3)}\boldsymbol{\epsilon}\,.
\end{align}
The tensor $N_{\mu\nu}$ is known as Bondi news tensor and will be briefly introduced below. Note that, as we will explain below, despite recycling a GR result, the meaning of $N_{\mu\nu}$ in the context of luminal Horndeski theory is not the same as in Einstein's theory.
The total flux associated with the luminal Horndeski Lagrangian in Eq.~\eqref{equ:general_form} is therefore given by a sum of Eqs.~\eqref{eq:FluxDensityGR} and \eqref{equ:flux_Horndeski_final}
\begin{equation}
    \bT^{\ell}
    =
    \bT_{\text{GR}}
    +
    \bT_{\text{Horndeski}}\,.
\end{equation}

For a finite portion $\Delta\scrip\subset\scrip$, following Eq.~\eqref{equ:Jann_Lfux}, the resulting flux associated with a Killing field $\xi$ is given by
\begin{equation}
        \mathcal  F_{\xi}[\Delta\scrip] = \int_{\Delta \scrip} \bT^\ell(\delta_\xi\varphi_1)\,.
\end{equation}
 For an arbitrary Killing vector field $\xi$, one finds $\Omega \delta \varphi_1 = \delta _\xi \hat \Phi = \Omega \lie_\xi \varphi_1 + \xi^\alpha n_\alpha \varphi_1$ and thus $\delta_\xi \varphi_1 = \lie_\xi\varphi_1 + K \varphi_1$ with $K= \xi^\alpha n_\alpha /\Omega$, such that 
\begin{widetext}
\begin{align}\label{equ:super_flux}
    \mathcal  F_{\xi}[\Delta\scrip]=-\frac{\bar G_4}{2\kappa_0} \int_{\Delta \scrip}\Bigg(&N_{\mu\nu}\big([\lie_\xi D_\alpha-D_\alpha\lie_\xi]\ell_\beta + 2\ell_{(\alpha}D_{\beta)}K\big)q^{\mu\alpha}q^{\nu\beta}\notag\\&+(\lie_\xi\varphi_1 +K\varphi_1)\dot\varphi_1 
    \Bigg[  3\frac{\bar G_{4\hat \Phi}^2}{\bar G_4^2}+\frac{ \bar G_{2\tilde{X}} -2\bar G_{3\hat \Phi}}{\bar G_4}
    \Bigg]\Bigg)\,.
\end{align}
\end{widetext}
The derivative $D_\mu$ here denotes the covariant derivative on $\scrip$ and $\ell_\mu$ denotes the second null normal with respect to $\scrip$ (in our choice of coordinates $\ell_\mu\sim \partial_ \Omega$ at $\scrip$). Note here that the sign of the flux agrees with the physical intuition that the flux ``carries'' charge away from the source. The first term in Eq.~\eqref{equ:super_flux} is exactly the GR flux for asymptotically flat spacetimes (up to the prefactor $\bar G_4$). Thus, the full flux across a section of $\scrip$ of the luminal Horndeski theory Eq.~\eqref{equ:general_form}, according to Eq.~\eqref{equ:super_flux}, is simply given by the sum of the GR flux plus a term purely derived from the scalar sector times an overall constant factor $\bar G_4$.

\section{Balance laws of luminal Horndeski gravity}
\label{subsec:balance_law_equation}

As elaborated in the introduction, in the context of compact binary
coalescences one is primarily interested in flux balance relations,
i.e., in fluxes describing changes in the corresponding ``conserved''
quantities. More precisely, as discussed in Sec.~\ref{subsec:WZ_II},
the generators of asymptotic symmetries define asymptotic charges
that are conserved up to the presence of a null flux through $\scrip$,
as given in Eq.~\eqref{equ:super_flux}. It is therefore essential to
fully characterize the symmetry generators $\xi$ at $\scrip$, which
act as asymptotic Killing fields of the metric and with respect to
which the fluxes are defined.


\subsection{BMS generators}

For asymptotically flat metrics, introduced in detail in
Sec.~\ref{ssSec:AsymptoticFlatness}, it is well known that the symmetry group at
$\scrip$ is the Bondi–Metzner–Sachs (BMS) group
\cite{Bondi:1962px,sachsGravitationalWavesGeneral1962}. The latter extends the
Poincar\'e group by an infinite tower of additional translation-type generators.
In total, the BMS group decomposes into supertranslation (ST) and Lorentz (L)
generators. The translational sector itself further splits into the ordinary
Poincar\'e spacetime translations and an infinite tower of proper
supertranslations.

In the same way that the asymptotically flat structure of spacetime in massless
Horndeski gravity is preserved, its asymptotic symmetry group remains unchanged,
and the generators assume the same form as in GR, as previously noted for specific
theories beyond GR \cite{Hou:2021bxz,hou_gravitational_2021,hou_gravitational_2021_2}.
The equations of motion are crucial in establishing this result, as they determine
the fall-off behavior of the individual metric components. Concretely, recall that
the additional scalar degree of freedom contributes only through an effective
stress-energy tensor in the equations for the physical metric
[Eq.~\eqref{equ:New_Einstein_II}]. As argued in
Appendix~\ref{app:asymptotic_Horn}, this extra term does not modify the asymptotic
behavior of the metric components near null infinity $\scrip$. Consequently, the
luminal Horndeski metric admits the same Killing vector fields at $\scrip$ as in
pure GR.

By explicit computation, one finds for the generators in Bondi coordinates
$\{u,y^A\}$ and on $\scrip$ (see, e.g., Ref.~\cite{Maibach:2025iku})
\begin{align}\notag
    \underline{\xi}_\text{BMS}
    &= \underline{\xi}_\text{ST} + \underline{\xi}_\text{L} \\
    &= \underbrace{\alpha(\theta,\phi)\partial_u}_{\underline{\xi}_\text{ST}}
    + \underbrace{\left(\frac{u}{2}D_A Y^A(\theta,\phi)\,\partial_u
    + Y_A(\theta,\phi)\partial_A\right)}_{\underline{\xi}_\text{L}} \, .
    \label{equ:original_ST_and_L}
\end{align}
For a fixed-$u$ cross section of $\scrip$, denoted by $\partial\scrip$,
$\underline{\xi}_\text{L}$ represents the infinitesimal Lorentz transformation.

In general, the representation of the BMS generators is not unique. Rather,
one works with equivalence classes of vector fields. Although this ambiguity is
irrelevant at the purely theoretical level, it affects the computation of
conserved charges, since in the Wald–Zoupas prescription the charges depend
explicitly on the choice of $\underline{\xi}$
\cite{waldGeneralDefinitionConserved2000}. A standard resolution is to select a
representative $\xi$ satisfying $\nabla_\mu \xi^\mu = 0$. Since this condition is
imposed in the conformally compactified spacetime, it applies not directly to
Eq.~\eqref{equ:original_ST_and_L}, but to its extension into the bulk
\cite{Flanagan_2017}.\footnote{In fact, it suffices to require
$\nabla_\mu \xi^\mu = \mathcal O(\Omega^2)$ to ensure uniqueness of the charges
\cite{Geroch:1981ut}.}
Imposing this condition yields the specific form
\begin{align}\label{equ:real_st}
    \xi_\text{ST}
    &= \left(\alpha(\theta,\phi)+\frac{u_0}{2}D_A Y^A\right)\partial_u \, ,\\
    \xi_\text{L}
    &= \left(\frac{u-u_0}{2}D_A Y^A(\theta,\phi)\,\partial_u
    + Y_A(\theta,\phi)\partial_A\right) \, .
    \label{equ:real_L}
\end{align}
The condition $\nabla_\mu \xi^\mu = 0$ in particular ensures that
$\xi_\text{L}$ is tangent to $\scrip$ on the chosen $u_0$-cross section,
which significantly simplifies the computation of conserved charges
\footnote{See, for instance, Eq.~(44) in \cite{waldGeneralDefinitionConserved2000}.}.
In the following, we will make explicit use of this property of $\xi_\text{L}$.

\subsection{Supertranslation flux balance laws}
While a balance flux equation can in principle be formulated for all symmetries that arise at $\scrip$, i.e., all symmetry generators of the BMS group, the most interesting case arises for supertranslation Killing vector fields. Thus, we start by setting $\xi_\mu=\alpha(\theta,\phi)\,n_\mu$, which implies that $K=0$ in Eq.~\eqref{equ:super_flux} due to $n_\mu$ being null. While not generally the case, for this selected subgroup of generators of the BMS group, $F_{\xi,\Delta\scrip}$ as defined in Sec.~\ref{subsec:WZ_II} is an exact $3$-form so that Eq.~\eqref{equ:flalal} indeed defines a valid balance law. To this end, we apply Stokes' theorem to each contribution in
Eq.~\eqref{equ:super_flux}.

\subsubsection{Asymptotic charges}

Consider first the scalar-sector contribution
arising from the Lagrangian~\eqref{equ:general_form}. The flux $3$-form
$\bT_\text{Horndeski}$ can be written as a total derivative,
\[
(\dd \bT_\text{Horndeski})_{\alpha\beta\gamma}
= 3 D_{[\alpha}\epsilon_{\beta\gamma]\mu} P^\mu \, .
\]
Contracting
with $\epsilon^{\alpha\beta\gamma}$ yields
\begin{align}
    \bT_\text{Horndeski}(\delta_{\alpha(\theta,\phi) n}\varphi_1)\epsilon^{\alpha\beta\gamma} &= -6\bar G_4 \alpha(\theta,\phi) \lie_{n}\varphi_1\dot\varphi_1\Bigg[ \nonumber 3\frac{\bar G_{4\hat \Phi}^2}{\bar G_4^2}\\& +\frac{ \bar G_{2\tilde{X}} -2\bar G_{3\hat \Phi}}{\bar G_4}
    \Bigg]=6D_\mu P^\mu\,.
\end{align}
Using
\[
\lie_n\varphi_1\dot\varphi_1 = \lie_n\varphi_1 \lie_n\varphi_1 = n^\mu D_\mu \varphi_1 n^\nu D_\nu \varphi_1 \,,
\]
and the Killing property $D_\mu n^\mu=0$, one finds
\begin{align}
    D_\mu\!\left(\varphi_1 n^\mu (n^\nu D_\nu \varphi_1)\right)
    = n^\mu D_\mu \varphi_1\, n^\nu D_\nu \varphi_1 \, .
\end{align}
Hence,
\begin{align}
    &D_\mu(-n^\mu\alpha(\theta,\phi) \varphi_1 \dot\varphi_1)\bar G_4\Bigg[ \nonumber 3\frac{\bar G_{4\hat \Phi}^2}{\bar G_4^2} +\frac{ \bar G_{2\tilde{X}} -2\bar G_{3\hat \Phi}}{\bar G_4}
    \Bigg]\\&\equiv D_\mu P^\mu\,.
\end{align}
Since the scalar contribution to the supertranslation flux
$\mathcal F_{\alpha(\theta,\phi)n}[\Delta\scrip]$ is thus a total
derivative, one can define the corresponding charge on an arbitrary
cross section $\partial\scrip$ at $u=u_0$,
\begin{align}\label{equ:charge_Horndeski}
  &\mathcal  Q^\text{Horndeski}_{\alpha(\theta,\phi) n} [\partial \scrip] = \frac{1}{2\kappa_0} \oint_{u=u_0} \epsilon_{\alpha\beta\gamma}P^{\gamma}\dd S^{\alpha\beta}\nonumber\\
    &=-\frac{\bar G_4}{2\kappa_0} \oint_{u=u_0}\alpha(\theta,\phi) \varphi_1 \dot \varphi_1\Bigg[  3\frac{\bar G_{4\hat \Phi}^2}{\bar G_4^2} +\nonumber\\
    &\quad\quad \frac{\bar  G_{2\tilde{X}} -2\bar G_{3\hat \Phi}}{\bar G_4}
    \Bigg] n^\mu\epsilon_{\mu\alpha\beta}\dd S^{\alpha\beta}\,.
\end{align}
This expression can be rewritten as a standard integral over the asymptotic $2$-sphere $\mathcal S^2$
\begin{align}\label{equ:charge_HorndeskiSimple}
   \mathcal Q^\text{Horndeski}_{\alpha(\theta,\phi) n} (u_0) =- &\frac{\bar G_4}{2\kappa_0} \oint_{\mathcal S^2}\alpha(\theta,\phi) \,\varphi_1 \dot \varphi_1\notag\\
    &\times\Bigg[  3\frac{\bar G_{4\hat \Phi}^2}{\bar G_4^2} +\frac{\bar  G_{2\tilde{X}} -2\bar G_{3\hat \Phi}}{\bar G_4}
    \Bigg] \,.
\end{align}
This expression is derived for a general supertranslation
\eqref{equ:original_ST_and_L}. Repeating the derivation for the
representative~\eqref{equ:real_st} yields the same result upon replacing
$\alpha\rightarrow\alpha + u_0/2 D_A Y^A$.

Turning to the first term in Eq.~\eqref{equ:super_flux}, an analogous
construction leads to the supermomentum charge for the GR-like sector of luminal Horndeski theory. The result is displayed in~\cite{Ashtekar:1981}, starting at Eq.~(5.1).\footnote{In Appendix~\ref{app:GR_mod}, we provide alternative formulations of the GR flux and charge that are commonly used in waveform data anaylsis.}
For our purposes, it is however sufficient to define the GR sector through the common parametrization via the Bondi mass aspect $M(u,\theta,\phi)$ through
\begin{equation}\label{equ:charge_GRSimple}
     \mathcal Q^\text{GR}_{\alpha(\theta,\phi) n}(u_0) = \frac{2\bar G_4}{\kappa_0} \oint_{\mathcal S^2}\alpha(\theta,\phi) M(u_0,\theta,\phi)
\end{equation}
Both contributions to the flux~\eqref{equ:super_flux} are thus expressed
as integrals over the unit $2$-sphere, so that the flux–balance relation
\eqref{equ:flalal} can be directly applied to the supertranslation flux.

\subsubsection{Shear and asymptotic strain in Horndeski theory}

The final step in the computation of the supermomentum flux balance consists in combining the charges with the asymptotic flux in Eq.~\eqref{equ:super_flux} following Eq.~\eqref{equ:flalal}.
However, to present a useful and complete version of the resulting expression, we first need to discuss the notion of asymptotic radiative information in beyond-GR theories.

Very generally, radiative information at $\scrip$ is encoded in the covariant derivative of the null normal $\ell_\mu$ from Sec.~\ref{ssSec:AsymptoticFlatness}. Concretely, one can show that the radiative degrees of freedom of the metric field at $\scrip$ are carried by the shear $\sigma$ defined as \cite{DAmbrosio:2022clk}
\begin{align}\label{eq:AsymptoticShear}
    \sigma \equiv - \lim_{r\rightarrow\infty}r^2(\hat m^\mu \hat m^\nu\hat \sigma_{\mu\nu})\,.
\end{align}
with
\begin{equation}
    \hat \sigma_{\mu\nu}= \hat \nabla_\mu \hat \ell_\nu - \frac{1}{2}\hat g_{\mu\nu}\hat\nabla^\rho\hat \ell_\rho
\end{equation}
where $\hat m_\mu,\hat{\bar m}_\mu$ are members of the Newman-Penrose null tetrad $\{\hat n,\hat\ell,\hat m, \hat{\bar m}\}$ constructed to follow luminal radiation to null infinity $\scrip$. Thus, the radiative shear is defined as the trace free part of the pullback $\newpb{\hat\nabla_\mu\hat\ell_\nu}$. 
The connection to radiation is thereby drawn from the relation to the Schouten tensor. 

\paragraph{Pure GR.} To set the stage, we quickly recall the notions of radiative degrees of freedom in pure GR. One can show that the concept of gravitaional radiation in terms of asymptotic shear is equivalent to the perhaps more prominent linearized version of gravitational radiation, captured by the leading order metric perturbation
\begin{equation}
    h_{\mu\nu}=\lim_{r\rightarrow\infty}(\hat g_{\mu\nu}-\eta_{\mu\nu})\,,
\end{equation}
with $\eta_{\mu\nu}$ the Minkowski metric in asymptotic Minkowski coordinates $\{t,x,y,z\}$. In the case of GR, the shear carries two tensor degrees of freedom, which can be expressed in the form of two polarizations $h_+,h_\times$ of the transverse-traceless component of metric perturbation \cite{Ashtekar_2017}
\begin{equation}
    h^\text{TT}_{\mu\nu}=\frac{1}{r}\left(h_+ e^+_{\mu\nu}+h_\times e^\times_{\mu\nu}\right)\,.
\end{equation}
where
\begin{align}
    e^+_{\mu\nu}&=m_\mu m_\nu+\bar m_\mu\bar  m_\nu\,,\\
     e^\times_{\mu\nu}&=-i(m_\mu m_\nu-\bar m_\mu\bar  m_\nu)\,.
\end{align}
Explicitly, we have at $\scrip$ 
\begin{align}\label{eq:defh}
    2\bar \sigma^\text{GR} \simeq h_+-ih_\times \equiv  h \,, 
\end{align}

This shear tensor of GR is also equivalent to the dynamical $1/r$ transverse-traceless metric component denoted as $ c_{AB}$,
where the capital Latin indices run over the angular coordinates only, arising in the definition of the general asymptotically flat metric within the Bondi framework \cite{Bondi:1962px, sachsGravitationalWavesGeneral1962} (see Appendix~\ref{app:Bondi_coord}).
 More precisely, this leading order angular component of the Bondi metric is equivalent to $\hat \sigma^\text{GR}_{\mu\nu}|_{\scrip}= \hat \Gamma\ud{u}{\mu\nu}|_{\scrip}$ at $\scrip$, where $\hat\Gamma$ is the Levi-Civita connection with respect to the Bondi metric. 
At the same time this dynamical asymptotic metric component naturally describes the transverse-traceless tensor space of the asymptotic metric
\begin{equation}
    c_{AB} = c_{\mu\nu} e^A_\mu e^B_\nu = \frac{r}{2} h^\text{TT}_{\mu\nu} e_A^\mu e_B^\nu
\end{equation}
where the tangent vectors to the unit $2$-sphere, $e_A^\mu$, provide the associated embedding of the tensor degrees of freedom. 
In GR, we thus have
\begin{equation}
    \hat\sigma^\text{GR}_{\mu\nu}|_{\scrip}\equiv \sigma^\text{GR}_{\mu\nu} \simeq c_{\mu\nu}\,
\end{equation}
where\footnote{In this context, it is also standard to define the Bondi news tensor $ N_{\mu\nu} = \partial_u c_{\mu\nu}$
describing the ``time derivative'' of the radiative information, a quantity that vanishes if no gravitational waves are present. Any passage of gravitational radiation is inevitably connected to a non-trivial Bondi news tensor for a finite $u$-interval.}
\begin{equation}\label{eq:defc}
    c_{\mu\nu}=\frac{r}{2}h^\text{TT}_{\mu\nu}=\frac{1}{2}\left( h_+ e^+_{\mu\nu}+h_\times e^\times_{\mu\nu}\right)\,,
\end{equation}

\paragraph{Luminal Horndeski.}

Let us now turn to an analog consideration for beyond-GR theories and in particular luminal Horndeski theory. The presence of additional luminally propagating degrees of freedom is a key structural property of theories beyond GR. Within a perturbative framework, it is well known that such extra propagating degrees of freedom can source additional polarizations of the metric \cite{Eardley:1973zuo,Eardley:1973zzz,poisson2014gravity,Will:2018bme,heisenberg2025unifyingordinarynullmemory,Zosso:2024xgy}. This phenomenon typically arises in the presence of non-minimal couplings in the Lagrangian.

Within the fully non-linear formulation this implies that the asymptotic shear defined in Eq.~\eqref{eq:AsymptoticShear} as the pullback of $\hat \nabla_\mu \hat\ell_\nu$ to $\scrip$, may no longer only describe two tensor degrees of freedom. Indeed, for luminal Horndeski theory, the asymptotic shear is related to the tensor degrees of freedom $ c_{\mu\nu}$ in Eq.~\eqref{eq:defc}
as 
\begin{align}\label{equ:shear_red}
    \sigma_{\mu\nu} \simeq  c_{\mu\nu} - \frac{\bar G_{4\hat\Phi}\varphi_1}{\bar G_{4}} \gamma_{\mu\nu}\,,
\end{align}
where $\gamma_{\mu\nu} $ is the embedding of the 2-sphere metric. As we explicitly show in Appendix~\ref{app:asymptotic_Horn} this shear tensor is still equivalent to the $1/r$ component of the angular sector of the Bondi metric within the Jordan-frame formulation. However, for a non-trivial coupling to the Ricci scalar, such that $\bar G_{4\hat\Phi}\neq 0$ this dynamical metric component acquires an additional polarization in terms of the scalar degree of freedom on top of the familiar tensor polarizations. 

Physically, this implies that the radiative part of the physical metric, which for instance influences the geodesic deviation of freely falling test masses within current GW experiments, would directly feel the presence of additional scalar radiation. On the other hand, for Horndeski theories with $\bar G_{4\hat\Phi}= 0$, the scalar degree of freedom does not excite an additional gravitational polarizations and GW experiments would not directly be sensitive to the presence of scalar GWs. This is for instance the case in scalar-Gauss-Bonnet gravity considered in Sec.~\ref{sSec:sGB} below. This does not mean, however, that in this case the presence of scalar radiation has no effect, since the detectable tensor radiation will indirectly be influenced by the presence of an additional emission channel.


\subsubsection{Flux balance laws in luminal Horndeski theory}
The observation that the shear carries additional degrees of freedom, i.e., Eq.~\eqref{equ:shear_red}, is crucial for the computation of a consistent flux balance.  
Starting with Eq.~\eqref{equ:super_flux} for supertranslations only, we find 
\begin{align}
    \label{equ:result_tot_flux}
    &\mathcal F_{\alpha(\theta,\phi) n}[\Delta\scrip]=-\frac{\bar G_4}{2\kappa_0} \int_{\Delta \scrip} \alpha (\theta,\phi)\Bigg(\frac{1}{2} \dot\sigma_{\mu\nu}\dot\sigma^{\mu\nu} + \nonumber\\&+D_\mu D_\nu \dot\sigma^{\mu\nu}+ \dot\varphi_1^2 
    \Bigg[  3\frac{\bar G_{4\hat \Phi}^2}{\bar G_4^2}+\frac{ \bar G_{2\tilde{X}} -2\bar G_{3\hat \Phi}}{\bar G_4}
    \Bigg]\Bigg)\notag\,,
\end{align}
which yields
\begin{align}
    &=-\frac{\bar G_4}{2\kappa_0} \int_{\Delta \scrip} \alpha (\theta,\phi)\Bigg(\frac{1}{2} \dot c_{\mu\nu}\dot c^{\mu\nu} + D_\mu D_\nu \dot c^{\mu\nu}\nonumber\\&+\frac{\bar G_{4\hat \Phi}}{\bar G_{4}}D_\mu D_\nu\dot \varphi_1+ \dot\varphi_1^2 
    \Bigg[  3\frac{\bar G_{4\hat \Phi}^2}{\bar G_4^2}+\frac{ \bar G_{2\tilde{X}} -2\bar G_{3\hat \Phi}}{\bar G_4}
    \Bigg]\Bigg)\,,
\end{align}
where we used that $c_{\mu\nu}\gamma^{\mu\nu}=0$, i.e., the shear is trace-free. This flux is balanced by the difference in asymptotic supermomentum charge 
\begin{align}
    \mathcal F_{\alpha(\theta,\phi) n}[\Delta\scrip]=-\Delta \mathcal Q_{\alpha(\theta,\phi) n}\,,
\end{align}
where $\Delta \mathcal Q \equiv \mathcal Q(u)-\mathcal Q(u_1)$, and the selected cross section $\Delta \scrip$ extends from a given $u_1$ to an arbitrary retarded time $u$. The asymptotic supermomentum charge on the other hand is given by the two contributions in Eqs.~\eqref{equ:charge_GRSimple} and \eqref{equ:charge_HorndeskiSimple}
\begin{align}
     \mathcal Q_{\alpha(\theta,\phi) n}(u)&= \mathcal Q^{\text{GR}}_{\alpha(\theta,\phi) n}(u)+\mathcal Q^{\text{Horndeski}}_{\alpha(\theta,\phi) n}(u)\nonumber \\
     &= \frac{2\bar G_4}{\kappa_0} \oint_{\mathcal S^2}\alpha(\theta,\phi) \mathfrak M(u,\theta,\phi)\,.
\end{align}
where we define the generalized Bondi mass aspect
\begin{align}
    \mathfrak M \equiv M - \frac{1}{4}\varphi_1\dot\varphi_1\Bigg[  3\frac{\bar G_{4\hat \Phi}^2}{\bar G_4^2}+\frac{ \bar G_{2\tilde{X}} -2\bar G_{3\hat \Phi}}{\bar G_4} \Bigg] \,.
\end{align}
which incorporates the GR-like Bondi mass aspect $M$, defined in Eq.~\eqref{equ:charge_GRSimple}, as well as the Horndeski contribution in Eq.~\eqref{equ:charge_HorndeskiSimple}, computed in the previous subsection.

By rearranging the total flux Eq.~\eqref{equ:result_tot_flux} we therefore arrive at the total supertranslation flux–balance law
\begin{widetext}
    \begin{align}\label{equ:the_BL}
       \int_{\Delta \scrip} \alpha (\theta,\phi) \text{Re}(\eth^2 \dot h) =- \underbrace{\oint_{\mathcal S^2} \alpha (\theta,\phi) \bigg[4 \Delta \mathfrak M - \frac{\bar G_{4\hat \Phi}}{\bar G_{4}}D_\mu D_\nu\Delta \varphi_1\bigg]}_{=:\Delta \mathcal Q_{\alpha(\theta,\phi)n}} + \underbrace{\int_{\Delta \scrip} \alpha (\theta,\phi) \bigg(\frac{1}{4} |\dot h|^2 + \dot\varphi_1^2 
    \Bigg\{  3\frac{\bar G_{4\hat \Phi}^2}{\bar G_4^2}+\frac{ \bar G_{2\tilde{X}} -2\bar G_{3\hat \Phi}}{\bar G_4}
    \Bigg\}\bigg)}_{=:\mathcal P _{\alpha(\theta,\phi)n}}
    \end{align}
\end{widetext}
where we have used Eq.~\eqref{eq:defc} as well as the definition in Eq.~\eqref{eq:defh}. The operator $\eth$ denotes the covariant derivative operator on the unit $2$--sphere. Through formulating this equation, we have achieved our goal of deriving a constraint equation for full, nonlinear luminal Horndeski theory. This expression explicitly reveals the dependence of the strain on the dynamics of the scalar-field sector. In particular, the scalar field contributes both to the variation of the supermomentum charge $\mathcal Q_{\alpha(\theta,\phi)}$, explicitly via the term proportional to $D_\mu D_\nu\Delta\varphi_1$ and implicitly via the modified Bondi mass aspect, and to the quadratic flux through null infinity, $\mathcal P_{\alpha(\theta,\phi)}$, which effectively describes the energy flux of the propagating degrees of freedom across $\Delta \scrip$.
We stress that Eq.~\eqref{equ:the_BL} constitutes a genuine constraint equation for the full strain as a function of the retarded time $u$. It provides a pointwise consistency condition at each instant of time and for each direction on the sky, and must therefore be satisfied by any physically admissible radiating system. 

In particular, note that by choosing $\alpha (\theta,\phi) =1$, which is equivalent to considering a standard Poincar\'e time translation, the left-hand side integrates to zero, such that the balance equation can be written as a generalized Bondi-massloss formula\footnote{As a cross-check, this evolution equation for the generalized Bondi mass aspect is equally obtained by solving Einstein's equations within the Bondi framework (see for instance for Brans--Dicke theory \cite{tahura_brans-dicke_2021}).}
\begin{align}\label{equ:evo_mass_aspect}
    \partial_u \mathfrak M =& \frac{1}{4}\Bigg(\frac{1}{4} |\dot h|^2 -\text{Re}(\eth^2 \dot h)+\frac{\bar G_{4\hat \Phi}}{\bar G_{4}}D_\mu D_\nu\dot \varphi_1 \notag\\&+ \dot\varphi_1^2 
    \Bigg[  3\frac{\bar G_{4\hat \Phi}^2}{\bar G_4^2}+\frac{ \bar G_{2\tilde{X}} -2\bar G_{3\hat \Phi}}{\bar G_4}
    \Bigg]\Bigg)\,.
\end{align}

\subsection{Extracting the gravitational memory offset}
Equation~\eqref{equ:the_BL} may also be viewed from a complementary perspective. To this end, let us assume that the cross section $\Delta \scrip$, over which both sides of Eq.~\eqref{equ:the_BL} are evaluated, extends over the entire null infinity $\scrip$, that is, we take the upper limit of integration to be $u \rightarrow \infty$. We find
\begin{align}\label{equ:BMSSupermomentumBalanceLawBD2}
  \oint_{S^2} d^2\Omega\,\alpha(\theta,\phi)\,\text{Re}[\eth^2\Delta{h}]= \mathcal{P}_{\alpha(\theta,\phi)n} - \Delta\mathcal Q_{\alpha(\theta,\phi)n}
\end{align}
with (as above)
\begin{equation}\label{equ:mem_def}
    \Delta{h}\equiv h(\infty)-h(-\infty)\,.
\end{equation}
With this definition, the late-time limit of Eq.~\eqref{equ:BMSSupermomentumBalanceLawBD2} yields a decomposition of the GW memory, namely, the permanent displacement of test masses induced by the passage of gravitational radiation.

To make this feature explicit, we isolate the strain appearing on the left-hand side of Eq.~\eqref{equ:BMSSupermomentumBalanceLawBD2}, which can be achieved as follows:\footnote{Equivalently, one may remove the angular integrations and employ the identity $\bar\eth^{2}\eth^{2} h = \mathcal D^{2}(\mathcal D^{2}-2)h$, with $\mathcal D^{2}=\bar\eth\eth$ denoting the spin-weighted Laplacian on the two-sphere, and then invert the resulting differential operators, as done, for example, in \cite{Mitman:2020bjf,Zosso:2026czc} in the case of GR.}
Since all terms involve integrals over the unit $2$-sphere, we expand their angular dependence in spin-weighted spherical harmonics (SWSHs).The GW
strain carries spin weight $s=-2$ and can thus be decomposed as
\begin{equation}
   h(u,\theta,\phi)
   =\sum_{l=2}^\infty \sum_{m=-l}^l
   h_{lm}(u)\,{}_{\scriptscriptstyle -2}Y_{lm}(\theta,\phi)\,.
\end{equation}
Using the identity
\begin{align}
    \eth^2\,{}_{\scriptscriptstyle -2}Y_{lm}
    =\sqrt{\frac{(l+2)!}{(l-2)!}}\,Y_{lm}\,,
\end{align}
and choosing $\alpha(\theta,\phi)=Y^*_{lm}(\theta,\phi)$, the orthogonality relations of the spherical harmonics imply that the left-hand side of
Eq.~\eqref{equ:BMSSupermomentumBalanceLawBD2} naturally projects onto the electric-parity
component of the spin-weighted multipole (see, e.g., Appendix~D.1 of
\cite{Zosso:2024xgy}),
\begin{equation}\label{eq:AHlmToUlmVlm}
    h_{lm}=\frac{1}{\sqrt{2}}\left[h^E_{lm}-ih^M_{lm}\right]\,,
\end{equation}
which can be inverted to yield
\begin{align}
    h^E_{lm}
    =\frac{1}{\sqrt{2}}
    \left[h_{lm}+(-1)^m h^*_{l,-m}\right]\,.
    \label{eq:AUVlmToHlm}
\end{align}
One then finds (see, e.g., Ref.~\cite{Heisenberg:2023prj})
\begin{align}\label{BMSSupermomentumBalanceLawBD3}
\boxed{
  \Delta h^E_{lm}
  =\sqrt{\frac{2(l-2)!}{(l+2)!}}
  \Big[\mathcal{P}_{Y^*_{lm}}-\,\Delta \mathcal Q_{Y^*_{lm}}\Big]\,,
}
\end{align}
where $\mathcal{P}_{Y^*_{lm}}$ and $\Delta \mathcal Q_{Y^*_{lm}}$ are defined in Eq.~\eqref{equ:the_BL} in which we replace
$\alpha(\theta,\phi)=Y^*_{lm}(\theta,\phi)$. The right-hand side can be further simplified by expanding all fields entering $\mathcal{P}$ and $\mathcal Q$ in SWSHs, allowing the angular integrals to be evaluated analytically. Note that Eq.~\eqref{BMSSupermomentumBalanceLawBD3} is only valid for $l\geq 2$, since 
\begin{equation}
    \sqrt{\frac{(l+2)!}{(l-2)!}}=\sqrt{(l+2)(l+1)l(l-1)},
\end{equation}
For $l=0$ and $l=1$, the left hand side in Eq.~\eqref{equ:BMSSupermomentumBalanceLawBD2} vanishes and the balance laws describe the Bondi mass loss and momentum loss formulas.

Eq.~\eqref{BMSSupermomentumBalanceLawBD3} constitutes the final form of the flux balance relation appearing in Eq.~\eqref{equ:the_BL}. It interpretation as a constraint equation for the 
(electric-parity sector of the)
asymptotic strain becomes most transparent when fixing the initial retarded time to $u_1\to-\infty$ and choosing the post-Newtonian Bondi frame, for which
\begin{equation}
    \lim_{u_1\rightarrow-\infty} h(u_1)=0\,.
\end{equation}
In this frame, the initial values of both sides of
Eq.~\eqref{BMSSupermomentumBalanceLawBD3} vanish, leaving a consistency condition for each mode $h^E_{lm}(u_2)$.\footnote{Using the Bianchi identities, one may equivalently rewrite Eq.~\eqref{BMSSupermomentumBalanceLawBD3} as a consistency relation for the full strain, see \cite{Mitman:2020bjf,Zosso:2026czc}.} In this form, the balance laws provide a stringent sanity check for state-of-the-art waveform models
\cite{Ashtekar:2019viz,Mitman:2020bjf,DAmbrosio:2024zok}.

On the other hand, taking also the limit $u_2\to+\infty$,
Eq.~\eqref{BMSSupermomentumBalanceLawBD3} yields an explicit expression for the permanent
memory offset,
\[
\lim_{u_{2,1}\rightarrow\pm\infty}\Delta h
= h(+\infty)-h(-\infty)\,.
\]
Each electric-parity strain mode acquires a net offset determined by the flux through
$\scri$ and by the change in the associated charges, while the magnetic parity contribution does not contain any offset. In particular, both the presence of an
asymptotic energy-momentum flux of gravitational radiation,
$\mathcal{P}_{\alpha(\theta,\phi)}$, and changes in the supermomentum charges,
$\Delta\mathcal Q^{\text{GR}}_{\alpha(\theta,\phi)}$ and
$\Delta\mathcal Q^{\text{Horndeski}}_{\alpha(\theta,\phi)}$, induce a nonvanishing difference
between the initial and final values of the electric-parity strain.

More concretely, the two contributions $\mathcal P$ and $\Delta\mathcal Q$ correspond to distinct
types of gravitational-wave memory. The flux term $\mathcal P$ is defined to give rise to the
\emph{null memory}, while the contribution associated with $\Delta \mathcal Q$ is known as
\emph{ordinary memory}. Restricting attention to supertranslations, both effects contribute
to displacement memory. Physically, null memory is sourced by an energy flux reaching
$\scri$, whereas ordinary memory is associated with changes in the bulk energy--momentum
distribution that do not propagate to null infinity, such as unbound massive matter or
black-hole remnant kicks.

One of the most noteworthy features of the theory under consideration is that the scalar degree of freedom directly influences the observed gravitational radiation. This is immediately apparent from Eq.~\eqref{equ:the_BL}, since both components of the memory effect receive corrections induced by the scalar field. The classification of scalar contributions to the different types of gravitational memory, however, is not entirely unambiguous.
A natural approach based on linearity is to associate the terms proportional to $(\partial_u \varphi_1)^2$ with a radiative flux of the scalar’s propagating degrees of freedom. By contrast, the contribution to the supermomentum charge is intrinsically linear in $\varphi_1$ (at least for non-trivial choices of $G_4$) and may therefore be identified as sourcing linear GW memory. At this stage, one might be tempted, for a specific choice of $G_4$, to interpret this term as a scalar GW memory that is simultaneously to the tensor-type memory constrained through both the flux and the variation of the modified mass aspect. Such an interpretation would, however, require explicit knowledge of the mass aspect as well as of the nonradiative scalar data ($\Delta \varphi_1$). Because the memory equation \eqref{BMSSupermomentumBalanceLawBD3} does not provide sufficient freedom to determine $\varphi_1$, this viewpoint cannot be consistently implemented.
Finally, we emphasize a key structural property of Eq.~\eqref{BMSSupermomentumBalanceLawBD3}: the scalar field contributes through both radiative data, $\partial_u \varphi_1$, and nonradiative data, $\Delta\varphi_1$, in close analogy with the gravitational sector, which enters via $\varsigma_{\mu\nu}$ and $\Delta \mathfrak M$, respectively. Consequently, even in the absence of scalar radiative modes, one generically expects a nonvanishing contribution to the gravitational memory.

\subsection{Lorentz flux balance laws}
So far, we have only been concerned with one subgroup of the full BMS group generators, Eqs.~\eqref{equ:real_L} and \eqref{equ:real_st}. As the Lorentz flux balance generated by \eqref{equ:real_L} is not of primary interest to this investigation, we will only comment briefly on its derivation: Following the arguments of \cite{waldGeneralDefinitionConserved2000}, the charge associated to the Lorentz generators is given by the Noether 2-charge, $\boldsymbol Q$, integrated over a $u=$const. cross section of $\scrip$, 
\begin{align}
    Q_{\xi_\text{L}}[\partial\scrip]= \oint_{u=u_0} \boldsymbol Q_{\xi_\text{L}}\,.
\end{align}
The Noether charge itself can be computed using the Noether current, Eq.~\eqref{equ:Noether_current}. We emphasize at this point that, because we chose $\xi_\text{L}$ to be tangent to an arbitrary cross section of $\scrip$, all quantities contracted with $\xi_\text{L}$ and integrated over a given cross section will trivialize. 

Another strategy for the computation of the charge assoicated to the Lorentz generators is, analogous to the above, to compute the modified angular momentum aspect. The evolution of the latter can be defined by explicitly computing the flux and replacing the definition of the Bondi news tensor with the time-derivative of the shear $\sigma_{\mu\nu}$. A straightforward but slightly more lengthy calculation then leads to an equation structurally similar to Eq.~\eqref{equ:the_BL}. With the charge computed for both the GR-like and the scalar sector, the construction of the flux balance follows analogously to the supermomentum balance equation. The result provides an additional constraint and (subleading displacement) memory equation that can be applied to (numerical) gravitational waveforms.




\section{Conjecture for full Hordeski theory}


While the previous section explicitly derived the BMS balance laws for the massless and luminal Horndeski subclass, we now argue that higher-order operators do not alter these results. This will also give us a chance to relate our findings to alternative methods for computing gravitational memory beyond GR. We present explicit evidence for scalar-Gauss-Bonnet gravity. 

\subsection{A quantitative comparison to Isaacson's approach}
\label{subsec:comparison_Isaacson}

Before discussing the extension of these results to a broader class of theories with the WZ formalism, we relate them to previous computations performed within the Isaacson viewpoint on gravitational memory \cite{Heisenberg:2023prj,heisenberg2025unifyingordinarynullmemory,Heisenberg:2025tfh,Heisenberg:2025roe,Zosso:2025ffy,Zosso:2024xgy}, which include results for memory in full Horndeski theory \cite{Heisenberg:2023prj,heisenberg2025unifyingordinarynullmemory}.

Although these previous works explicitly compute only the memory, rather than the complete balance laws as a constraint equation for the full strain, we can directly compare results in the form of Eq.~\eqref{BMSSupermomentumBalanceLawBD3} within the memory limit $u_{2,1}\to \pm\infty$. Disregarding the subdominant ordinary memory, we obtain from Eq.~\eqref{BMSSupermomentumBalanceLawBD3}
\begin{align}\label{equ:MemNull}
  \lim_{u_{2,1}\rightarrow\pm\infty} \Delta h^E_{lm}
  =\,&\sqrt{\frac{2(l-2)!}{(l+2)!}} \int_{-\infty}^{\infty}\dd u\oint\dd^2\boldsymbol\Omega\,Y^*_{lm} \Bigg[|\dot{h}|^2\nonumber\\&+ (\dot\varphi_1)^2\bigg( 3\frac{\bar G_{4\hat \Phi}^2}{\bar G_4^2}+\frac{ \bar G_{2\tilde{X}}}{\bar G_4} -\frac{ 2\bar G_{3\hat \Phi}}{\bar G_4}
    \bigg)\Bigg]\,.
\end{align}
This result agrees with Eqs.~(90), (92) and (100) in \cite{Heisenberg:2023prj} in the limit $u\to \infty$ and for a vanishing vector field sector. Note, however, that in contrast to the flux balance laws in Eq.~\eqref{BMSSupermomentumBalanceLawBD3}, which characterize the total memory offset, the Isaacson result provides a formula for an isolated time-dependent memory rise $\delta h(u)$. This is achieved in the Isaacson approach through the additional physical input of a separation of scales between the GWs and the background spacetime. This assumption naturally introduces a spacetime averaging $\langle...\rangle$ over oscillatory scales, which can isolate a time-dependent memory contribution as a perturbative change of the background \cite{Zosso:2025ffy,Zosso:2026czc}. 

On the other hand, the Isaacson picture provides access to the ordinary memory, which can be computed straightforwardly using the bulk matter energy-momentum tensor $\hat T_{\mu\nu}$, as shown in \cite{heisenberg2025unifyingordinarynullmemory}. What remains to be clarified is how the Isaacson framework accounts for the additional ordinary memory contribution arising from the higher SWSH modes of the scalar memory. This is left for future work.

While the correspondence of the Isaacson results to the flux balance laws at the level of the total null memory is not surprising\footnote{Although the two approaches rely on very distinct mathematical frameworks, this correspondence can be viewed as a manifestation of the universality of gravitational memory within the asymptotic radiation. Indeed, in the limit of flat asymptotics there exists a direct correspondence between a perturbative treatment and full theory of radiation fields \cite{Ashtekar_2017}.} it is at first sight striking that the Isaacson results for the \emph{full} Horndeski theory exactly match with the presented computation in the subset of luminal Horndeski theory. Indeed, it appears that all operators that were disregarded in the full theory in Eq.~\eqref{eq:fullHorndeski} to define luminal Horndeski in Eq.~\eqref{equ:luminal_horn}, in particular the full $L_5$ term as well as the derivative of $G_4$, do not contribute to gravitational memory.

From the Isaacson perspective, the absence of higher-derivative operator couplings in the
gravitational memory formula can be understood in light of a general theorem formulated
in \cite{Heisenberg:2023prj}. This theorem characterizes the asymptotic
energy--momentum tensor as a genuinely nonlinear object which, nevertheless, depends
exclusively on the second-order action of the theory. As a consequence, only those
operators that enter the leading-order linearized equations of motion can contribute to
the gravitational null memory. 
It follows that for any theory of gravity with higher-derivative field equations, thus even including theories beyond the Horndeski class, which nevertheless admits decoupled second-order linearized equations around asymptotically flat spacetimes, the memory formula splits into independent sectors associated with each propagating degree of freedom. Moreover, the resulting memory expressions involve only operators containing at most two spacetime derivatives acting on the fields. This conclusion applies in particular to theories whose quadratic higher-curvature operators combine into a topological invariant, such as scalar--Gauss--Bonnet gravity, discussed below, or dynamical Chern-Simons gravity \cite{Jackiw:2003pm,Alexander:2009tp,hou_gravitational_2022}.\footnote{More generally, for quadratic gravity theories which naively lead to higher-order linearized equations of motion, the theorem continues to hold in the small-coupling
approximation. This approximation is required to render the theory predictive by
eliminating destabilizing ghost degrees of freedom
\cite{Heisenberg:2023prj,Zosso:2024xgy}.}


\subsection{Scalar Gauss Bonnet Gravity}\label{sSec:sGB}

To further shed light on the importance of higher derivative operators, we additionally explicitly perform the computation for the very popular higher order operator theory of scalar-Gauss-Bonnet (sGB) gravity \cite{Zwiebach:1985uq,Gross:1986iv,Moura:2006pz,Pani:2009wy}
\begin{align}
    S=\frac{1}{2\kappa_0}\int d^4x\sqrt{-g}(R-\frac{1}{2}(\nabla\hat\Phi)^2+\lambda f(\hat\Phi)\mathcal{G})\,,
\end{align}
where the Gauss-Bonnet curvature scalar is defined as 
\begin{equation}
    \mathcal{G}\equiv -\tilde R\ud{\mu\nu}{\rho\sigma} \tilde R\ud{\rho\sigma}{\mu\nu} =R^{\mu\nu\rho\sigma}R_{\mu\nu\rho\sigma}-4R^{\mu\nu}R_{\mu\nu}+R^2,
\end{equation}
with Hodge dual $\tilde R\ud{\mu\nu}{\rho\sigma}\equiv\frac{1}{2}\epsilon^{\mu\nu\alpha\beta}R_{\alpha\beta\rho\sigma}$,
$\lambda$ a coupling constant with dimensions of $[\text{mass}]^{-2}$ and $f(\hat\Phi)$ an arbitrary function of the scalar field $\hat\Phi$. Note that for constant values of the scalar field, the theory reduces locally to GR because the Gauss-Bonnet term is topological. For this theory, the balance laws 
Eq.~\eqref{equ:BMSSupermomentumBalanceLawBD2} can be written as 
\begin{align}\label{equ:BMSSupermomentumBalanceLaw SGB}
  \oint_{S^2} d^2\Omega\,\alpha(\theta,\phi)\,\text{Re}[\eth^2\Delta{h}]&= \mathcal{P}^\text{sGB}_{\alpha(\theta,\phi)n }-  \Delta\mathcal Q^\text{sGB}_{\alpha(\theta,\phi)n}\,.
\end{align}
where compared to Eq.~\eqref{equ:the_BL} we have
\begin{align}\label{equ:PPPsGB}
    \mathcal{P}^\text{sGB}_{\alpha(\theta,\phi)n }\equiv  \int_{\Delta\scrip}\dd u\dd^2\boldsymbol\Omega\, \,\alpha(\theta,\phi) \Big[\frac{1}{4}|\Dot{h}|^2+ (\dot\varphi_1)^2\Big]\,,
\end{align}
while
\begin{equation}
    \mathcal Q^\text{sGB}_{\alpha(\theta,\phi)n}=4\oint_{S^2} \dd^2\boldsymbol\Omega\,\alpha(\theta,\phi)\bigg(\underbrace{M-\frac{1}{4}\mathcal \varphi_1\dot \varphi_1}_{=:\mathcal M}\bigg),
\end{equation}
where $M(u,\theta,\phi)$ represents here the GR-like mass aspect, which solely depends on the tensor metric degrees of freedom.\footnote{Note, however, that due to the presence of the scalar field, its associated averaged Bondi mass is no longer a strictly decreasing quantity.}\footnote{In contrast to luminal Horndeski theory, sGB gravity does not give rise to additional propagating degrees of freedom in the metric sector, i.e., $\sigma_{\mu\nu}\sim\varsigma_{\mu\nu}$. This can be explicitly shown by computing the decay behavior for an asymptotically flat spacetime metric (as it is done for luminal Horndeski in Appendix~\ref{app:asymptotic_Horn}). One finds that both the $1/r$--term of the angular sector of the asymptotic metric as well as the evolution equation of the Bondi mass aspect remain unaffected by the Gauss-Bonnet term. The scalar's effective energy--stress tensor only contributes to the evolution of the mass aspect. Thus, there are no additional scalar-induced polarizations in the metric.} For the modified mass aspect in sGB gravity, $\mathcal M$, one finds an evolution equation reading
\begin{align}
    \partial_u \mathcal M = \frac{1}{4}\Bigg(\frac{1}{4} |\dot h|^2 -\text{Re}(\eth^2 \dot h)+ \dot\varphi_1^2 \Bigg)\,.
\end{align}

The result for sGB could have been anticipated from the expression of luminal Horndeski, since SGB theory is related to Eq.~\eqref{eq:fullHorndeski} by choosing \cite{Kobayashi:2011nu,Kobayashi:2019hrl}
\begin{equation}\label{eq:CorrespondencesGBHorndeski}
    \begin{split}
        G_2&=\hat X+8f^{(4)}(\hat \Phi)\hat X^2(3-\ln \hat X)\,,\\ G_3&=4f^{(3)}(\hat \Phi)\hat X(7-3\ln \hat X)\,,\\
        G_4&=1+4f^{(2)}(\hat \Phi)\hat X(2-\ln \hat X)\,,\\
        G_5&=-f^{(1)}(\hat \Phi)\ln \hat X\,,
    \end{split}
\end{equation}
where $f^{(n)}(\hat \Phi)\equiv\partial^n f/\partial\hat \Phi^n$. Indeed, one can verify that these definitions of $G$-functionals evaluated in the limit of an asymptotically flat spacetime satisfy \cite{Heisenberg:2023prj} 
\begin{align}\label{equ:sGB_bracket}
    \bigg( 3\frac{\bar G_{4\hat \Phi}^2}{\bar G_4^2}+\frac{ \bar G_{2\tilde{X}}}{\bar G_4} -\frac{ 2\bar G_{3\hat \Phi}}{\bar G_4}
    \bigg)\Bigg\lvert_\text{sGB}=1.
\end{align}
Hence, the higher-order sGB term does not modify the memory formula and the theory simply contributes through the canonical scalar term within the action.

\subsection{General absence of higher order operators}
\label{subsec:fullHorndeski}
Based on the above considerations of the explicit example of sGB gravity, together with the direct comparison to results obtained via Isaacson’s approach, we want to formulate the following conjecture
\begin{remark}
    For the full class of massless Horndeski theories characterized by the Lagrangian in Eq.~\eqref{eq:fullHorndeski} admitting an asymptotically flat solution as defined in Sec.~\ref{ssSec:AsymptoticFlatness}, the flux across null infinity $\scrip$ and the associated flux balance equation are given by Eqs.~\eqref{equ:super_flux} and~\eqref{equ:the_BL}, respectively. 
\end{remark}
In the following, we present further indicative evidence supporting the validity of this statement. Our argument is distilled to an analysis of the derivative operators appearing in the relevant expression and the constraints imposed by their structure. We thereby assume that for the full Horndeski theory the asymptotically flat structure of spacetime including the associated asymptotic symmetries is still BMS. For luminal Horndeski theory this has been shown in Appendix~\ref{app:asymptotic_Horn}. Generalizing the assumption to the full theory is motivated by the fact that the extra terms, i.e., $G_{4,\hat X}$ and $L_5$, do not contribute to the relevant lowest order equation of motions. Thus, for the crucial order in the expansion in $1/r$, the equations remain GR-like with an effective energy-stress tensor of the scalar field yielding only next-to-leading order contributions. 

Recall that Eqs.~\eqref{equ:super_flux} and~\eqref{equ:the_BL} were derived explicitly in the context of luminal Horndeski theory defined in Eq.~\eqref{equ:luminal_horn}, that is, in the absence of an $\hat X$–dependent $G_4$ and without contributions from $L_5$. Yet, within the WZ framework we observe a restriction on the number of derivative operators appearing in the contributing terms that closely parallels the theorem established in \cite{Heisenberg:2023prj} for the Isaacson prescription. Specifically, at the level of the Lagrangian, the terms contributing to the final flux contain at most two derivatives acting on the fields, up to derivative operators hidden in the Horndeski functionals $G_i$.\footnote{Since the Horndeski functionals depend on $\hat\Phi$ or $\hat X$, both of which approach either a fixed value $\varphi_0$ or zero at $\scrip$, they are irrelevant for the present discussion under the assumptions on the $X$ dependence stated in Appendix~\ref{app:assumptionsonG}.} Moreover, the flux itself is only composed of operators with at most two derivatives acting on the fields.
In Appendix~\ref{app:GB}, the effect of higher-derivative terms on the pullback to $\scrip$ in the WZ approach is demonstrated explicitly using the Gauss–Bonnet term as an example. Guided by this result, one can formulate a heuristic argument in favor of the conjecture that applies to all possible terms arising in higher-derivative extensions of luminal Horndeski theory:

First, consider a term involving the scalar field and containing more than two derivatives (at the Lagrangian level and up to the Horndeski functionals). Since the overall expression must again be a scalar, the number of derivatives can only be increased by an even integer. We therefore examine terms of the schematic form $\mathcal O(\nabla^4)$ acting on one or more scalars. In the symplectic potential, such terms reduce to $\mathcal O(\nabla^3)$, with the remaining free index contracted with the null normal $n_\mu$ (c.f. Appendix~\ref{app:GB}). For each scalar field involved, the expression acquires one power of $\Omega$ (if a derivative or variation is applied), while the contractions of the derivative operators necessarily involve at least two metric tensors, each scaling as $\Omega^2$. Altogether, this yields a contribution of at least $\mathcal O(\Omega^5)$, which, in view of $\tilde{\boldsymbol{\epsilon}}=\Omega^{-4}\boldsymbol{\epsilon}$, is sufficient to ensure a vanishing pullback to $\scrip$. Including additional derivatives necessarily increases the number of metric contractions and thus leads to an even faster decay toward $\scrip$.

Next, consider terms with more than two derivatives acting exclusively on the metric, such as those appearing in sGB gravity. An argument analogous to the previous one shows that such contributions behave at least as $\mathcal O(\nabla^4)$ at the Lagrangian level and as $\mathcal O(\nabla^3)$ in the symplectic potential. The combined metric factors then decay as $\mathcal O(\Omega^4)$, independently of the specific index contractions. At least one of these factors becomes a variation in the symplectic potential. Since $\delta g_{\mu\nu}$ vanishes at $\scrip$ proportionally to $\Omega$ (note here that $\gd$ refers to the conformally compactified metric), the full expression, including the overall factor of $\tilde{\boldsymbol{\epsilon}}$, scales as $\mathcal O(\Omega)$ and therefore vanishes at $\scrip$. As before, the inclusion of further derivatives only strengthens this suppression.

Finally, consider mixed terms containing more than two derivatives acting on both scalar and metric fields. At the level of the symplectic potential, these again give rise to contributions of order $\mathcal O(\nabla^3)$. Analog arguments to the two previous cases then lead to the same conclusion, i.e., the combinations of metric tensors and scalar fields inevitably introduce an $\Omega$--scaling ensuring that the resulting terms vanish upon pullback to $\scrip$.

One is thus led to the conclusion that, because of particular behavior of the metric tensor in the conformal compactification and its limit to $\scrip$, additional metric contractions necessary in higher-order operators yield a vanishing contribution of the corresponding terms to the final flux formula in the WZ approach. Following the general discussion above, it then follows directly that the terms in $L_5$ of Eq.~\eqref{eq:fullHorndeski}, as well as the contributions derived from them--such as the double-dual Riemann-type couplings--cannot contribute to the flux at $\scrip$. The same applies to terms proportional to $G_{4,\hat X}$ in said Lagrangian. The conjecture can therefore be regarded as confirmed.

A natural extension of this analysis would be to construct a similar heuristic argument for theories that include an additional vector field, for example scalar–vector–tensor (SVT) theories. Such a generalization, however, entails essential differences, including the admissible number of derivatives, the asymptotic behavior of the fields, and the structure of the associated Killing vectors. These issues require a separate and careful treatment, and we therefore defer a discussion of the full SVT case to future work.

\section{Discussion and Conclusion} 
\label{sec:discuss}

Our main results can be summarized as follows. Using the covariant phase space approach, we explicitly establish a supermomentum flux balance law for a large subclass of Horndeski theories. We further decompose the supermomentum flux into integrable charges and establish a direct connection to displacement memory. A consistency check is provided by explicitly relating our results to gravitational memory computations within the alternative Isaacson framework, as well as through a comparison with previous results obtained in the Brans–Dicke subclass. Finally, we present a conjecture for the flux balance laws of the full Horndeski theory class and provide a tentative ``proof'' thereof.

Future space-based detectors will provide high-precision measurements of gravitational waveforms from across the cosmos. In this new observational regime, interesting effects such as gravitational memory \cite{Inchauspe:2024ibs,Zosso:2026czc,Cogez:2026frh} or echoes \cite{Deppe:2024fdo} may become accessible, ultimately allowing for stringent constraints on modifications of GR. The equations derived in this work find direct application in this context: they enable the systematic incorporation of memory effects into beyond GR gravitational waveform models \cite{Mitman:2020bjf} and provide quantitative diagnostics for the validation of associated numerical waveform simulations \cite{Ashtekar:2019viz,DAmbrosio:2024zok}. By translating rigorous analytical balance laws into concrete constraints on waveforms, this work lays the groundwork for exploiting future gravitational-wave data as robust tests of GR and its extensions.

At a technical level, we have shown that even in the presence of non-minimal couplings between the metric and scalar field operators, the supermomentum flux balance laws separate into two distinct sectors: a GR-like contribution associated with the physical Jordan-frame metric, and an additional luminal Horndeski contribution. This separation arises from the observation that, upon pullback to null infinity $\scri$, the symplectic potential generically admits such a decomposition, with all cross-terms vanishing systematically. As a result, one obtains both a purely Horndeski supermomentum flux and a corresponding supermomentum charge. These additional contributions admit a clear physical interpretation. The scalar field directly contributes to the radiated energy flux through combinations of the Horndeski functions $G_2$, $G_3$, and $G_4$. In contrast, the Horndeski contribution to the change in supermomentum charge, $\Delta \mathcal Q_{\alpha(\theta,\phi) n}$, computed as the difference of two $2$-sphere integrals on distinct cross-sections of $\mathcal{I}^+$ at retarded times $u_1$ and $u_2$, encodes bulk dynamics that permanently modify the scalar field configuration, which additionally balances the emitted energy flux. In the limit $u_{2,1}\to\pm\infty$, the higher-harmonic contribution of the scalar flux and the change in supermomentum charge induce an additional null, respectively ordinary tensor memory component in the metric strain. The latter is sourced by a lasting offset in the asymptotic scalar field value itself. In scalar–tensor theories with sufficiently strong non-minimal coupling between the scalar field and the physical metric, this scalar field offset further induces a direct scalar memory effect in the scalar polarization of the asymptotic radiation \cite{du_gravitational_2016,heisenberg2025unifyingordinarynullmemory}.

In order to place these flux balance computations in a broader theoretical context, we additionally discussed in Sec.~\ref{subsec:comparison_Isaacson} the correspondence between the fully nonlinear covariant phase space approach employed here and the perturbative Isaacson framework. This comparison not only serves as a non-trivial consistency check, but also points toward a deeper conceptual connection between Hamiltonian methods in covariant phase space and the interpretation of energy–momentum fluxes as leading-order quantities determined entirely by the second-order Lagrangian of a gravitational theory \cite{Heisenberg:2023prj}. From the Isaacson point of view, a separation into distinct GR and scalar field sectors is therefore a direct consequence of the diagonalizability of the linear equations of motion. An important open question that emerges from this comparison concerns the precise role of scalar-induced ordinary memory within the Isaacson formalism, which remains to be fully clarified.

Guided by the comparison and explicit computations of the flux laws for sGB gravity, we formulated a conjecture extending our result to the full class of Horndeski theories described by the Lagrangian \eqref{eq:fullHorndeski}. A tentative justification was obtained by examining the general structure of the terms that may appear in the symplectic potential and current of an arbitrary scalar–tensor theory. This analysis led us to conclude that contributions involving more than two derivatives at the level of the Lagrangian--apart from scalar-field derivatives entering through the Horndeski functionals $G_i$--cannot survive the pullback to $\scrip$ within the adopted framework of asymptotically flat spacetimes and the WZ approach. Consequently, on the basis of the heuristic arguments presented above, we have strong grounds to regard the conjecture as valid also definite proof remains to be provided.

Finally, this work provides a practical roadmap for computing constraint equations beyond GR at the level of the full, nonlinear theory. We have emphasized an intuitive understanding of the covariant phase space formalism and demonstrated its concrete implementation in theories beyond GR through explicit calculations in luminal Horndeski gravity. In particular, we showed that the asymptotic BMS symmetry structure is preserved despite the failure of the standard peeling theorem. As a consequence, many results derived in GR can be directly repurposed for a broad class of modified gravity theories, provided a suitable notion of asymptotic flatness is maintained. An interesting direction for future work is especially the extension of this analysis to theories involving gravitational vector fields \cite{Heisenberg:2023prj,Heisenberg:2025tfh,Heisenberg:2025roe}.

\section*{Acknowledgements}
We thank Fabio D'Ambrosio for countless discussions on the generalization of balance flux laws, and for arousing our interest in the approach by Wald \& Zoupas utilized in this work. We thank Francesco Gozzini for useful discussions on the covariant phase space formalism. We further thank Fabio D’Ambrosio, Shaun Fell, Lavinia Heisenberg, Francesco Gozzini and Stefan Zentarra for valuable conversations on asymptotic symmetries and BMS flux balance laws.
DM is supported by the Deutsche Forschungsgemeinschaft (DFG,
German Research Foundation) under Germany’s Excellence Strategy EXC 2181/1 - 390900948 (the Heidelberg STRUCTURES Excellence Cluster). JZ is supported by funding from the Swiss National Science Foundation
(Grant No. 222346) and the Janggen-Pöhn-Foundation. The Center of Gravity is a Center of Excellence funded by the Danish National Research Foundation under Grant No. 184.


\appendix

\numberwithin{equation}{section}

\section{General Relativity \`a la Wald--Zoupas}
\label{app:GR_mod}
In this appendix, we sketch the computation of the WZ flux $\bT$ for GR and highlight steps where modifications to Einstein's theory become relevant:

Applying the notation of Sec.~\ref{sec:intuition} and \ref{sec:Horndeski}, we consider the Lagrangian 4-form that contains a Einstein-Hilbert term
\begin{align}\label{equ:EH_wald}
    \boldsymbol{L}_\text{tot}\supseteq\boldsymbol{L}_\text{GR} = \frac{1}{16 \pi} \hat R \,\,{}^{(4)}\hat {\boldsymbol\epsilon}\,.
\end{align}
Beyond this term, the total Lagrangian $\boldsymbol{L}_\text{tot}$ may contain additional (pseudo) tensors constructed out of the metric or (non-)minimal coupling to a set of matter fields $\hat \phi \in \hat \psi = \{\hat g_{\mu\nu}, \hat \phi\}$. We denote them collectively as $\boldsymbol{L}_\text{+}=\boldsymbol{L}_\text{+}(\hat\phi)$. Considering the generalized Lagrangian $\boldsymbol{L}_\text{tot}= \boldsymbol L _\text{GR}+\boldsymbol{L}_\text{+}$, one finds its variation to result in
\begin{align}
    \boldsymbol E(\hat\psi) &= \boldsymbol E_\text{GR}(\gd)+ \boldsymbol E_\text{+}(\hat\psi)\,, \label{equ:eoms_1} \\ \bt(\hat\psi) &=\bt_\text{GR}(\gd, \delta\gd) + \bt_+(\hat\psi,\delta\hat\psi)\,.\label{equ:eoms_2} 
\end{align}
The boundary term $\bt_\text{GR}(\gd, \delta\gd)$ directly results from variation of Eq.~\eqref{equ:EH_wald} while all other terms encompass beyond-GR terms, including $\boldsymbol E_\text{GR}(\gd)$ which denotes Einstein's equation (in vacuum), $\boldsymbol E_\text{Einstein}(\gd)$, with an additional stress-energy tensor, $ \boldsymbol T_\text{matter}[\hat \psi]=-2/\sqrt{-g}\,\partial(\sqrt{-g}\,\boldsymbol{L}_+)/\partial g^{\mu\nu}$, and corrections, $\boldsymbol{E}_\text{corr}(\hat \psi)$, due to the presence of higher-order derivative interactions of the metric besides the Einstein-Hilbert term \eqref{equ:EH_wald}, $ \boldsymbol E_\text{GR}(\hat{\psi})\simeq \boldsymbol T_\text{matter}[\hat \psi] +  \boldsymbol E_\text{Einstein}(\gd)+\boldsymbol{E}_\text{corr}(\hat \psi)$.\footnote{Note that as long as \eqref{equ:EH_wald} can be isolated in the Lagrangian, so can the right-hand side of Einstein's equations, i.e., $\boldsymbol E_\text{Einsteins} = R_{\mu\nu}-\frac{1}{2}\gd R$.} In the absence of such metric interactions, the boundary term $\bt_+$ does not depend on metric perturbations $\bt_+(\hat \psi, \delta  \hat\phi)$. Regarding the equations of motion, Eq.~\eqref{equ:eoms_1}, the additional contribution $\boldsymbol E_\text{+}(\hat\psi)$ contains solely the equation of motion of the remaining degrees of freedom (the remaining fields), i.e., $ \boldsymbol E_+(\hat{\psi})\simeq  \boldsymbol E_\text{matter}(\hat\psi)$ and is obtained by the respective variation of the Lagrangian $\delta_{\hat\phi}\boldsymbol{L}_+$.

Let us first focus on the plain GR flux, $\bT_\text{GR}$, assuming that $\boldsymbol{L}_\text{tot}= \boldsymbol L _\text{GR}$. One can straightforwardly compute the symplectic potential $\bt_\text{GR}$ as [cf. Eq. (44) in \cite{waldGeneralDefinitionConserved2000}]
\begin{align}\label{equ:idk_KKKK}
    \theta^\text{GR}_{\mu\nu\rho} = \frac{1}{16\pi}\hat\epsilon_{\sigma\mu\nu\rho}\hat g^{\mu\alpha}\hat g^{\beta\gamma}(\hat \nabla_\beta \delta \hat g_{\alpha\gamma} - \hat \nabla_\alpha \delta \hat g_{\beta\gamma})\,,
\end{align}
and the corresponding current $3$-form $\bw_\text{GR}$, i.e., 
\begin{align}\label{equ:symp_current_GR}
    \omega^\text{GR}_{\mu\nu\rho} = &\frac{1}{16\pi}\hat\epsilon_{\sigma\mu\nu\rho}P^{\sigma\alpha\beta\gamma\delta\tau}\nonumber\\&\times(\delta_2\hat g_{\alpha\beta}\hat \nabla_\gamma \delta_1 \hat g_{\delta \tau} - \delta_1\hat g_{\alpha\beta}\hat \nabla_\gamma \delta_2 \hat g_{\delta \tau} )\,,
\end{align}
where 
\begin{align}
    P^{\alpha\beta\gamma\delta\mu\nu} = &\hat g^{\alpha\mu} \hat g^{\nu\beta}\hat g^{\gamma\delta}-\frac{1}{2}\hat g^{\alpha\delta}\hat g^{\beta\mu}\hat g^{\nu\gamma}-\frac{1}{2}\hat g^{\alpha\beta}\hat g^{\gamma\delta}\hat g^{\mu\nu}\notag\\-&\frac{1}{2}\hat g^{\beta\gamma}\hat g^{\alpha\mu}\hat g^{\nu\delta}+\frac{1}{2}\hat g^{\beta\gamma}\hat g^{\alpha\delta}\hat g^{\mu\nu}\,.
\end{align}
We continue by computing the pullback $\newpb{\bw}_\text{GR}$ of the symplectic current to the desired hypersurface, $\scrip$. We apply the convention (note especially the notation of different metrics) and coordinate choices of Subsec.~\ref{subsec:null_infty_shit}. For better readability, we restate relevant identities from \ref{subsec:null_infty_shit}. As we make heavy use of the tensor $\tau_{\mu\nu}$ defined via $\delta \gd = \Omega \tau_{\mu\nu}$, where $\tau_{\mu\nu}$ extends smoothly to $\scrip$ and is non-vanishing, it is worth to also define its trace as $\tau=\tau\ud{\mu}{\mu}$. In addition to the identities derived in \ref{subsec:null_infty_shit}, one further needs that $n_{(\mu}\tau_{\nu)\rho}n^\rho|_{\scrip}=0$ and $\tau_{\nu\rho}n^\rho= \Omega \tau_\nu$ where $\tau_\nu$ is smooth and non-vanishing at $\scrip$, implying that $\delta n^\mu = -\Omega^2\tau^\mu $ (while $\delta n_\mu=0$). Using $\boldsymbol{\epsilon}= \Omega^4\hat{\boldsymbol{\epsilon}}$, where $\epsilon_{\mu\nu\rho\sigma}= 4 \epsilon_{[\mu\nu\rho}n_{\sigma]}$, and $\hat \nabla_\mu \delta \hat g _{\nu\rho} = \nabla_\mu \delta g_{\nu\rho} + 2 C\ud{\sigma}{\mu(\nu}\delta g_{\rho)\sigma}$ with $C\ud{\rho}{\mu\nu}= 2\Omega^{-1}\delta^\rho_{(\mu}n_{\nu)} - \Omega^{-1}n^\rho g_{\mu\nu}$, a tedious calculation leads to \cite{waldGeneralDefinitionConserved2000}\footnote{The identities used in the derivation do generally hold only for GR in asymptotically Minkowskian spacetimes and are of purely geometrical nature. They are, therefore, independent of the precise nature of the Lagrangian of the underlying theory \textit{as long as} the underlying metric theory obeys the definition given in Subsec.~\ref{subsec:null_infty_shit}, concretely, the asymptotic metric fall-off conditions. While we discuss this requirement for luminal Horndeski theory in detail in Appendix \ref{app:asymptotic_Horn}, we henceforth assume that it is satisfied for $\boldsymbol{L}_\text{tot}= \boldsymbol{L}_\text{GR} + \boldsymbol{L}_+$.}
\begin{align}\label{equ:pullback_symp_current_GR}
    \newpb{\bw}_\text{GR} = -\frac{1}{16\pi} \Omega^{-4}n_\mu w^\mu{}^{(3)}\newpb{\boldsymbol{\epsilon}}\,,
\end{align}
where ${}^{(3)}\newpb{\boldsymbol{\epsilon}}$ establishes a positively orientated pullback of the volume form to $\scrip$ and 
\begin{align}\label{equ:pullback_symp_current_GR_I}
    \Omega^{-4}n_\mu w^\mu|_{\scrip} = \frac{1}{2}&\left([\tau'^{\nu\rho}n^\mu\nabla_\mu \tau_{\beta \rho} + \tau' n^\mu\nabla_\mu \tau_1 + \tau' n^\mu \tau_{\mu}]\notag\right. \\
    &+\left. [\delta \leftrightarrow \delta']\right)\,.
\end{align}
In Eq.~\eqref{equ:pullback_symp_current_GR_I}, the $\tau'^{\mu\nu}, \tau'$ refer to $\delta'\gd$ respectively. 

To connect quantities appearing in Eq.~\eqref{equ:pullback_symp_current_GR} with radiative degrees of freedom of the metric sector $\boldsymbol{L}_\text{GR}$, one uses the well-established connections between Ricci, Schouten, and Bondi News tensor (see \cite{DAmbrosio:2022clk, Maibach:2025iku} for exhaustive reviews). Concretely, Eq.~\eqref{equ:pullback_symp_current_GR_I} is simplified by an additional set of identities obtained from the Schouten tensor. The latter can be computed once using its definition in terms of the Ricci tensor, $S_{\mu\nu}= R_{\mu\nu}-\frac{1}{6} R \gd$, with
\begin{align}
    \delta R_{\mu\nu}|_{\scrip} = -n_{(\mu}\nabla_{\nu)}\tau - n^\rho \nabla_\rho \tau_{\mu\nu}+n_{(b}\nabla^\rho \tau_{\nu)\rho} + n_{(\mu}\tau_{\nu)}\,,
\end{align}
and once using the conformal transformation behavior of $\hat R_{\mu\nu}$, 
\begin{align}\label{equ:ricci_conformal}
    \hat R_{\mu\nu} =  &R_{\mu\nu} + 2 \Omega^{-1} \nabla_\mu\nabla_\nu\Omega+\Omega^{-1}\gd \nabla^\rho\nabla_\rho \Omega\notag\\
    &-3\Omega^{-2}\gd n^\rho \nabla^\rho \Omega \nabla_\rho \Omega\,.
\end{align}
To compute the Schouten tensor based on \eqref{equ:ricci_conformal}, one inserts the equations of motions \eqref{equ:eoms_1} to replace $\hat R_{\mu\nu}$, leading to
\begin{align}\label{equ:important_Sab}
    \delta S_{\mu\nu}|_{\scrip} = 4n_{(\mu}\tau_{\nu)} - n^\alpha \nabla_\alpha\tau_{\mu\nu} - n^\alpha \tau_\alpha g_{\mu\nu}\,,
\end{align}
in the absence of $\boldsymbol{L}_+$. Comparing the latter with 
$\delta S_{\mu\nu}|_{\scrip}$ computed based on \eqref{equ:ricci_conformal}, one can infer new identities for $\tau_{\mu\nu}$ that help putting \eqref{equ:pullback_symp_current_GR_I} into a more instructive format. Concretely, in the vacuum case, $\boldsymbol E(\phi)=\boldsymbol{E}_\text{GR} = \hat R_{\mu\nu}=0$, one finds $(\nabla^\nu \tau_{\mu\nu}-\nabla_\mu \tau-3\tau_\mu)|_{\scrip}=0$ and $(n^\nu\nabla_\nu \tau + 2n^\nu\tau_\nu)|_{\scrip}=0$. Thus, one rewrites Eq.~\eqref{equ:pullback_symp_current_GR} as
\begin{align}
    \newpb{\bw}_\text{GR} = -\frac{1}{32\pi}[\tau_2^{\mu\nu}\delta_1N_{\mu\nu}-\tau_1^{\mu\nu}\delta_2 N_{\mu\nu}]\,,
\end{align}
where a replacement of the Schouten tensor by Bondi news tensor is justified in \cite{waldGeneralDefinitionConserved2000}. This result suggests that a GR-flux at null infinity, $\scrip$, is given by
\begin{align}\label{equ:last_step}
    \bT_\text{GR} =  -\frac{1}{32\pi}\tau^{\mu\nu}N_{\mu\nu} {}^{(3)}\boldsymbol{\epsilon}\,.
\end{align}
Let us now extend the theory as aforementioned to $\boldsymbol{L}_\text{tot}= \boldsymbol L _\text{GR}+\boldsymbol{L}_\text{+}$ such that $\boldsymbol{E}_\text{GR}$ may contain contributions form other fields $\hat \phi$ as well as the corrections to Einstein's equations, if higher derivatives are present in the Lagrangian besides $\boldsymbol{L}_\text{GR}$.
In this case, the flux, $\bT_\text{GR}$, is modified as well and is no longer given by Eq.~\eqref{equ:last_step}. 

We note that for the GR flux, namely Eq.~\eqref{equ:last_step} inserted into Eq.~\eqref{equ:Jann_Lfux}, there are multiple equivalent representation established in the literature. Following for instance Sec. 2.4.3 in \cite{Maibach:2025iku}, the supertranslation flux can be purely expressed in terms of Bondi News $N_{\mu\nu}$, i.e., 
\begin{align}\label{equ:some_equ_GR}
    &\mathcal F_{\alpha(\theta,\phi) n}[\Delta\scrip] =\nonumber\\ &-\frac{1}{2\kappa_0} \int_{\Delta \scrip} \alpha (\theta,\phi)\bigg(\frac{1}{2} N_{\mu\nu}N^{\mu\nu} + D_\mu D_\nu N^{\mu\nu}\bigg)\, {}^{(3)}\boldsymbol \epsilon\,.
\end{align}
Similar statement is true for the associated charge. The latter is, however, not fully described by the Bondi news tensor. Instead it is composed of the shear tensor, here denoted as $\sigma_{\mu\nu}$, and the second Newman Penrose scalar $\Psi_2$,
\begin{align}\label{equ:charge_GR_II} 
    Q^\text{GR}_{\alpha(\theta,\phi) n} [\partial \scrip] =& \frac{1}{2\kappa_0} \oint_{u=u_0}\dd ^2\boldsymbol \Omega \alpha(\theta,\phi)\text{Re}\Big[\Psi_2 + \bar \varsigma \dot \varsigma\Big]
\end{align}
where the shear can be defined in terms of Bondi News as $N_{\mu\nu} =  \lie_n \sigma_{\mu\nu}=-(\Dot{\bar \sigma} m_\mu m_\nu + \Dot{\sigma}\bar m_\mu\bar m_\nu)$, where $m_\mu,\Bar{m}_\nu$ are the frame fields spanning the $2$-sphere metric at $\scrip$. A detailed derivation and thorough explanation of involved quantities is provided, e.g., in \cite{Maibach:2025iku}. 

\subsubsection{Absence of higher derivative metric interactions}
First, consider the case where the Lagrangian solely contains terms depending on other fields and no higher derivative metric couplings are present aside from the Einstein-Hilbert action.\footnote{An instance of such theory is coincidentally provided by Einstein-frame Lagrangian \eqref{equ:einstein_frame_lagrangian}.} Then, $\boldsymbol E_\text{GR}(\hat{\psi})$ can be written as $\boldsymbol E_\text{GR}(\hat{\psi})\simeq \boldsymbol T_\text{matter}[\hat \phi] +  \boldsymbol E_\text{Einstein}(\hat\psi)$. The Schouten tensor, in this case, becomes
\begin{align}\label{equ:Shouten_EST}
    S_{\mu\nu}=& -2\Omega^{-1}\nabla_\mu n_\nu + \Omega^{-2}n_\alpha n^\alpha g_{\mu\nu} \nonumber\\
    &+ 8\pi\Omega^2 \left( T^\text{matter}_{\mu\nu}-\frac{1}{3}g_{\mu\nu}g^{\alpha\beta} T^\text{matter}_{\alpha\beta}\right)\,,
\end{align}
which results from substituting $\boldsymbol E_\text{GR}(\hat{\psi})\simeq \boldsymbol T_\text{matter}[\hat \phi] +  \boldsymbol E_\text{Einstein}(\gd)$ into Eq.~\eqref{equ:ricci_conformal} instead of just $  \boldsymbol E_\text{Einstein}(\gd)$. The factor of $\Omega^2$ arises due to $ T^\text{matter}_{\mu\nu} := \Omega^{-2}\hat{ T}^\text{matter}_{\mu\nu}$. If $ T^\text{matter}_{\mu\nu}$ is now smooth and finite at $\scrip$ (given the smoothness of the involved field content), any contribution from $ T_{\mu\nu}$ in Eq.~\eqref{equ:Shouten_EST} trivializes in the limit to $\scrip$, when $\Omega \rightarrow0$. 

For the luminal Horndeski theory, for instance, this is certainly the case. We note here, that strictly speaking, to bring the Lagrangian considered in the main text to the form $\boldsymbol{L}_\text{tot}= \boldsymbol L _\text{GR}+\boldsymbol{L}_\text{+}$, one has to switch to the Einstein frame. Yet, the argument still applies as transforming between Einstein and Jordan frame corresponds to shifting a scaling function within Einstein-equations if the rescaling factor $\Sigma$ has a smooth inverse and is finite at $\scrip$, such that $\widetilde g_{\mu\nu}= \Sigma ^2 g_{\mu\nu}$, yields $\widetilde T_{\mu\nu}=\Sigma^{-2}T_{\mu\nu}$. Thus, the Schouten tensor remains unaffected by the additional stress-energy tensor provided it is finite at $\scrip$.

It can be concluded that for beyond-GR theories for which the Einstein-Hilbert Lagrangian can be isolated (even if it requires an Einstein-frame transformation), such that $\boldsymbol L_\text{tot} = \boldsymbol{L}_{GR}+\boldsymbol{L}_+$, where $\boldsymbol{L}_+$ does not contain metric derivatives, the results for the flux associated to $\boldsymbol{L}_{GR}$ carry over from \cite{waldGeneralDefinitionConserved2000} unchanged. The total flux contains the GR part and a term describing the flux due to additional (dynamical) fields, i.e., 
\begin{align}
    \bT_\text{tot} = \bT_\text{GR} + \bT_+ \,.
\end{align}

\subsubsection{Including higher derivative metric interactions}
Including higher-order metric derivative terms reduces the predictability of resulting flux. The boundary term $\bt_+$ may obtain an additional contribution which does not only depend on field perturbations $\delta \hat \phi$ but includes metric perturbations as well. If they survive the pullback to $\scrip$, these novel terms may appear in the total flux. The precise nature of their physical intuition can thereby only be speculated. We note, however, that various higher metric derivative interaction terms commonly used in literature, such as the Gauss-Bonnet scalar or the double dual Riemann tensor, do not admit a non-trivial pullback.

This, however, is not the only modification to the previous case. While we can still isolate the GR-like boundary term, $\bt_\text{GR}$, its associated flux, $\bT_\text{GR}$, may be additionally modified by virtue of the Schouten tensor (as demonstrated above), which itself depends on the modified Einstein's equations $ \boldsymbol E_\text{GR}(\hat{\psi})\simeq \boldsymbol T_\text{matter}[\hat \psi] +  \boldsymbol E_\text{Einstein}(\gd)+\boldsymbol{E}_\text{corr}(\hat \psi)$. The term $\boldsymbol{E}_\text{corr}(\hat \psi)$ thereby accounts for the modifications due to higher metric derivative terms.

Finally, we remark that in either of the above cases \textit{a.} or \textit{b.}, it remains to check is whether the flux contribution $\bT_+$ is affected by metric perturbations. These will manifest in the beyond-GR flux if cross terms of the symplectic current, $\bw_\times(\hat \psi, \delta \hat \psi)$, have non-vanishing pullback to $\scrip$. These cross terms appear even if, as in case \textit{a.}, the boundary term $\bt_+(\hat \psi, \delta  \hat\phi)$ does not depend on metric perturbations. For a concrete example, we refer to the luminal Horndeski theory treated in the main text. 

\section{Assumptions on Horndeski functionals}\label{app:assumptionsonG}
In this Appendix, we derive constraints for the Horndeski functionals $G_i$ to assure well-definedness of relevant symplectic quantities in the asymptitic limit.

The Horndeski functionals $G_i$ in Eq.~\eqref{equ:general_form} are, by construction, such that their dependence on the scalar field $\hat \Phi$ and or its derivatives (in form of $\tilde X$) does not break Lorentz invariance. Yet, their explicit functional dependence is often left unspecified in literature although constrained experimentally \cite{ATLAS_2019, Lombriser:2015sxa, Bettoni:2016mij,Creminelli:2017sry,Sakstein:2017xjx}. In this work, we adapt a similar standpoint and do not restrict the exact functional form of the $G_i$'s with only two crucial exceptions: First, to ensure masslessness of $\hat \Phi$, we constrain the derivative of $G_2$ (see main text below Eq.~\eqref{equ:general_form}). Second, to ensure the finiteness of symplectic quantities in the computations of the main text, the functional dependence of the $G_i$'s with respect to $\tilde{X}$ is limited. 

Why this necessity arises becomes immediately clear if one considers the asymptotic limit to $\scrip$: For large radii (or equally small $\Omega$, where $\Omega\sim 1/r$ is the conformal factor), $\tilde{X}$ scales at least as $\Omega^3$. Thus, one may encounter divergencies if $G_i(\tilde X) \sim \mathcal O(\tilde{X}^{-\epsilon})$ with $\epsilon>0$ are allowed in the limit $\Omega \rightarrow 0$. We, thus, assume that for all functions $G_i(\tilde{X})$ the first derivatives are free of poles in the limit $\Omega\rightarrow0$, i.e., they behave as 
\begin{align}\label{equ:assumption}
    G_{i\tilde{X}} = \frac{\partial}{\partial \tilde{X}}G_i\sim \mathcal{O}(\tilde{X}^\varepsilon)&&\text{with}&&\varepsilon\geq0\,,
\end{align} 
for $\tilde{X}\rightarrow0$. With condition \eqref{equ:assumption} one can readily check that all terms part of the symplectic current and potential $\bw$, $\bT$ computed in the main text remain finite in the limit to $\scrip$.

We can now proceed and ask whether similar assumptions are necessary for the functionals' dependence on the scalar field $\hat \Phi$. Ignoring for a second the masslessness assumption for $\hat \Phi$, it is equally desirable to avoid divergences of the symplectic quantities in the limit $\hat \Phi \rightarrow 0$ for the scalar field dependence of the Horndeski functions. Whether or not this limit is relevant is, however, more subtle since the asymptotic value of the scalar field depends sensitively on the structure of the scalar potential encoded in $G_2(\hat \Phi,\tilde X)$, in particular on whether the scalar field is massless or massive. 

In the massless case, one may expand the scalar field as $\hat \Phi = \varphi_0 + \Omega \varphi_1 + \mathcal{O}(\Omega^2)$, where the constant $\varphi_0$ can be set to zero in the absence of a potential term.\footnote{Note that this expansion is connected to assuming that the field is regular at $\scrip$ which relies heavily on the smoothness of null infinity itself. We highlight, that in the mathematical community, the latter aspect has caused some debate (see for instance \cite{Kehrberger_2021} and following works).}
For instance in GR coupled to a massless scalar field, the action is independent of $\varphi_0$ as in the Lagrangian the scalar $\hat \Phi$ appears only with a derivative. Thus, this constant may therefore be chosen to vanish without loss of generality, since neither the stress-energy tensor nor the equations of motion are affected. A similar situation arises in scalar Gauss–Bonnet gravity, where $\varphi_0 = 0$ is then often imposed for convenience.
By contrast, in the presence of a non-minimal coupling between the scalar field and the Ricci scalar, as it is the case for the Lagrangian \eqref{equ:general_form}, i.e., a term such as $\hat \Phi R$, setting $\varphi_0 = 0$ has physical consequences, although its precise numerical value remains irrelevant.\footnote{The constant part $\varphi_0$ establishes an overall factor, see for instance the results of this work, which does not affect physical predictions as these rely on relative values.}
Therefore, it is reasonable to assume $\varphi_0 \neq 0$ in luminal Horndeski theory. Given a non-trivial lowest order expansion of $\hat \Phi$, it immediately follows that, no matter the functional dependence, $G_i(\hat \Phi)$ remains finite since $\hat\Phi\rightarrow\varphi_0\neq 0$ close to null infinity $\scrip$. Thus, potential divergences in the Horndeski functionals $G_i$ are, by definition of the scalar field expansion, automatically circumvented. 

For a massive scalar field, an expansion as for the massless case with a non-trivial constant value $\varphi_0$ is more troublesome. After all, it is well-established that massive modes do not propagate to null infinity by virtue of the dispersion relation. Further, a non-trivial $\varphi_0$ field value for a scalar with a ``classical'' mass term $\sim m^2 \hat \Phi^2$ would certainly violate the finite energy condition. One could however imagine a non-trivial potential giving rise to an effective mass that vanishes asymptotically, resulting in a massless scalar in the vicinity of $\scrip$. A particular realization of this scenario would be given by an effective mass coupled to the vacuum expectation value (VEV) $\varphi_0^*$ for a self-interacting scalar field. If the VEV is such that the mass vanishes at large radii close to $\scrip$, the scalar field becomes effectively massless and the massless case discussed above applies. Since this scenario is rarely discussed in the literature, we do not pursue it further here and instead restrict our attention to theories with massless scalar fields although the assumption of a massless scalar field is not strictly required in the presented analysis.

\section{Asymptotic flatness in luminal Horndeski theory}
\label{app:asymptotic_Horn}
In this Appendix, we focus on the asymptotic flatness of the considered luminal Horndeski theory \eqref{equ:general_form}. Concretely, we sketch the derivation of asymptotic behavior of the individual metric components by which the asymptotic structure of the metric unfolds. This step is crucial for the analysis of the main text as we make heavy use of identities derived under the premise of asymptotic flatness of $\gd$.

\subsection{Asymptotic flatness in Bondi coordinates}
\label{app:Bondi_coord}
It is a well-established \cite{bondiGravitationalWavesGeneral1962, Barnich:2009se} that, near null infinity and using retarded Bondi coordinates\footnote{Technically, Bondi coordinates (or Bondi-Metzner-Sachs coordinates) refer to the choice $(u,r,y_1,y_2)$ at $\scrip$ where the tetrad is adapted to the null geodesic congruence of $\ell^\mu\partial_\mu=\partial_r$ such that $n^\mu\partial_\mu = \partial_u$ and $q_{\mu\nu}=\newpb{\gd}$ is the unit $2$-sphere metric.} $(u,r,y_1,y_2)$, the line element can be written as \cite{Maibach:2025iku}
\begin{align}\label{equ:le_metric}
    d s^2 = &-\frac{V}{r}e^{2\beta}d u^2-2e^{2\beta}dudr\nonumber\\&+r^2h_{AB}(dy^A- U^Adu)(dy^B- U^Bdu)\,,
\end{align}
where $A,B\in \{1,2\}$ index the angular sector of the metric, and $V, \beta, U_A, h_{AB}$ are functions of $u,r, y^A$. Gauge conditions can be imposed by $g_{rr}=0, g_{rA}=0$ and $\det(h_{AB})=q(x^C)$, where $q(x^C)$ is the determinant of the metric on the 2-sphere, where $h_{AB}=\gamma_{AB}+\mathcal{O}(1/r)$. The free functions are determined by Einstein's equations, which in the case of luminal Horndeski theory amount to
\begin{align}\label{equ:New_Einstein}
    E_{\mu\nu}:=\hat R_{\mu\nu}-\frac{1}{2}\hat g_{\mu\nu}\hat R - \underbrace{\frac{\kappa_0}{G_4(\hat \Phi)}\hat{\mathcal{T}}_{\mu\nu}}_{:= \kappa_0 \hat \Sigma_{\mu\nu}}=0\,,
\end{align}
with $\hat{\mathcal{T}}_{\mu\nu}$ denoting the effective scalar field stress-energy tensor. For computational ease, we absorbe the prefactor $1/G_4(\hat\Phi)$ and define a new tensor $\hat \Sigma_{\mu\nu}$.

Let us now try to solve \eqref{equ:New_Einstein} with ansatz \eqref{equ:le_metric} to the degree that we obtain the scaling behavior of the metric components at large radii.
Using the contracted Bianchi identities one can separate Einsteins equations into \textit{hypersurface} and \textit{evolution equation} \cite{M_dler_2016}, 
\begin{align}\label{equ:BMS_equations}
    E_{u\mu}=0\,,&& E_{AB}-\frac{1}{2}g_{AB}E\ud{\mu}{\mu}=0\,.
\end{align}
The first set of equations can be used to determine the functions $\beta, U^A, V$ via $E_{ur}, E_{rA},E_{AB}$ respectively. The evolution equation determines the retarded time derivative of the two degrees of freedom of the conformal 2-metric $h_{AB}$. To estimate how the presence of the effective scalar field stress-energy tensor $\hat \Sigma_{\mu\nu}$ affects the asymptotic behavior of the metric tensor, we trace back $\hat \Sigma_{\mu\nu}$'s contribution to the individual metric functions by virtue of Einstein's equation \eqref{equ:New_Einstein}.\footnote{We focus on the stress-energy tenor from the scalar field alone, which can be explicitly computed. If one would include an unspecified matter, additional conditions regarding its asymptotic scaling in $1/r$ need to be imposed, see, e.g.,  \cite{hou_conserved_2021, Flanagan_2017}.} To assure asymptotic flatness in the sense that the metric approaches the Minkowski metric in the Bondi frame at null infinity, $\scrip$, one requires
\begin{align}\label{equ:metric_decay}
    \lim_{r\rightarrow \infty}\beta = \lim_{r\rightarrow \infty} U^A = 0\,,&& \lim_{r\rightarrow \infty}\frac{V}{r}=1\,, &&\lim_{r\rightarrow \infty}h_{AB}= \gamma_{AB}\,,
\end{align}
where $\gamma_{AB}$ is the unit 2-sphere metric. Inserting \eqref{equ:metric_decay} into \eqref{equ:le_metric}, one can straightforwardly show that it approaches a Minkowskian spacetime for large $r$. Thus, if the scalar field stress-energy does not obscure this decaying behavior, the metric satisfies the flatness-at-null-infinity condition and various identities derived from this property.

\subsubsection{Asymptotic metric decay}
Solving \eqref{equ:BMS_equations} for the metriv functions $V,\beta,U,h_{AB}$, we need to compute $\hat \Sigma_{\mu\nu}$ explicitly. In the Jordan frame, the stress-energy tensor is given by (using the notation of the main text)
\begin{widetext}
    \begin{align}\label{equ:stress_energy_Horndeski}
         \hat \Sigma_{\mu\nu}=& \frac{1}{G_4(\hat \Phi)}\Bigg[-\frac{1}{2}\left(G_2(\hat \Phi, \tilde  X)+ (G_{3\hat \Phi}(\hat \Phi, \tilde  X)\hat \nabla_\alpha\hat \Phi - G_{3\tilde X}(\hat \Phi, \tilde  X)\hat \nabla_\alpha\hat \nabla_\beta \hat \Phi \hat \nabla^\beta \hat \Phi)\hat \nabla^\alpha \hat \Phi\right)\gd-\frac{1}{2}G_{2\tilde X}(\hat \Phi, \tilde  X)\hat \nabla_\mu\hat \Phi \hat \nabla_\nu \hat \Phi\nonumber \\&+ G_{3\hat \Phi}(\hat \Phi, \tilde  X)\hat \nabla_\mu\hat \Phi\hat \nabla_\nu\hat \Phi-G_{3\tilde X}(\hat \Phi, \tilde  X)\bigg[ \hat \nabla_\mu\hat \nabla^\alpha\hat \Phi \hat \nabla_\alpha \hat \Phi \hat \nabla_\nu \hat \Phi + \hat \nabla_\alpha \hat \nabla_\mu \hat \Phi \hat \nabla_\nu \hat\Phi \hat \nabla^\alpha \hat \Phi\bigg] + \gd \bigg[G_{4\hat\Phi\hat\Phi}(\hat \Phi)\hat \nabla_\alpha \hat \Phi \hat \nabla^\alpha\hat \Phi\nonumber \\& + G_{4\hat\Phi}(\hat \Phi) \hat \nabla_\alpha \hat \nabla^\alpha \hat \Phi\bigg] + G_{4\hat\Phi\hat\Phi}(\hat \Phi)\hat \nabla_\mu\hat\Phi\hat \nabla_\nu\hat\Phi + G_{4\hat \Phi}(\hat \Phi)\hat \nabla_\mu\hat \nabla_\nu\hat\Phi\Bigg]\,.
    \end{align}
\end{widetext}
Inserting \eqref{equ:stress_energy_Horndeski} into Eq.~\eqref{equ:New_Einstein}, we can solve the corresponding equations order by order in $1/r$ by using an expansion ansatz for scalar field\footnote{Note that in case of mass terms or other relevant potentials being present in $G_2$, the expansion behavior has to be adapted accordingly, see discussion in the main text.} and $\gamma_{AB}$, namely 
\begin{align}
    \hat \Phi = \varphi_0 + \frac{\varphi_1}{r} + \mathcal{O}\left(\frac{1}{r^2}\right)\,, &&h_{AB}=\gamma_{AB}+\mathcal{O}\left(\frac{1}{r}\right)\,.
\end{align}

For an the scalar field stress-energy tensor $\hat\Sigma_{\mu\nu}$, the $E_{ur}$ component reads\footnote{Technically, the computation can be done in the Einstein frame first. The Jordan frame result then follows from transforming the Einstein-frame-solution with an adequate transformation, see \cite{tahura_brans-dicke_2021}. Here, however, we continue with the Jordan frame.}
\begin{align}
    \partial_r\beta = \frac{r}{16}h^{AC}h^{BD}\partial_rh_{AB}\partial_r h_{CD}+ 2\pi r \hat\Sigma_{rr}\,,
\end{align}
which can be solved to obtain $\beta$ in terms of $\hat \Phi$, its derivatives, and $h_{AB}$. Solving the latter order by order, we find a first contribution of the scalar field at $1/r$. Applying the constraints on the Horndeski functionals \eqref{equ:assumption} postulated in Appendix \ref{app:assumptionsonG}, it appears that terms proportional to $\hat\nabla_\mu\hat\nabla_\nu\hat\Phi$, computed using the Christoffel tensor associated with metric \eqref{equ:le_metric}, can result in a novel $1/r$ term for $\beta$, compared to the GR result. This extra term appears if the prefactor of $\hat\nabla_\mu\hat\nabla_\nu\hat\Phi$ is such that it does not contribute any additional powers of $1/r$. Otherwise, it is shifted to a higher order equation, and the GR result for $\beta$ is recovered at least to order $1/r^2$. Concretely, for terms such as $G_i\hat\nabla_\mu\hat\nabla_\nu\hat\Phi$, with $G_i=G_i(\tilde X)$, the first contribution form the scalar field arises only at $\mathcal O(1/r^2)$ if $G_i(\tilde X)\sim\tilde X^n $, for $n\geq1$. Taking for instance Brans-Dicke theory, where $G_4\sim \hat \Phi$ and therefore $\sim \tilde X^0$, we see that the last term in \eqref{equ:stress_energy_Horndeski} is proportional to $\hat\nabla_\mu\hat\nabla_\nu\hat\Phi$ without any additional power of $1/r$. Thus, in this theory, $\beta=\mathcal O(1/r)$ instead of $\mathcal O(1/r^2)$, confirm \cite{hou_conserved_2021} (see also \cite{Tahura_2025} for similar computations).

Once $\beta$ is computed, we can continue with the $E_{Ar}$ component of Einstein's equations and find an expression for $U^A$ in terms of the same quantities by solving
\begin{align}
    \partial_r \left(r^4 e^{-2\beta}h_{AB}\partial_r U^B\right)&=2r^4\partial_r \left(\frac{1}{r^2}D_A\beta\right)\nonumber \\&-r^2h^{EF}D_E\partial_rh_{AF}+16\pi r^2 \hat\Sigma_{rA}\,,
\end{align}
where $D_A$ is the covariant derivative with respect to $h_{AB}$.
Upon inserting \eqref{equ:stress_energy_Horndeski} and checking the highest order in $1/r$, one finds that $U^A=\mathcal O (1/r^2)$, just as in GR, independent of the choice of functions $G_i$ within the constraints \eqref{equ:assumption}.

Given these results, there are multiple ways to determine $V$. We choose to compute the trace of the $E_{AB}$ equations. Compared to GR, the additional contribution to $E\ud{A}{A}$ from the scalar field is given by a term $\sim h_{AB}\hat\Sigma^{AB}$. One finds that this term contributes at most at $\mathcal O(r)$ such that the GR decay behavior $V=\mathcal O(r)$ is recovered. 

We conclude that, given the assumptions outlined in this appendix and the main text, the luminal Horndeski metric satisfies the metric conditions \eqref{equ:metric_decay} which qualifies it to be asymptotically Minkowkian (at null infinity). On a superficial level, there is thus no obstruction to introduce the null tetrad, foliate $\scrip$ and parametrize its cross sections as it has been done in the main text. The choice of the conformal metric and its perturbation is thus also justified and ensures the validity of the computations presented in this analysis above. 

\subsubsection{Shear component and its correction}
As aforementioned, all metric components retain their asymptotic scaling necessary to diagnose asymptotic flatness, including the angular part $h_{AB}$. In GR and to first order in $1/r$, it is well-established that this metric component carries the radiative information of the metric tensor (see explicitly \cite{Maibach:2025iku}). What has been overlooked so far in this appendix is that in beyond-GR theories, the other sectors can ``source'' additional degrees of freedom in the metric (in the Jordan frame), manifesting by corrections appearing in the first order term. From a perturbative point of view, this is a well-understood phenomena and one finds additional metric polarizations being sourced through, for instance, non-minimal coupling to a scalar field in Brans-Dicke theory. For the full theory, one can equally show that the shear of the metric obtains corrections by non-minimally coupling fields. At the heart of this conjecture lies the realization that the shear tensor on $\scrip$, $\sigma_{AB}$, is given by the trance-free part of 
\begin{align}
    \newpb{\nabla_\mu\ell_\nu}
\end{align}
defined based on the Newman-Penrose null tetrad. Explicit computation then demonstrates that $\sigma_{AB}\sim c_{AB}$ (Sec. 2.3.1 in \cite{Maibach:2025iku}) where $h_{AB}= \gamma_{AB} + \frac{1}{r}c_{AB} + \mathcal O (\frac{1}{r^2})$. Thus, if the metric carries additional radiative degrees of freedom, they must appear in $c_{AB}$. 

Let us now explicitly focus on luminal Horndeski theory: To obtain $c_{AB}$ explicitly, we can circumvent the full computation of all metric components and instead consider only the gauge conditions on the angular part. In GR, the gauge choice affecting $h_{AB}= \gamma_{AB} + \frac{1}{r}c_{AB} + \mathcal O (\frac{1}{r^2})$ amounts to $\det(h_{AB})= \sin{\theta}^2$. This remains true even in the presence of an additional field as long as it is minimally coupled. Thus, for the Einstein frame metric $\tilde g_{\mu\nu}$ of the main text, we can impose $\det(\tilde h_{AB})= \sin^2{\theta}$. Switching to the Jordan frame now amounts to the replacement $\tilde g_{\mu\nu} = G_4 (\hat \Phi) g_{\mu\nu}$ where $\gd$ is the conformally completed Jordan frame metric. Similar statement is true for the angular components. Thus,
\begin{align}\label{equ:condition_metric}
    \det(\hat h_{AB}) &= \sin^2{\theta}\, G_4(\hat\Phi)^{-2}\notag\\
    &=\frac{1}{\bar G_4}\sin^2{\theta}\left(1-2\frac{\bar G_4’ \varphi_1}{\bar G_4}\frac{1}{r} + \mathcal{O}\left(\frac{1}{r^2}\right)\right)\,,
\end{align}
where we used that, by the assumptions made in Appendix~\ref{app:assumptionsonG}, $G_4$ is regular at $\scrip$ and can be expanded as 
\begin{align}\label{equ:expansion_G_4}
    G_4 = \bar G_{4}+ \frac{\bar G_{4\hat \Phi} \varphi_1}{r} + \mathcal O (r^{-2})\,.
\end{align}
For Eq.~\eqref{equ:condition_metric} to be true, one finds that 
\begin{align}
    h_{AB} &= \gamma_{AB}\left(1- \frac{1}{r}\frac{\bar G_4’ \varphi_1}{\bar G_4}\right) + \frac{1}{r}c_{AB} + \mathcal{O}\left(\frac{1}{r^2}\right)\notag\\
    &= \gamma_{AB} + \frac{1}{r}\left(c_{AB}- \gamma_{AB}\frac{\bar G_4’ \varphi_1}{\bar G_4}\right)+ \mathcal{O}\left(\frac{1}{r^2}\right)\,.
\end{align}
Thus, the radiative degrees of freedom of the metric are now captured by
\begin{align}\label{equ:crucial}
    \sigma_{AB}\sim c_{AB}- \gamma_{AB}\frac{\bar G_4’ \varphi_1}{\bar G_4}\,,
\end{align}
The scalar thus mixes into the propagating degrees of freedom of the metric tensor, where $c_{AB}$ are the two ``pure GR'' ones. For the computation in the main text, Eq.~\eqref{equ:crucial} is crucial to obtain a correct result for the ordinary (or linear) GW memory. We emphasize at this point, the a trivial $G_4$ implies no mixing, thus, the latter is a direct ramification of the non-minimal coupling in luminal Horndeski theory.

\subsection{Killing vector fields at $\scrip$}

Finally, let us comment on the validity of the chosen Killing vector fields, \eqref{equ:real_st} and \eqref{equ:real_L}, in luminal Horndeski theory. 

With the heuristic derivation of the asymptotic condition for the metric sector above, we established that the metric functions $V,\beta,U_A,h_{AB}$ follow the same asymptotic behavior as in GR, the only exception being $\beta$. This decay, Eq.~\eqref{equ:metric_decay}, can be translated into the asymptotics of individual metric components, i.e., 
\begin{align}\label{equ:Killing_cond}
    g_{uu}=1+\mathcal{O}(1/r)\,, && g_{ur}=-1+\mathcal O(1/r)\,,&& g_{uA}= \mathcal{O}(1)\nonumber \\
    g_{rr }= g_{rA}=0\,,&& h_{AB}= \gamma_{AB}+\mathcal{O}(1/r)\,,
\end{align}
where novel terms in $\beta=\mathcal{O}(1/r)$ lead to $g_{ur}=-1+\mathcal O(1/r)$ instead of $g_{ur}=-1+\mathcal O(1/r^2)$, as in GR. 

For GR, the Killing vector fields of the asymptotic metric preserve this decay behavior of the metric components. This is analogous to preserving the universal structure of the metric, which was historically first used to define the asymptotic symmetry group (the BMS group) \cite{gerochAsymptoticStructureSpaceTime1977}. One can show \cite{Maibach:2025iku}, that the universal structure follows directly from writing the metric as \eqref{equ:le_metric} and imposing fall of conditions such as \eqref{equ:Killing_cond}.

For luminal Horndeski theory, we impose the same requirement: Killing vectors of the metric sector are required to preserve \eqref{equ:Killing_cond} but for the new decay behavior of the $g_{ur}$ component. By explicit computation one finds that the scaling of $g_{ur}$ up to $\mathcal O(1/r)$ is irrelevant for the Killings equations and \eqref{equ:real_st} and \eqref{equ:real_L} (more precisely, their continuation off $\scrip$) are recovered as Killing fields of luminal Horndeski theory as well. Those vectors leave the asymptotic conditions \eqref{equ:Killing_cond} intact (see \cite{hou_conserved_2021} for analogous computations). It can therefore be concluded that the presence of the scalar field in luminal Horndeski theory, given its asymptotic expansion, preserves the universal structure of the metric if the pure metric sector is itself asymptotically Minkowskian as defined in, e.g., Sec.~2.1.3 of \cite{Maibach:2025iku}.

\section{Brans-Dicke gravity as a concrete instance of luminal Horndeski theory}
\label{sec:BDT}
To provide an additional consistency check, we fall back on a well-studied instance of luminal Horndeski theory, i.e., Brans-Dicke gravity (\cite{Brans:1961sx, Dicke:1961gz}; see \cite{tahura_brans-dicke_2021, koyama_testing_2020, tahura_gravitational-wave_2021, Hawking:1972bb, hou_gravitational_2021_2, hou_gravitational_2021, hou_conserved_2021, Heisenberg:2023prj} for more recent analyses). For Brans-Dicke theory, both the WZ formalism \cite{hou_conserved_2021} as well as perturbative approaches \cite{Heisenberg:2023prj} have been applied to analyze the theory. It, thus, establishes a suitable candidate for cross checking the computed flux law formulated for the general class of luminal Horndeski theories. 

Brans-Dicke theory is recovered from the general Lagrangian \eqref{equ:general_form} by choosing
\begin{align}\label{equ:Brans_Dicke_selection}
    G_2 &= \frac{2\omega }{\hat \Phi} X\,,\notag\\
    G_4 &=\hat \Phi\,,\notag\\
    G_i &= 0\,\,\, \text{for }i\neq 2,4\,.
\end{align}
The resulting theory is summarized in
\begin{align}
    {L}^\text{BD} = \frac{1}{2\kappa_0}\left(\hat \Phi  R - \frac{\omega}{\hat \Phi} g^{\mu\nu}\nabla_\mu\hat \Phi\nabla_\nu \hat \Phi\right)\,.
\end{align}
While for computing the flux, we could technically just use the end result, Eq.~\eqref{equ:super_flux}, for completeness we go through the full derivation as for the general theory: 

We start by applying similar rescaling as for Horndeski Gravity, i.e., \begin{equation}
     g_{\mu\nu}(x)=\frac{1}{{\hat \Phi}}\tilde{g}_{\mu\nu}(x)\,,
\end{equation}
to find, in the Einstein frame, 
\begin{align}\label{equ:Brans_Dicke_conformal}
    \tilde {L}^\text{BD} = \frac{1}{2\kappa_0}\left(\tilde  R - \frac{2\omega+3}{2\hat \Phi^2} \tilde g^{\mu\nu}\tilde \nabla_\mu\hat \Phi\tilde \nabla_\nu \hat \Phi\right)\,.
\end{align}
Just as for luminal Horndeski theory, the Lagrangian in the Einstein frame therefore separates into the rescaled Ricci scalar and a ``matter'' part depending on the scalar fields. Varying the Lagrangian \eqref{equ:Brans_Dicke_conformal}, one obtains the symplectic potential 
\begin{align}
    \tilde \bt_\text{BD}=\tilde \epsilon_{\mu\alpha\beta\gamma} \delta \hat \Phi \tilde \nabla^\mu \hat \Phi\Bigg[-\frac{2\omega +3}{\hat \Phi^2}\Bigg]\,.
\end{align}
For the current, it follows that that
\begin{align}
    \tilde \bw_\text{BD}+ \tilde \bw_\times&=\tilde \epsilon_{\mu\alpha\beta\gamma} \Bigg[-\frac{2\omega + 3}{\hat \Phi^2}\Bigg]\Bigg(\frac{1}{2}\tilde g^{\sigma \rho}\tilde g^{\mu\lambda} \tilde\nabla_\lambda \hat \Phi \delta'\tilde g_{\sigma \rho}\delta \hat \Phi \nonumber\\&+ \tilde\nabla_\alpha\hat \Phi \delta \hat \Phi \delta' \tilde g^{\alpha \mu}+ \tilde g^{\alpha\mu}\tilde\nabla_\alpha\delta \hat \Phi\delta'\hat \Phi \Bigg) - \braket{\delta'\leftrightarrow\delta}
\end{align}
Here, note that terms proportional to $\delta \hat \Phi\delta'\hat \Phi$ vanish because of the anti-symmetry of the current with respect to the variations. Such a term arises for instance when varying $\hat \Phi^{-2}$ cancel out.

We proceed by showing that the pullback of $\tilde \bw_\times$ vanishes at $\scrip$. 
\begin{align}\label{equ:current_limit_BD_cross}
    \tilde\bw_\times= &-\epsilon_{\alpha\beta\gamma}\Omega^{-4}n_\mu\Bigg[-\frac{2\omega +3}{(\varphi_0 + \Omega\varphi_1)^2}\Bigg]\nonumber \\&\Bigg(\Omega^3\nabla_\alpha(\Omega\varphi_1)\delta'g^{\alpha\mu}\delta\varphi_1\nonumber \\&+\Omega^3 \frac{1}{2}g^{\sigma\rho}g^{\mu\lambda}\nabla_\lambda(\Omega\varphi_1)\delta'g_{\sigma \rho}\delta \varphi_1\Bigg)-\braket{\delta'\leftrightarrow\delta}
\end{align}
where $\hat \Phi$ is expanded up to first order in $\Omega$ only, as higher order expansions immediately lead to a scaling of $\mathcal O(\Omega)$, which vanishes at $\scrip$. While $\delta \varphi_1$ and $g^{\mu\nu}$ are smooth at $\scrip$, for the metric perturbation we have  $\delta \gd\equaldot 0$ at $\scrip$ such that $\delta \gd = \Omega \tau_{\mu\nu}$ where $\tau_{\mu\nu}$ has a smooth limit to $\scrip$. Thus, what remains to show is that $\nabla_\mu(\Omega\varphi_1)\sim \mathcal O(\Omega)$. Indeed, $\nabla_\mu(\Omega\varphi_1) = \Omega \nabla_\mu \varphi_1 + n_\mu\varphi_1$ which, contracted with $g^{\mu\lambda}$ or $\delta g^{\mu\lambda}$ (as $\tau_{\mu\nu}n^\nu = \Omega \tau_\mu$) leads to $\tilde \bw_\times \sim \mathcal O(\Omega)$ which vanishes as $\scrip$. Therefore, we are concerned solely with 
\begin{align}\label{equ:current_limit_BD}
    \tilde\bw_\text{BD}= &\epsilon_{\alpha\beta\gamma}\Omega^{-4}n_\mu\Bigg[-\frac{2\omega +3}{(\varphi_0 + \Omega\varphi_1)^2}\Bigg]\nonumber \\&\Bigg(\Omega^3g^{\alpha\mu}\nabla_\alpha(\Omega\delta \varphi_1)\delta'\varphi_1\Bigg)-\braket{\delta'\leftrightarrow\delta}\,,
\end{align}
from which we obtain
\begin{align}
    \newpb{\tilde\bw}_\text{BD}=-\epsilon_{\alpha\beta\gamma}\Bigg[-\frac{2\omega+3}{\varphi_0^2}\Bigg]\left(\delta \varphi_1 \partial_u\delta'\varphi_1- \braket{\delta'\leftrightarrow\delta}\right)\,.
\end{align}
The WZ flux can be read off the latter equation and one finds
\begin{align}
    \tilde\bT_\text{BD}(\delta_\xi \varphi_1) = \epsilon_{\alpha\beta\gamma}\Bigg[\frac{2\omega+3}{\varphi_0^2}\Bigg]\left(\delta_\xi \varphi_1 \dot \varphi_1\right)\,.
\end{align}
Double-checking this result, it is apparent that the same result can be directly obtained from Eq.~\eqref{equ:flux_Horndeski} by inserting the explicit functions $G_i$ into $\bar \Xi[\varphi_0]$.

For a supertranslation Killing vector field, $\lie_{\xi_\text{ST}} \varphi_1 \equaldot \alpha(\theta,\phi)\dot \varphi_1$ at $\scrip$. Thus, the corresponding Brans-Dicke flux across a section $\Delta \scrip$ of $\scrip$ is, after back-transforming into the Jordan frame, 
\begin{align}\label{equ:half_result_BD}
     \mathcal{F}^{\text{BD}}_{\alpha(\theta,\phi) n} =-\frac{\varphi_0}{2\kappa_0} \int_{\Delta\scrip} \dd u \dd^2\boldsymbol\Omega\, \alpha(\theta,\phi) (2\omega+3)\left(\frac{\dot \varphi_1 ^2}{\varphi_0^2}\right)\,.
\end{align}
The flux formula \eqref{equ:half_result_BD} does only display the scalar contribution to the full flux. Again, we neglected, for now, the GR-like term in Lagrangian \eqref{equ:Brans_Dicke_conformal}. As elaborated in the previous section however, this term contributes the know flux of GR sketched in Appendix \ref{app:GR_mod}, here including a constant prefactor $\varphi_0$. Thus, for the full Brans-Dicke theory, we obtain
\begin{align}
    \mathcal{F}_{\alpha(\theta,\phi) n}(u_1,u_2)&\equiv -\,\frac{\varphi_0}{2\kappa_0} \int_{u_1}^{u_2}\dd u'\int_{S^2}\dd^2\boldsymbol\Omega\,\alpha(\theta,\phi)\Bigg[ \frac{1}{4}|\dot h|^2\notag\\&-2\text{Re}[\eth^2h] + \eth^2\frac{\dot\varphi_1}{\varphi_0}+(2\omega+3)\left(\frac{\dot\varphi_1}{\varphi_0}\right)^2\Bigg]\,,
\end{align}
where we inserted Eq.~\eqref{equ:result_tot_flux} for the GR sector. A derivation of the respective memory formula follows analogously the treatment in Sec.~\ref{sec:BL_all}. The result agrees with flux formulas computed by \cite{hou_conserved_2021} and recovers the same memory for Brans-Dicke theory as in \cite{hou_gravitational_2021_2, Heisenberg:2023prj}. 

\section{Asymptotics of the Gauss-Bonnet term}
\label{app:GB}
The goal of this Appendix is the explicit demonstration of how higher derivative terms involved in the Lagrangian fail to produce non-vanishing contributions to the overall flux in the WZ approach. To that end we consider sGB gravity. Concretely, we show that the coupling of a scalar field to the Gauss-Bonnet term does not affect the flux computation. Consider a coupling such as 
\begin{align}\label{equ:GB_term}
    \int d^4x \sqrt{-g}\,\left(\epsilon^2 f(\hat\Phi) \hat{\mathcal{G}}\right)\,,
\end{align}
where $f(\hat\Phi)$ is an arbitrary function. For a scalar defined as $\hat{\mathcal{G}} = \hat R^2 -4\hat R_{\mu\nu}\hat R^{\mu\nu} + \hat R^{\mu\nu\alpha\beta}\hat R_{\mu\nu\alpha\beta}$, the variation of the term \eqref{equ:GB_term} yields a symplectic potential 
\begin{align}
    \bt_\text{GB} = &\hat \epsilon_{\mu \eta\zeta\vartheta}f(\hat \Phi)\Bigg[\hat g^{\lambda \xi}\hat R_{\lambda \xi}\left(\hat g^{\mu\nu}\delta\hat\Gamma\du{\alpha\nu}{\alpha}-\hat g^{\nu\alpha}\delta\hat \Gamma\du{\nu\alpha}{\mu}\right)\nonumber\\
    &+\left(2\hat g^{\alpha\mu}\hat g^{\beta\nu}\hat R_{\alpha\beta}\delta \hat \Gamma\du{\lambda\nu}{\lambda}- 2\hat g^{\alpha \lambda}\hat g^{\beta\nu}\hat R_{\alpha\beta}\delta \hat \Gamma\du{\lambda\nu}{\mu}  \right)\nonumber\\
    &+\Big(2\hat g_{\sigma\lambda}\hat g^{\rho \nu}\hat g^{\mu\alpha}\hat g^{\delta\beta}\hat R\ud{\lambda}{\nu\alpha\beta}\delta \hat \Gamma\ud{\sigma}{\rho\delta}\nonumber\\
    &-2\hat g_{\sigma\lambda}\hat g^{\rho\nu}\hat g^{\gamma\alpha}\hat g^{\mu\beta}\hat R\ud{\lambda}{\nu\alpha\beta}\delta \hat \Gamma\ud{\sigma}{\rho\gamma}\Big)
    \Bigg]\,.
\end{align}
The symplectic potential above captures all boundary terms associated to the Gauss-Bonnet term. Note here that it only depends on variations of the metric as $f(\hat \Phi)$ does not depend on $\hat X$. One has to acknowledge, however, that by computing the symplectic current from $\bt_\text{GB}$ one obtains a cross term $\bw_\times$ and a ``pure'' metric contribution $\bw_\text{GB}$ from Eq.~\eqref{equ:GB_term}. The former contains variations of both metric and scalar field, while the latter only contains variations of the metric. Both vanish in the limit to $\scrip$ although not immediately apparent.

Explicitly inserting the variation of the Levi-Civita connection, the potential can be rewritten as 
\begin{align}
    &\bt_\text{GB} = \hat \epsilon_{\mu\eta \zeta\vartheta}f(\hat \Phi)\Bigg[\underbrace{\hat g^{\lambda \xi}\hat R_{\lambda \xi}\left(\hat \nabla_\nu \delta \hat g^{\mu\nu}-\hat g_{\sigma\alpha}\hat g^{\mu\beta}\hat \nabla_\beta\delta \hat g^{\sigma\alpha}\right)}_{=:\,\text{I}}\nonumber\\
    &+\underbrace{\Big( 4\hat g^{\alpha\sigma}\hat g^{\beta\nu} \hat R_{\alpha\beta} \hat g_{\lambda \sigma} \hat \nabla_\nu \hat g^{\lambda\mu}- \hat R_{\alpha\beta}\hat g^{\mu\lambda}\hat\nabla_\lambda\delta \hat g^{\alpha\beta}}_{=:\,\text{II}}\nonumber\\
    &\underbrace{-\hat g^{\alpha\mu}\hat g^{\beta\lambda}\hat R_{\alpha\beta} \hat g_{\sigma\nu}\hat \nabla_\lambda \delta \hat g^{\sigma \nu}\Big)}_{=:\,\text{II}}\nonumber\\
    &+\underbrace{\Big(2 \hat g^{\rho\nu}\hat g^{\mu\alpha}\hat g^{\delta \beta}\hat R\ud{\lambda}{\nu\alpha\beta}\left(\hat \nabla_\rho\delta\hat g_{\lambda\delta}+\hat\nabla_\delta \delta\hat g_{\lambda\rho}-\hat \nabla_\lambda \delta \hat g_{\rho\sigma}\right)\Big)}_{=:\,\text{III}}
    \Bigg]\,.
\end{align}
In the following, we abbreviate the term inside the large square brackets with $\theta^\mu_\text{GB}$. Computing the current, one then has $\bw_\text{GB}= \hat\epsilon_{\mu\eta \zeta\vartheta}f(\hat \Phi)\left(\frac{1}{2}\hat g^{\rho\sigma}\delta' \hat g_{\rho\sigma} \theta^\mu_\text{GB}+\delta'\theta^\mu_\text{GB}\right)-\braket{\delta'\leftrightarrow\delta}$ and $\bw_\times = \hat \epsilon_{\mu\eta \zeta\vartheta}f'(\hat \Phi)\delta'\hat \Phi\theta^\mu_\text{GB}-\braket{\delta'\leftrightarrow\delta}$.
When expressing the physical metric $\hat g_{\mu\nu}$ and its related physical tensors in terms of unphysical properties\footnote{To compute the scaling of $\hat R_{\mu\nu\alpha\beta}$, $\hat R_{\mu\nu}$, and $\hat R$ explicitly, we refer to their transformation behavior under Weyl rescaling, see for instance Appendix A in \cite{Maibach:2025iku}, in combination with the above notation of asymptotic quantities such as $n_\mu$ and $\tau_{\mu\nu}$.}, one can easily check that $\theta^\mu_\text{GB}$ scales as $\mathcal O(\Omega^4)$ while $\hat{\boldsymbol{\epsilon}}= \Omega^{-4}\boldsymbol{\epsilon}$. Thus, naively, non of the above currents vanishes. We observe however that in $\bw_\times$, the factor $f'(\hat \Phi)\delta\hat \Phi \sim \Omega$ and thus, this term vanishes at $\scrip$. Similarly, in $\bw_\text{GB}$, we have $\hat g^{\rho\sigma}\delta' \hat g_{\rho\sigma} \sim \Omega$ such that the term proportional to $\delta' \hat g_{\rho\sigma}$ vanishes was well. What is left to analyze is the contribution from $\delta \theta^\mu_\text{GB}$. Here, we go through the expression term by term. 

For I, if the variation acts on any metric, an additional power of $\Omega$ arises since $\delta g_{\mu\nu}\sim \Omega$. Thus, in these cases, the contribution from I to $\delta \theta^\mu_\text{GB}$ vanishes. For the variation acting on $\hat R_{\mu\nu}$, one finds a smooth and finit limit (compare Eq. (63) in \cite{waldGeneralDefinitionConserved2000}) and thus cannot use this argument again. Note however, that since $\epsilon_{\mu\alpha\beta\gamma}= 4 \epsilon_{[\alpha\beta\gamma}n_{\mu]}$, one has an extra factor of $n_\mu$ to contract with the metric variations. Since $ \nabla_\mu n_\nu|_{\scrip}=0$, $n_\mu n^\mu = \mathcal{O}(\Omega^2)$, and $n_\mu\delta \hat g^{\mu\nu}= \Omega^2 n_\mu \Omega \tau^{\mu\nu}=\Omega^4 \tau^\nu$, one finds that 
\begin{align}
   n_\mu \hat \nabla_\nu \delta \hat g^{\mu\nu}= \hat \nabla_\nu \Omega^4\tau^\nu = \Omega^3 n_\nu\tau^\nu + \Omega^4 \hat \nabla_\nu \tau^\nu
\end{align}
and 
\begin{align}
    &n_\mu\hat g_{\sigma\alpha}\hat g^{\mu\beta}\hat \nabla_\beta\delta \hat g^{\sigma\alpha} =  g_{\sigma\alpha}n^\beta\hat \nabla_\beta (\Omega^3 \tau^{\sigma\alpha} ) \nonumber \\ &= \Omega^3 g_{\sigma\alpha}n^\beta\hat \nabla_\beta \tau^{\sigma\alpha} + \Omega^2 g_{\sigma\alpha}\underbrace{n^\beta n_\beta}_{\sim \mathcal{O}(\Omega^2)}\tau^{\sigma\alpha}\,.
\end{align}
Both terms scaling at least with $\Omega^3$ in combination with $\hat g^{\lambda \xi}\hat R_{\lambda \xi}\sim \Omega^2$ and $\hat{\boldsymbol{\epsilon}}= \Omega^{-4}\boldsymbol{\epsilon}$, one thus finds that I scales at least as $\mathcal{O}(\Omega)$ and therefore is irrelevant for the flux computation. 

For II, one can use analogous arguments, i.e., $\delta g$ yielding an extra $\Omega$, to reason that every variation besides the one acting on the Ricci tensor vanishes. In those cases, the above computations allow us to ignore the first two contributions of II. In the last one, i.e., $-\hat g^{\alpha\mu}\hat g^{\beta\lambda}\delta\hat R_{\alpha\beta} \hat g_{\sigma\nu}\hat \nabla_\lambda \delta \hat g^{\sigma \nu}$, however, $n_\mu$ contracts with $\hat R_{\mu\nu}$. One finds that 
\begin{align}\label{equ:NR_I}
    &n_\mu\hat g^{\alpha\mu}\hat g^{\beta\lambda}\delta\hat R_{\alpha\beta} \,\hat g_{\sigma\nu}\hat \nabla_\lambda \delta \hat g^{\sigma \nu}= \Omega^2 g^{\beta\lambda}n^{\alpha}\delta R_{\alpha\beta}\, g_{\sigma\nu}\hat \nabla_\lambda (\Omega^3  \tau^{\sigma \nu})\nonumber \\ &=\Omega^5 g^{\beta\lambda}n^{\alpha}\delta R_{\alpha\beta} \,g_{\sigma\nu}\hat \nabla_\lambda \tau^{\sigma \nu} + \Omega^4 n^{\beta}n^{\alpha}\delta R_{\alpha\beta}\, g_{\sigma\nu}\hat \tau^{\sigma \nu}\,.
\end{align}
In the latter, the last term would have a smooth and finite limit to $\scrip$, one finds however that 
$n^\mu n^\nu \delta R_{\mu\nu}|_{\scrip} = 0$ since $\delta R_{\mu\nu}=-n_{(\mu}\nabla_{\nu)}\tau-n^\alpha\nabla_\alpha \tau_{\mu\nu}+n_{(\nu}\nabla^\beta \tau_{\mu)\beta} + n_{(\mu}\tau_{\nu)}$. The contraction with $n_\mu n_\nu$ yields $n^\mu n^\nu \delta R_{\mu\nu}|_{\scrip}$ (most of the resulting terms are proportional to $n_\alpha n^\alpha$) which is of order $\Omega$ and thus, the term in Eq.~\eqref{equ:NR_I} scales at least as $\Omega^5$. The full term II therefore decays towards $\scrip$ with $\mathcal{O}(\Omega)$ and thus does not contribute to the flux computations as well. 

Finally, similar discussion can be applied to III, where $\hat \epsilon_{\mu\eta\zeta\vartheta}\hat g^{\mu\alpha}\delta \hat R\ud{\lambda}{\nu\alpha\beta} \sim n^\alpha\delta  R\ud{\lambda}{\nu\alpha\beta}$ is the deciding factor. Here, it is trivial to show that $n^\alpha\delta  R\ud{\lambda}{\nu\alpha\beta} = n^\alpha\left(\nabla_\alpha \delta \Gamma\ud{\lambda}{\beta\nu}-\nabla_\beta\delta \Gamma\ud{\lambda}{\alpha\nu}\right)\sim \mathcal O (\Omega)$, and thus all variations of III vanish at $\scrip$. 

To summarize, we find that the sGB term has no relevant contribution to the flux when computed following the WZ prescription. 
\newpage

\bibliographystyle{apsrev4-2-no-names}
\bibliography{reference.bib}

\end{document}